\documentclass[fleqn,usenatbib]{mnras}

\usepackage[dvipsnames]{xcolor}

\usepackage[T1]{fontenc}
\usepackage{anyfontsize}
\DeclareRobustCommand{\VAN}[3]{#2}
\let\VANthebibliography\thebibliography
\def\thebibliography{\DeclareRobustCommand{\VAN}[3]{##3}\VANthebibliography}
\usepackage{graphicx}	
\usepackage{amsmath}	
\usepackage{amssymb}	
\usepackage{natbib}
\usepackage{tablefootnote}
\usepackage{booktabs} 
\usepackage{caption}
\usepackage{float}
\usepackage{comment}

\newcommand{\dcorr}{$f_{\Delta\nu}$}

\newcommand{\teff}{$T_{\mathrm{eff}}$}
\newcommand{\muhz}{$\mu$Hz}
\newcommand{\numax}{$\nu_{\mathrm{max}}$}

\newcommand{\dnu}{$\Delta\nu$}  

\newcommand{\msol}{M$_\odot$}
\newcommand{\rsol}{R$_\odot$}
\newcommand{\lsol}{L$_\odot$}
\newcommand{\kepler}{\textit{Kepler}}

\newcommand{\amlt}{$\alpha_{\mathrm{MLT}}$}
\newcommand{\dnusurf}{$\Delta\nu_{\mathrm{surf}}$}
\def\orcid#1{\kern .08em\href{https://orcid.org/#1}{\includegraphics[keepaspectratio,width=0.7em]{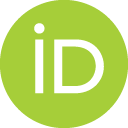}}}

\title[Asteroseismology of M67]{Asteroseismic Study of Subgiants and Giants of the Open Cluster M67 using Kepler/K2: Expanded Sample and Precise Masses.}
\author[C. Reyes et al.]{
Claudia Reyes$^{1,3}$\orcid{0000-0001-9632-2706}, 
Dennis Stello$^{2,1}$\orcid{0000-0002-4879-3519}, 
Marc Hon$^{4,5}$\orcid{0000-0003-2400-6960}, 
Yaguang Li$^{5}$\orcid{0000-0003-3020-4437}, 
Timothy R. Bedding$^{2}$\orcid{0000-0001-9632-2706}, \newauthor
Enrico Corsaro$^{6}$\orcid{0000-0001-8835-2075},
Lauren Taylor,
Andrew Vanderburg$^{4}$\orcid{0000-0001-7246-5438},
Eric Sandquist$^{7}$\orcid{0000-0003-4070-4881}, and
Robert D. Mathieu$^{8}$\orcid{0000-0002-7130-2757}.
\\
$^{1}$School of Physics, University of New South Wales, NSW 2052, Australia\\
$^{2}$Sydney Institute for Astronomy (SIfA), School of Physics, University of Sydney, NSW 2006, Australia\\
$^{3}$Research School of Astronomy \& Astrophysics, Australian National University, Canberra ACT 2611, Australia\\
$^{4}$ Department of Physics and Kavli Institute for Astrophysics and Space Research, Massachusetts Institute of Technology, 77 Massachusetts Ave, Cambridge, \\ MA 02139, USA.\\
$^{5}$Institute for Astronomy, University of Hawaii, 2680 Woodlawn Drive, Honolulu, HI 96822, USA\\
$^{6}$INAF – Osservatorio Astrofisico di Catania, Via S. Sofia, 78, 95123 Catania, Italy\\
$^{7}$Department of Astronomy, San Diego State University, San Diego, CA 92182, USA\\
$^{8}$Department of Astronomy, University of Wisconsin-Madison, 475 North Charter Street, Madison, WI 53706, USA\\
}

\usepackage{newtxtext,newtxmath}
\date{Accepted XXX. Received YYY; in original form ZZZ}

\pubyear{2024}

\begin{document}
\label{firstpage}
\pagerange{\pageref{firstpage}--\pageref{lastpage}}
\maketitle

\begin{abstract}
Sparked by the asteroseismic space revolution, ensemble studies have been used to produce empirical relations linking observed seismic properties and fundamental stellar properties. 
Cluster stars are particularly valuable because they have the same metallicity, distance, and age, thus reducing scatter to reveal smoother relations. 
We present the first study of a cluster that spans the full evolutionary sequence from subgiants to core helium-burning red giants using asteroseismology to characterise the stars in M67, including a yellow straggler.
We use Kepler/K2 data to measure seismic surface gravity, examine the potential influence of core magnetic fields, derive an empirical expression for the seismic surface term, and determine the phase term $\epsilon$ of the asymptotic relation for acoustic modes, extending its analysis to evolutionary states previously unexplored in detail. Additionally, we calibrate seismic scaling relations for stellar mass and radius, and quantify their systematic errors if surface term corrections are not applied to state-of-the-art stellar models. Our masses show that the Reimers mass loss parameter can not be larger than $\eta\sim0.23$ at the $2\sigma$ level.
We use isochrone models designed for M67 and compare their predictions with individual mode frequencies. We find that the seismic masses for subgiants and red giant branch stars align with the isochrone-predicted masses as per their luminosity and colour. However, our results are inconsistent with the mass of one of the stellar components of an eclipsing binary system near the TAMS. 
We use traditional seismic $\chi^2$ fits to estimate a seismic cluster age of $3.95 \pm\, 0.35\, \mathrm{Gyrs}$.
\end{abstract}


\section{Introduction}
Open star clusters are important laboratories for studying stellar astrophysics due to the shared properties among stellar members.
The open cluster M67 is particularly important because of its solar-like age and composition, making it ideal for exploring the physics of stars that are only somewhat different to the Sun, which is the anchor point of most stellar models. This cluster is well-populated, providing a diverse and extensive sample of stars to study, which allows for thorough testing of stellar evolution theory \citep{1993AJ....106.2441G}.

Asteroseismology, the study of stellar oscillations, has firmly established itself as a powerful tool to study stars \citep{2021RvMP...93a5001A}. The technique provides several advantages, such as the ability to quickly and accurately determine bulk properties, including surface gravity of large stellar samples across the Galaxy, which has provided a dramatic boost to the field of Galactic archaeology \citep{2016MNRAS.455..987C, 2021NatAs...5..640M, 2023MNRAS.524.1634S}. Additionally, it has been suggested that asteroseismology can provide evidence of magnetic fields in the cores of red giants. Many observed red giants exhibit a low visibility of dipole oscillation modes compared to radial modes. This "mode suppression" is theorised to result from the dissipation of dipole mode energy, caused by strong magnetic fields in the core. \citep{2015Sci...350..423F, 2016Natur.529..364S}.

However, asteroseismology is often equated to just the application of seismic scaling relations \citep{1991ApJ...368..599B, 1995A&A...293...87K}. 
These relations, anchored to the Sun, rely on only two seismic observables (\numax\ and \dnu) to describe the global characteristics of the oscillations, providing estimates of stellar radius and mass, and, with the aid of models, stellar age. Despite their widespread use, the accuracy of these scaling relations, which are partly empirical in nature, remains not fully established \citep{2011ApJ...743..143H, 2012MNRAS.419.2077M, 2013ASPC..479...61B, 2020FrASS...7....3H, 2018MNRAS.476.3729B}.


Beyond the scaling relations, the accuracy can be considerably improved, particularly of mass and age estimates, by measuring and performing detailed modelling of individual oscillation mode frequencies \citep{2020MNRAS.495.3431L, 2021A&A...645A..85M, 2022MNRAS.509.4344A}.  Such `full' asteroseismic analysis does, however, require corrections to the model frequencies because of improper prescriptions of the near-surface layers in current stellar models. As pointed out by 
\citet{2013ARA&A..51..353C}, these corrections should also be applied to the \dnu\ scaling relation. Empirical methods exist to perform these corrections on a star-by-star basis \citep{2014A&A...568A.123B}, but unless very good seismic data are available, the corrections can introduce star-to-star biases.
If these biases can be overcome, one interesting -- and still unexplored -- aspect of a full asteroseismic analysis is to investigate whether individual mode frequencies of both subgiants and red giants within a single stellar cluster can be consistently aligned with isochrone models that are matched to the classical observables.

\begin{figure*}
\includegraphics[width=\textwidth]{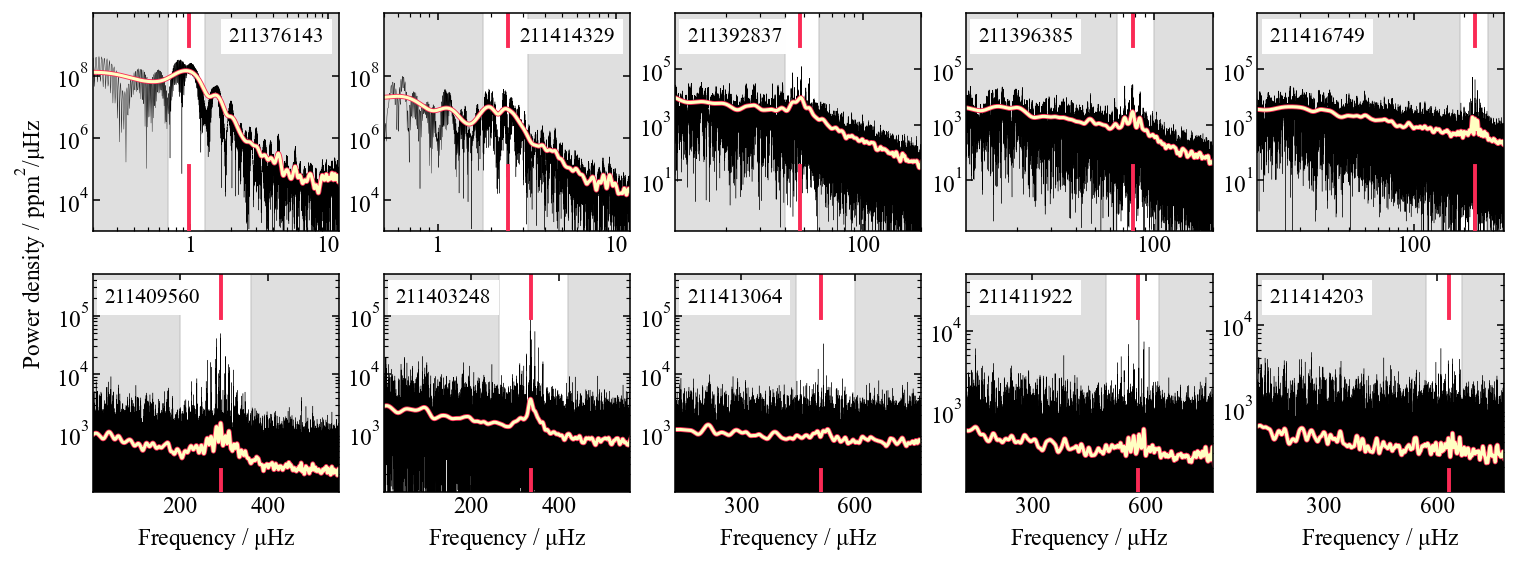}
\caption{Examples of the characteristic "near Gaussian"-shaped excess power observed in the power spectra of M67 stars from near the tip of the red-giant branch (EPIC 211376143) to the subgiant branch (EPIC 211414203). The yellow curves represent the smoothed power spectrum, the white area represents the portion of the spectrum where the excess power is located, and the red vertical lines indicate the frequency of maximum power, \numax.}
\label{fig:excesspower}
\end{figure*}

M67 presents a unique opportunity to investigate the key aspects highlighted above. In particular, 
(1) To determine if seismic $\log g$ aligns with state-of-the-art isochrones;
(2) To assess to what degree mode suppression in 
this cluster follows the field population and whether clump stars are more or less susceptible to mode suppression than red giant branch stars; 
(3) To measure individual mode frequencies to obtain a more robust surface correction that evolves smoothly, and map how the oscillation phase term $\epsilon$ (of the asymptotic relation for acoustic modes) evolves during the poorly studied transition between subgiants and red giants; 
(4) To use individual frequencies to calibrate the scaling relations used to determine mass and radius; and 
(5) To employ individual frequencies for isochrone fitting.

In this study, we tackle all these aspects. We build upon the asteroseismology-based work on M67 presented by \citet{2016ApJ...832..133S} and its follow-up study modelling three stars by \citet{2024MNRAS.530.2810L}. Both studies were based on single-campaign light curves from \textit{K2} \citep{Howell_2014} and focused solely on red giants. We expand on their research by including data from two additional \textit{K2} campaigns. The resulting improvement in signal-to-noise ratio led to the discovery of additional cluster members showing solar-like oscillations down to the subgiant branch and allowed us to identify individual frequencies for a large fraction of the seismic sample.
We use the results of \citet{2024MNRAS.tmp.1617R} (R24), particularly their M67-specific isochrone models, distance modulus, and membership sample. This incorporates four additional oscillating red giants into the cluster membership over the membership sample used by \citet{2016ApJ...832..133S} (The \citet{2015AJ....150...97G} sample). We focus on integrating seismic data from our extended sample with the isochrone models from R24 to derive highly precise stellar properties.
In the following sections, we detail our methodology, present our findings, and discuss their implications.

\section{Seismic analysis}
\label{sec:seis}

\subsection{Seismic data}
\label{sec:seismicdata} 
We performed seismic analysis using time series photometry from \textit{K2} campaigns 5, 16, and 18, 
although not all stars have data in all three campaigns. Most stars were observed in long-cadence mode ($\sim 30\,\mathrm{minutes}$) while the least evolved stars were observed in short cadence ($\sim 1\,\mathrm{minute}$). Long- and short-cadence light curves were created as described in \citet{2014PASP..126..948V, 2016ApJ...832..133S}, and \citet{2016ApJS..222...14V}, except for four stars: EPIC IDs 211443624, 211414351, 211384259, and 211406144, which we downloaded from MAST (\url{https://archive.stsci.edu/}) as PDC-SAP flux corrected lightcurves.

The lightcurve preparation was performed as in \citet{2016ApJ...832..133S} and included gap filling of up to three consecutive data points and high-pass filtering with a cutoff frequency of 3\muhz, except for stars with \numax~$<8$~\muhz.
We combined short cadence time series from different campaigns using statistical weights 
that take into account the varying data quality as a way to give lower weight to noisier parts of the time series. To calculate the weights, we used the rms point-to-point variability of the light curves in a 24-hour rolling window. We then scaled the weights to be consistent with the noise measured in the amplitude spectrum, as described in \citet{2006MNRAS.373.1141S}.
Next, we calculated the weighted power spectrum as described by \citet{1995A&A...301..123F} and \citet{2013PhDT.......408H}. 
For short-cadence data, the weighting scheme led to an average 12\% improvement in the signal-to-noise ratio in power. However, when we applied the same scheme to the long-cadence data, we did not see a significant improvement. We therefore decided to proceed without weighting the long-cadence lightcurves.

\begin{figure}
\includegraphics[width=\columnwidth]{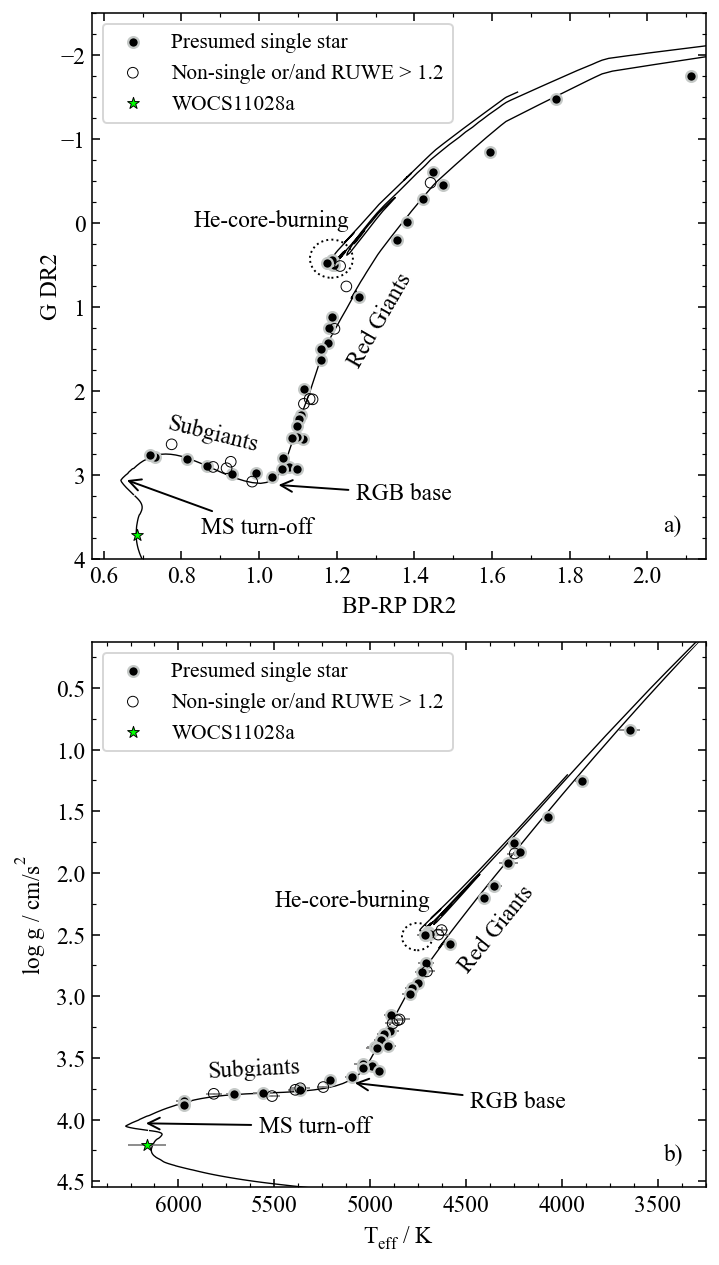}
\caption{Isochrone A and the 50 M67 stars for which we could measure \numax. Open circles show non-single stars and/or stars with $\mathrm{RUWE}>1.2$. The sample spans three evolutionary phases, which have been labelled on both panels. Additionally we show the primary component of the eclipsing binary WOCS1028. The arrows indicate the main sequence turnoff and the base of the red giant branch for reference in later figures. (a) M67 colour-magnitude diagram in {\it Gaia} DR2 absolute magnitudes after correcting for the distance modulus, extinction, and reddening, following R24.  
(b) Kiel diagram with seismic $\log g$ and \teff\ obtained using the iterative InfraRed Flux Method as described in Section \ref{sec:temperatures}.}
\label{fig:CMDKIEL}
\end{figure}

Upon analysing the oscillation spectra, we found that 51 stars showed solar-like oscillations, as indicated by a characteristic broad hump of excess power.
As expected \citep{1995A&A...293...87K}, the power excess is less pronounced in the highest \numax\ cases, leading to a reduced signal-to-noise ratio. Despite this, we remain confident in our detections, as the power excess appears at frequencies consistent with predictions based on the stars’ positions in the colour-magnitude diagram\footnote{Strictly speaking, in this work we only present colour-"absolute magnitude" diagrams, but for brevity, we refer to them simply as colour-magnitude diagrams.} 
Ten example stars, representative of the range of frequencies detected, are shown in \autoref{fig:excesspower}. In Section \ref{sec:peak}, we outline our detection criteria for individual mode identification.

Among the 51 detections is EPIC 211415650 (S1072 from the \citet{1977A&AS...27...89S} catalogue, or WOCS 2008), a known evolved blue straggler, also known as a yellow giant, which is a member of a binary system \citep{1986AJ.....92.1364M, 2008MNRAS.390..665L,  2015AJ....150...97G, 2018MNRAS.480.4314B}. Evolved blue stragglers can be the product of stellar mergers or of mass-transfer episodes due to Roche-lobe overflow of close binaries, and do not follow the same standard evolutionary path as most stars in the cluster. EPIC 211415650, in particular, has been suggested to have formed through Kozai cycles \citep{2009ApJ...697.1048P} in a hierarchical triple system \citep{2018MNRAS.480.4314B}. Therefore, we excluded EPIC 211415650 from the sample to be matched to our model isochrone, but we present its power spectrum and a complete seismic characterisation in Appendix \ref{sec:blue} for future reference.

We used \textsc{pySYD}\footnote{the python implementation of the \textit{SYD} pipeline \citep{2009CoAst.160...74H}.} \citep{2021arXiv210800582C} to measure \numax\ and \dnu\ for our remaining sample of 50 M67 stars that show solar-like oscillations. In the following, we present results based on those \numax\ and \dnu\ measurements.

\subsection{Temperatures and seismic log g}
\label{sec:temperatures} 

Figure~\ref{fig:CMDKIEL}a shows Isochrone A from R24 and {\it Gaia} DR2 photometry \citep{2018A&A...616A...1G} of the stars in our seismic sample. The observed magnitudes have been converted to absolute magnitudes using a distance modulus of 9.614 calculated from {\it Gaia} DR3 \citep{2023A&A...674A...1G} parallaxes with zero-point corrections from \citet{2021A&A...649A...4L}, and star-by-star corrections calculated from reddening dust maps \citep{2016A&A...594A..13P} (see R24).

To test whether the seismology agrees well with Isochrone A and therefore, with external observables of the cluster, we first calculated effective temperatures using the iterated InfraRed Flux Method (IRFM) and colour-$T_{\mathrm{eff}}$ relations from \citet{2021MNRAS.507.2684C} using visible and near-infrared photometry from {\it Gaia} DR2 and 2MASS \citep{2006AJ....131.1163S}. We adopted a metallicity of $\mathrm{[Fe/H]}=0.05$ dex for all stars, to be consistent with Isochrone A. We then derived seismic $\log g$ values using the expression $g \simeq g_{\odot}  (\nu_{\mathrm{max}}/\nu_{\mathrm{max},\odot}) \sqrt(T_{\mathrm{eff}}/T_{\mathrm{eff}, \odot})$ \citep{1991ApJ...368..599B, 1995A&A...293...87K}, 
with solar values of $\log g_{\odot}$ = 4.4383 \citep{2000asqu.book.....C} and $\nu_{\mathrm{max}, \odot}$ = 3090 \muhz, and a first-iteration \teff\ from APOGEE DR17. In cases where the APOGEE temperature was not available, we used the GALAH DR3 \citep{2021MNRAS.506..150B} or {\it Gaia} DR2 temperatures in the first iteration. 
We obtained stable temperatures after three iterations and minimised their difference to Isochrone A by applying a $-36.4 \mathrm{K}$ offset determined through least-squares optimisation.
It is not unusual to find systematic offsets of such magnitude when estimating effective temperatures. For example, \citet{2017MNRAS.472..979H} reported offsets of $+34, +15, \mathrm{and} -45\,\mathrm{K}$ between their IRFM temperatures of the open cluster NGC 6819 and the averages of three other studies.

We compare our seismic $\log g$ and IRFM \teff\ with Isochrone A in the Kiel diagram in Figure~\ref{fig:CMDKIEL}b, which shows a remarkable good agreement. We indicate non-single stars and stars with {\it Gaia} renormalised unit weight error (RUWE) exceeding 1.2 \citep{2019A&A...628A..35K} with an open symbol. Non-single stars also include those identified by \citet{2015AJ....150...97G} as binary, likely binary, stragglers, or possibly contaminated by a nearby source, and stars from the {\it Gaia} DR3 non-single star catalogue \citet{2022arXiv220605595G}. These stars are not necessarily expected to closely follow the isochrone sequence, although most of them do. Additionally we show the primary component of the eclipsing binary WOCS1028 \citep{2021AJ....161...59S}.
The temperatures and $\log g$, as well as \numax\ and \dnu\ for the full M67 sample, are presented in Table~\ref{tab:pysyd_teff}.

\subsection{Mode suppression}
\label{sec:sup}

Performing our initial seismic analysis, we found several stars whose $\ell$=1 and/or $\ell$=2 modes were undetectable or barely detectable. Therefore, we decided to measure their visibilities following \citet{2016Natur.529..364S}, who found that the prevalence of mode suppression, thought to be due to interior magnetic fields \citep{2015Sci...350..423F}, is a strong function of stellar mass. Because our stars are in a cluster, we can unambiguously identify which of them belong to the helium-core burning clump; a task that is challenging for field stars with suppressed dipole modes. This difficulty led \citet{2016Natur.529..364S} to exclude red clump stars and focus solely on red giant branch stars observed by \kepler\ in long-cadence mode. In contrast, we measured visibility in all stars within our seismic sample with $8.5<\nu_{\mathrm{max}}<600\,\mu \mathrm{Hz}$ as this frequency interval offered both the resolution and signal-to-noise ratio necessary for a reliable measurement of the phase term $\epsilon$ and therefore, a meaningful analysis of the visibilities. See Figures \ref{fig:suppression1} and \ref{fig:suppression2} for examples of $\ell$=1 suppressed and non-suppressed red-giant-branch stars in M67.

\begin{figure}
\begin{center}
    \includegraphics[width=\linewidth]{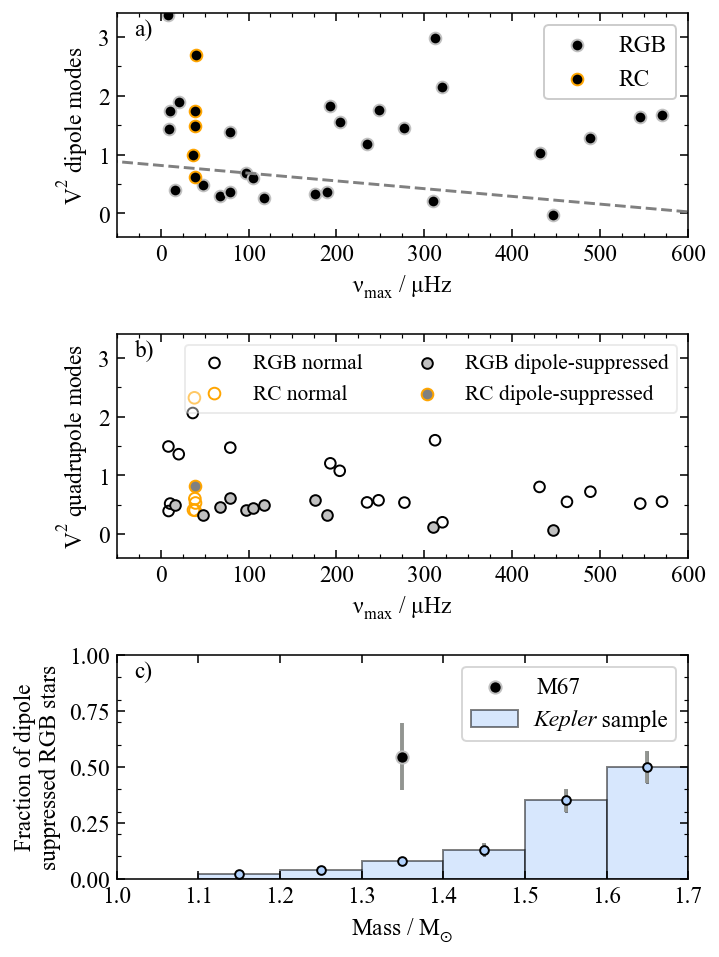}
    \caption{(a): The visibilities of dipolar ($\ell$=1) modes from the M67 seismic sample and the fiducial line from \citet{2016Natur.529..364S} that separates dipole-suppressed stars from normal visibilities stars. (b) The visibilities of quadrupole ($\ell$=2) modes where filled symbols represent the $\ell$=1 suppressed stars. (c): The black circle represents the fraction of M67 dipole-suppressed RGB stars with \numax\ between 70 and 250 \muhz, and the wide bars represent mass bins of comparable fractions from the \kepler\ field-stars \citep{2016Natur.529..364S}.}
    \label{fig:vis}
\end{center}
\end{figure}

\autoref{fig:vis}a shows $\ell$=1 visibilities from our M67 sample, where 11 of 29 red-giant-branch stars, and 1 of 7 red clump stars, are below the fiducial line from \citet{2016Natur.529..364S} that was found to separate the population of dipole suppressed red giants from the normal visibility stars. \autoref{fig:vis}b shows the $\ell$=2 visibilities, with an open symbol for stars with normal $\ell$=1 visibilities and a filled symbol for $\ell$=1 suppressed stars.
We found that $\ell$=1 suppressed red giant branch stars are more likely to also have lower $\ell$=2 visibilities than normal red giant branch stars, in agreement with \citet{2016PASA...33...11S}. We do not observe the same in red clump stars, but with only one dipole-suppressed star, the sample is too small to draw any conclusion in this regard. The low dipole suppression rate of $1/7$ that we observe in red clump stars seems to corroborate the "most likely scenario" proposed by \citet{2016ApJ...824...14C} for magnetic mode suppression in clump stars: that dipole suppression is rare in stars less massive than 2.1\msol\ because strong core magnetic fields tend not to survive helium flashes.

To be able to compare the $\ell$=1 suppression rate in M67 directly with that of the \kepler\ field stars measured by \citet{2016Natur.529..364S}, we selected only red-giant branch stars with $70<\nu_{\mathrm{max}}<250 \mu\mathrm{Hz}$. From this selection, 6 of the 11 stars were $\ell$=1 suppressed. This results in a fraction of $0.55 \pm 0.15$, with uncertainty following binomial statistics: $\sigma=\sqrt{p(1-p)/N}$, where $p$ is the fraction of dipole-suppressed stars and $N$ is the size of the sample (see \autoref{fig:vis}c). 
Our results tend to indicate that the suppression rate in M67 exceeds that found in comparable field stars. This remains true if we remove the two suppressed stars closest to the fiducial line in \autoref{fig:vis}a, ($0.44 \pm 0.14$), or if we count them as non-suppressed stars ($0.36 \pm 0.15$). It would be interesting to study other Galactic clusters to find whether suppression rates are larger in cluster stars and what the implication might be for the magnetic activity of stellar cores in a cluster environment.

\subsection{Individual radial mode frequencies}
\label{sec:peak} 
We expand the seismic analysis beyond just \numax\ and \dnu\ for both subgiant and red giant branch stars.
We identified individual radial mode frequencies using the \textsc{FAMED} pipeline, which is based on Bayesian multimodal fitting and significance tests obtained from a model comparison process \citep{2020A&A...640A.130C}. 
We note that \textsc{famed} is not optimised for our M67 \textit{K2} data, including aspects such as highly evolved stars, complex spectral windows, relatively short light curves, K2-specific noise characteristics and, for some stars, low signal-to-noise ratio. As a result, the Background tool facility \textsc{Diamonds} + \textsc{Background} \citep{2014A&A...571A..71C} meant to be used with \textsc{FAMED}\footnote{Software available for download at \url{https://github.com/EnricoCorsaro/DIAMONDS}, \url{https://github.com/EnricoCorsaro/Background}, and \url{https://github.com/EnricoCorsaro/FAMED}} often failed to converge, so we used the background models obtained by \textsc{pySYD} instead. The procedure used by \textsc{pySYD} to subtract the background of the power spectrum involves testing background models composed of one, two, or three Harvey components \citep{1985ESASP.235..199H}, with a white noise component either fixed or variable. It then computes the Bayesian Information Criterion (BIC, \citet{1978AnSta...6..461S}) for each model using the smoothed power spectrum and selects the best-fitting model based on the BIC value. 
On occasions \textsc{FAMED} mislabelled the degree of the modes or its process got aborted before all frequencies were scanned. In those cases, we corrected the degree or identified missing modes visually with the help of the echelle diagram. 
For the modes identified visually, we adopted the uncertainties given by \textsc{FAMED} to modes of similar signal strength from the same star or from one of the other stars in a similar phase of evolution. 
To ensure sample quality, we removed peaks with signal-to-noise ratio $\mathrm{SNR}<3.0$. However, we did not do this for stars with low overall signal-to-noise ratio; these are stars that we want to keep in the sample because their radial ridge was clear and their phase term $\epsilon$ followed the cluster sequence, as we describe in Section~\ref{sec:surface}. 
From our seismic sample of 50 stars, we excluded 4 highly evolved giants due to insufficient frequency resolution, 1 subgiant due to failure to identify oscillation modes confidently, 
and the 7 helium-burning stars because modelling this evolutionary phase is beyond the scope of this work. 
In \autoref{tab:modefreqs}, we present the radial mode frequencies of the resulting sample of 38 stars. We used an asterisk ($^*$) to mark stars with a low overall SNR.

\subsection{Surface correction}
\label{sec:surface}

\begin{figure}
\begin{center}
    \includegraphics[width=\columnwidth]{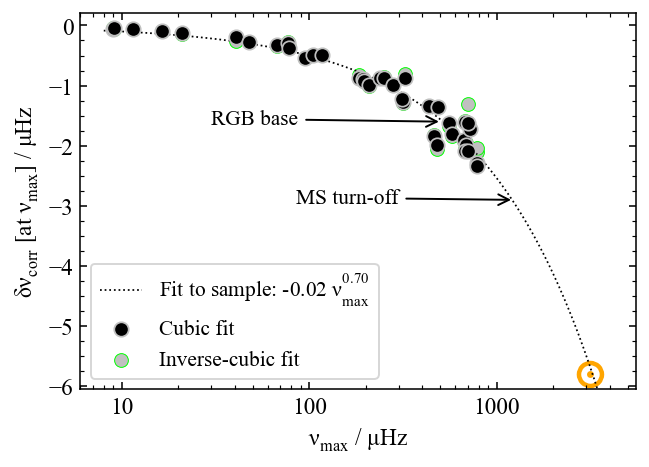} 
    \caption{Surface corrections applied to M67 giants and subgiants. Black circles show \numax\ from the 38 stars with individual mode frequencies from \autoref{tab:modefreqs} and their cubic surface term correction in \muhz, evaluated at \numax. The grey-lime circles represent the same, but using the inverse-cubic form of the correction. The dotted line shows a power-law fit to the black symbols. 
    The surface correction to a model of the Sun at $\nu_{\mathrm{max}, \odot}$ (orange symbol) is shown for reference and was not included in the fit. Arrows indicating the main sequence turnoff and the base of the red giant branch following Figure \ref{fig:CMDKIEL}.}
    \label{fig:surfcorr}
\end{center}
\end{figure}

\begin{figure*}
\begin{center}
    \includegraphics[width=\textwidth]{{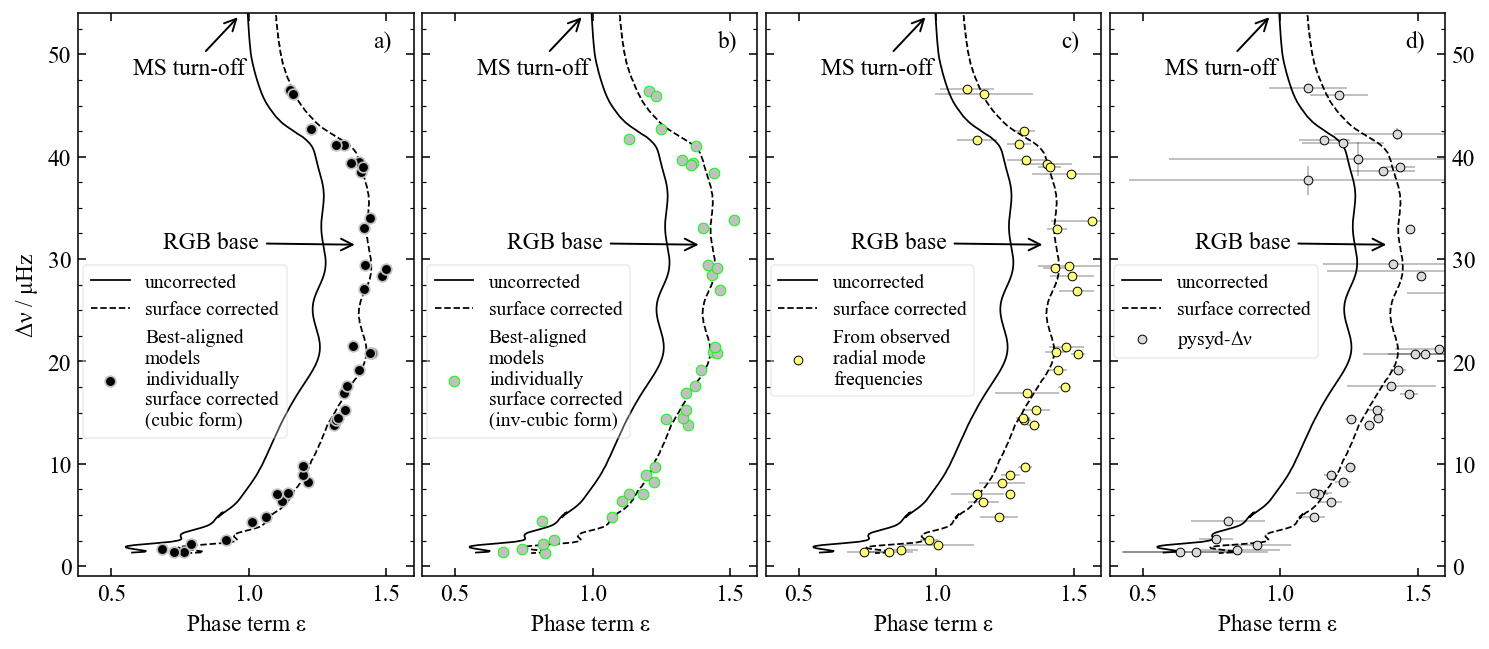}}
    \caption{
    In all panels the black solid curve shows phase-term and \dnu\ from models along the isochrone, and the dashed line shows the same, but after applying the correction given by the power-law fit from Figure \ref{fig:surfcorr}. (a) The black symbols show \dnu\ and $\epsilon$ of individual models found to align best with (lowest radial orders of) peakbagged stars, after applying cubic surface corrections based on each star's frequencies. (b) As a), but corrections calculated with the inverse-cubic form. (c) The yellow symbols indicate \dnu\ (and associated $\epsilon$) calculated from observed radial mode frequencies (Table \ref{tab:modefreqs}).  (d) grey circles represent observed \dnu\ (and associated $\epsilon$) obtained by the \textsc{pySYD} pipeline. In all panels, the main sequence turn-off and the base of the red giant branch are indicated with arrows following Figure~\ref{fig:CMDKIEL}.}

    \label{fig:phaseterm}
\end{center}
\end{figure*}

We followed the works by \citet{2008ApJ...683L.175K} and \citet{2014A&A...568A.123B}, where the authors investigate the systematic difference, known as the surface term,  between observed solar oscillation frequencies and those derived from a solar model. Instead of stellar models calibrated to the Sun, we use Isochrone A, which was calibrated to M67. And instead of a focus on how the surface term varies across radial mode orders of the Sun, we looked at the correction at \numax\ and how that varies across many stars; 
those from Section \ref{sec:peak}.

The surface term is known to be caused by poorly modelled physical processes in stellar atmospheres \citep{1988Natur.336..634C, 1996Sci...272.1286C, 2015LRSP...12....8H}. Because real and model atmospheres change smoothly along the isochrone, we can expect a corresponding smooth change in the surface term \citep{2023MNRAS.523..916L}. This suggests that it should be possible to parameterise the M67 surface term, similar to the approach taken for the Sun by \citet{2008ApJ...683L.175K} and \citet{2014A&A...568A.123B}.
To accomplish this, we obtained atmosphere-inclusive structure profiles from models along Isochrone A,
with initial masses ranging between 1.30 and 1.38 \msol, which span the masses from the main-sequence turn-off to the tip of the red giant branch.

We calculated adiabatic frequencies from the structure profiles using the oscillation code \textsc{GYRE} \citep{2013MNRAS.435.3406T} version 6.0.1, 
scanning for radial modes up to the acoustic cutoff frequency $\nu_{\mathrm{ac}} = \nu_{\mathrm{ac},\odot} \cdot (g/g_{\odot})/\sqrt{T_{\mathrm{eff}}/T_{\mathrm{eff},\odot}}$, taking $\nu_{\mathrm{ac},\odot}= 5 \mathrm{mHz}$.
For each star in \autoref{tab:modefreqs}, we selected the model that aligned best with the lowest order observed radial mode. Due to the low number of excited orders in some stars, we performed this selection manually noting that the lowest-order mode is already likely affected by the surface effect in stars with few excited modes. 
We corrected the model frequencies for the surface effect by fitting the individual mode frequencies with the cubic expression described by Equation 3 from \citet{2014A&A...568A.123B}: $\delta\nu_{\mathrm{corr}}(\nu) = a_{3}(\nu/\nu_{\mathrm{ac}})^{3}/I(\nu)$, where $a_{3}$ is calculated as in their Equations 8 and 9; and with the inverse-cubic expression $\delta\nu_{\mathrm{corr}}(\nu) = [a_{-1}(\nu/\nu_{\mathrm{ac}})^{-1} + a_{3}(\nu/\nu_{\mathrm{ac}})^{3}]/I(\nu)$, where $a_{-1}, a_{3}$ are calculated as in their Equations 10 and 11. In both cases, $\nu$ is the mode frequency, $\nu_{\mathrm{ac}}$ is the model's acoustic cutoff frequency, and $I(\nu)$ is the mode inertia.

\autoref{fig:surfcorr} shows in black symbols the resulting cubic surface corrections evaluated at \numax, as a function of \numax\ across the M67 sample.
A power-law fit to the cubic trend results in 
\begin{equation}
  \delta\nu_{\mathrm{corr}}(\mathrm{at}\,\nu_{\mathrm{max}}) = -0.02(\nu_{\mathrm{max}})^{0.7} 
\label{eq:powerlaw}
\end{equation}
\noindent as shown 
with a dotted line. The data point for the Sun was not included in the fit, but turns out to match our functional form when extrapolated. The resulting fit given by \autoref{eq:powerlaw} was almost identical whether we used the cubic or the inverse-cubic expressions (grey-lime circles) from \citet{2014A&A...568A.123B}, but the scatter around the fit was somewhat smaller in the cubic case.  

Different model physics and in particular, different atmosphere choices can lead to drastically different values for the surface effect at \numax\ \citep{1996Sci...272.1286C, 2023MNRAS.523..916L}. Therefore we note that our \autoref{eq:powerlaw} is expected to be valid for models such as the ones described in detail in R24, which use the generalised Hopf-functions for atmosphere by \citet{2014MNRAS.442..805T} as implemented by \citet{2021RNAAS...5....7B}, 
with a modified radiative gradient that follows this T$(\tau)$ relation.

In the phase-term diagrams of Figure~\ref{fig:phaseterm}, the solid black curves show \dnu\ and $\epsilon$ of the models along Isochrone A, where we calculated (\dnu, $\epsilon$) as in \citet{2011ApJ...743..161W}. 
We obtained the surface-corrected version of the isochrone models (dashed curves) by first deriving the surface term at \numax\ from \autoref{eq:powerlaw}. We then substitute this value into the expression for $\delta\nu_{\mathrm{corr}}(\nu)$ at $\nu=\nu_{\mathrm{max}}$ to obtain the cubic coefficient $a_{3}$. This coefficient allows us to calculate corrections for all radial modes in a given model. We repeated the procedure for all models along the isochrone to obtain the corrected (\dnu, $\epsilon$). The correction method had the effect of shifting $\epsilon$ toward higher values relative to the uncorrected isochrone. 

In Figure~\ref{fig:phaseterm}a the black circles represent cubic-expression surface-corrected models found to be the best aligned to each of the stars, as described above.
These individually corrected models show $\epsilon$ values largely consistent with the surface-corrected isochrone, and their scatter mirroring their scatter pattern seen around the fitted power law in Figure~\ref{fig:surfcorr}. The dashed curve reflects the smoothing effect of the surface correction procedure implemented using the power-law fit.
The extremely tight `evolution' of $\epsilon$ seen here has never been mapped out before due to the previous lack of very similar stars among field star samples, and more generally the lack of stars with measured $\epsilon$ in the transition between main sequence/early subgiant \citep{2017ApJ...835..172L} and red giant branch stars \citep{2012ApJ...757..190C}, as nicely spanned by this M67 sample. 

Figure~\ref{fig:phaseterm}b follows the same concept, but with the grey-lime circles representing the inverse-cubic form of the surface correction to the same best-aligned models.
Comparing Figures~\ref{fig:phaseterm}a and b, it becomes clear that the cubic expression led to a smoother progression of the phase term $\epsilon$ with \dnu\ when compared to the inverse-cubic expression. 
The extra scatter observed with the inverse-cubic expression could be reflecting a natural scatter around the power-law fit if in fact the inverse-cubic corrections are more accurate. However, if the surface correction using the cubic expression is more accurate, this extra scatter could suggest that the inverse-cubic expression slightly overfits the data when, for example, one or more radial modes deviate from their expected frequency due to actual structural discontinuities or if the error in the peakbagged frequencies has been underestimated. In \autoref{fig:cubic_vs_invcubic} we present the four stars whose surface-corrrected $\epsilon$ show the largest differences between using the cubic vs. inverse-cubic expressions. Given that a smoother progression of the phase-term aligns better with the expectation from a uniform stellar sample such as ours, going forward we use the cubic fit. 

Next, we calculated \dnu\ and $\epsilon$ using the observed radial mode frequencies (Table \ref{tab:modefreqs}) and their uncertainties as weights, and we also used the observed radial modes along with \textsc{pySYD}-derived \dnu\ to obtain \textsc{pySYD}-derived $\epsilon$ values. These results are shown in Figures~\ref{fig:phaseterm} c and d, respectively. 
We observe a greater scatter in \textsc{pySYD} based $\epsilon$ compared to the scatter in Figures~\ref{fig:phaseterm} a, b and c, which we attribute to the effects of strong radial modes dominating over weaker ones, and to the effects of mixed modes, both of which influence \dnu\ measurements made by autocorrelation-based pipelines like \textsc{pySYD}. These effects are amplified in $\epsilon$ due to its high sensitivity to even small changes in \dnu. In Figures~\ref{fig:phaseterm} c and d, we observe a trend toward larger $\epsilon$ compared to the model-derived \dnu\ in Figures~\ref{fig:phaseterm} a and b, particularly for stars on the red giant branch. This discrepancy may result from the curvature of the radial ridge known to be more pronounced on red giant branch stars (see Figure 4 of \citet{2013A&A...550A.126M}). The curvature's influence is more apparent in \dnu\ from models, where there are many more orders available. This suggests that the weight function used to obtain \dnu\ from models might not be narrow enough.

\begin{figure}
\begin{center}
    \includegraphics[width=\columnwidth]{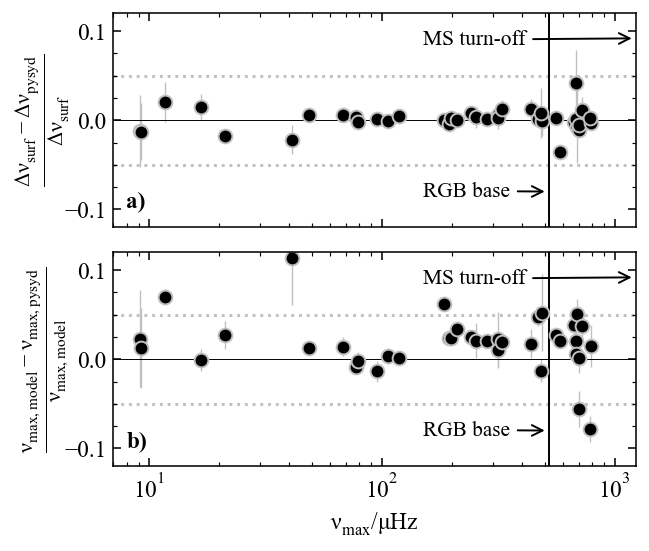}
    \caption{a) and b) show the fractional differences between observed \dnu\ and \numax\ values obtained from \textsc{pySYD} and the ones obtained through model-matching. Symbols have the same colour coding as Figures~\ref{fig:phaseterm}a and b. The main sequence turn-off and the base of the red giant branch are indicated with arrows following Figure~\ref{fig:CMDKIEL}.}
    \label{fig:frac}
\end{center}
\end{figure}

Figure~\ref{fig:frac} illustrates the fractional differences between the observed \dnu\ and \numax\ values obtained from \textsc{pySYD} and those obtained through model-matching. The figure suggests that observational features can influence \dnu\ and/or \numax, but not necessarily both at the same time or to the same extent. If we assume that the model values are accurate, the figure shows that \textsc{pySYD} uncertainties are sometimes underestimated, not always reflecting the actual measurement errors. We also see that the fractional difference between the observed and model values tends to be higher for \numax\ than for \dnu, suggesting that \numax\ estimations based on the Gaussian fitting to the smoothed spectrum tend to be less robust than \dnu\ estimation when the power distribution in the spectra is uneven, which can happen when the light curves are short.

\begin{figure}
\begin{center}
    \includegraphics[width=\columnwidth]{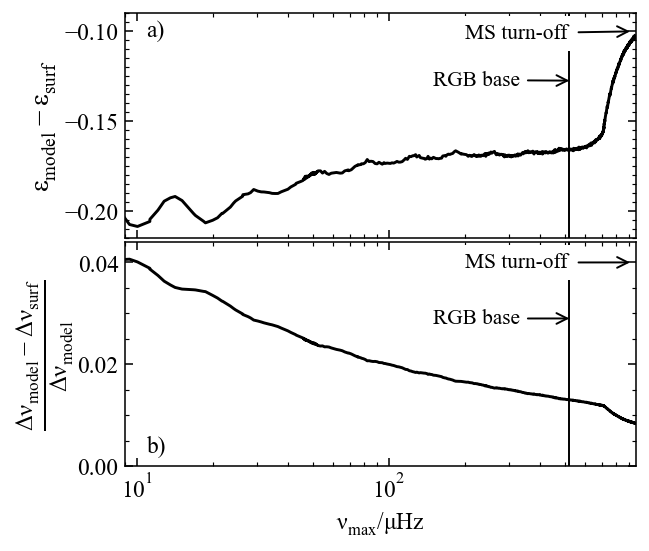}
    \caption{ Impact of the surface effect on isochrone models as a function of \numax. a) Differences in the asymptotic phase term, $\epsilon$, before and after applying surface corrections. (b) Fractional differences in \dnu\ before and after applying surface corrections. The main sequence turn-off and the base of the red giant branch are indicated with arrows following Figure~\ref{fig:CMDKIEL}.}
    \label{fig:frac_sc}
\end{center}
\end{figure}

The observed offset in $\epsilon$ between the uncorrected isochrone and the surface-corrected isochrone shown in Figure~\ref{fig:phaseterm}, and more explicitely in Figure~\ref{fig:frac_sc}a, confirms previous observations made by \citet{2011ApJ...743..161W} from stellar tracks and solar-metallicity field stars. Furthermore, our \autoref{eq:powerlaw} provides a recipe for estimating such an offset in models of other stars, at least those with masses and metallicity comparable to M67, and provided that models similar to those of R24 are used.
This was possible due to the size and ample \numax\ range of the sample and the fact that the stars differ only (gradually) by small amounts in mass.

Perhaps more significant is the effect of surface corrections on \dnu. To quantify its impact on our isochrone models, we calculate the fractional difference between \dnu\ from the uncorrected and surface-corrected models. This effect is particularly pronounced at low \numax. Given that in the widely used solar-scaling relation for mass (Equation \ref{eq:eq4}) \dnu\ is raised to the power of -4, the fractional error shown in Figure~\ref{fig:frac_sc}b becomes considerable, even for higher \numax. This underscores the importance of accurately accounting for surface corrections when interpreting modelled \dnu, particularly in the context of the mass scaling relations, which we examine in detail in the following section.

\subsection{Calibrating mass and radius scaling relations}
\label{sec:mass}
The asteroseismic scaling relations for $\nu_{\mathrm{max}} \propto (M/R^{2})/\sqrt{T_{\mathrm{eff}}}$ and $\Delta\nu \propto \sqrt{M/R^3}$) offer a simple way to determine stellar mass and radius \citep{2008ApJ...674L..53S, 2010A&A...509A..77K}. However, the relations are known to be only approximations \citep{2009MNRAS.400L..80S, 2011ApJ...743..161W, 2013EPJWC..4303004M}, and therefore corrections are needed to improve their accuracy \citep{2016ApJ...822...15S, 2017MNRAS.467.1433R, 2018A&A...616A.104K, 2020FrASS...7....3H}. Our cluster stars provide a means to benchmark and potentially correct these relations, particularly those related to mass, which has been called into question in the past. For example, \citet{2017MNRAS.467.1433R} suggested that we may have been overestimating scaling-based seismic masses, and \citet{2018A&A...616A.104K} said that scaling relations could not be applied directly to red clump stars, in agreement with \citet{2012MNRAS.419.2077M} who had previously stated that due to the different temperature profiles between the red giant branch and the red clump stars, a relative correction has to be considered between both groups if we intend to use the \dnu\ scaling to estimate the stellar mean density.

Following past cluster studies \citep{2012MNRAS.419.2077M, 2016MNRAS.461..760M, 2017MNRAS.472..979H, 2022MNRAS.515.3184H}, we investigate scaling-based stellar masses using four combinations of seismic (\dnu\ and \numax) and classic ($L$ and \teff) observables, 
following the notation of \citet{2016ApJ...822...15S}:
\begin{equation}
    \frac{M}{M_{\odot}} \simeq    \left(\frac{\nu_\mathrm{max}}{f_{\nu_{\mathrm{max}}}\nu_{\mathrm{max}, \odot}} \right)  \left(\frac{L}{L_{\odot}} \right) \left(\frac{T_{\mathrm{eff}}}{T_{\mathrm{eff}, \odot}} \right)^{-7/2}   
    \label{eq:eq1}
    \end{equation}
\begin{equation}
    \frac{M}{M_{\odot}} \simeq   \left(\frac{\Delta\nu}{f_{\Delta\nu}\Delta\nu_{\odot}} \right)^{2} \left(\frac{L}{L_{\odot}} \right)^{3/2} \left(\frac{T_{\mathrm{eff}}}{T_{\mathrm{eff}, \odot}} \right)^{-6}  
    \label{eq:eq2}
    \end{equation}
\begin{equation}
    \frac{M}{M_{\odot}} \simeq   \left(\frac{\nu_\mathrm{max}}{f_{\nu_{\mathrm{max}}}\nu_{\mathrm{max}, \odot}} \right)^{12/5} \left(\frac{\Delta\nu}{f_{\Delta\nu}\Delta\nu_{\odot}} \right)^{-14/5} \left(\frac{L}{L_{\odot}} \right)^{3/10}
    \label{eq:eq3}
    \end{equation}
\begin{equation}
    \frac{M}{M_{\odot}} \simeq  \left(\frac{\nu_\mathrm{max}}{f_{\nu_{\mathrm{max}}}\nu_{\mathrm{max}, \odot}} \right)^{3} \left(\frac{\Delta\nu}{f_{\Delta\nu}\Delta\nu_{\odot}} \right)^{-4} \left(\frac{T_{\mathrm{eff}}}{T_{\mathrm{eff}, \odot}} \right)^{3/2}
    \label{eq:eq4}
\end{equation}
\\
\noindent where $f_{\Delta\nu}$ and $f_{\nu_{\mathrm{max}}}$ are correction factors. For instance, $f_{\Delta\nu}$ has been calculated using model grids \citep{2016ApJ...822...15S, 2017MNRAS.467.1433R, 2017ApJS..233...23S}, and an expression for $f_{\nu_{\mathrm{max}}}$ based on mean molecular weights has been suggested \citep{2017ApJ...843...11V}. 

Similarly, radii will in the following be calculated with the seismic scaling relation:
\begin{equation}
    \frac{R}{R_{\odot}} =  \left(\frac{\nu_\mathrm{max}}{f_{\nu_{\mathrm{max}}}\nu_{\mathrm{max}, \odot}} \right) \left(\frac{\Delta\nu}{f_{\Delta\nu}\Delta\nu_{\odot}} \right)^{-2} \left(\frac{T_{\mathrm{eff}}}{T_{\mathrm{eff}, \odot}} \right)^{1/2}
    \label{eq:eq5}
    \end{equation}
\noindent except for the brightest star, EPIC 211376143, for which we adopt a radius of $73.8\, \mathrm{R}_{\odot}$, as predicted by Isochrone A for the star's \numax, 
because its \dnu\ is unavailable due to insufficient frequency resolution of the data.     

\begin{figure*}[ht]
\begin{center}
    \includegraphics[width=0.95\textwidth]{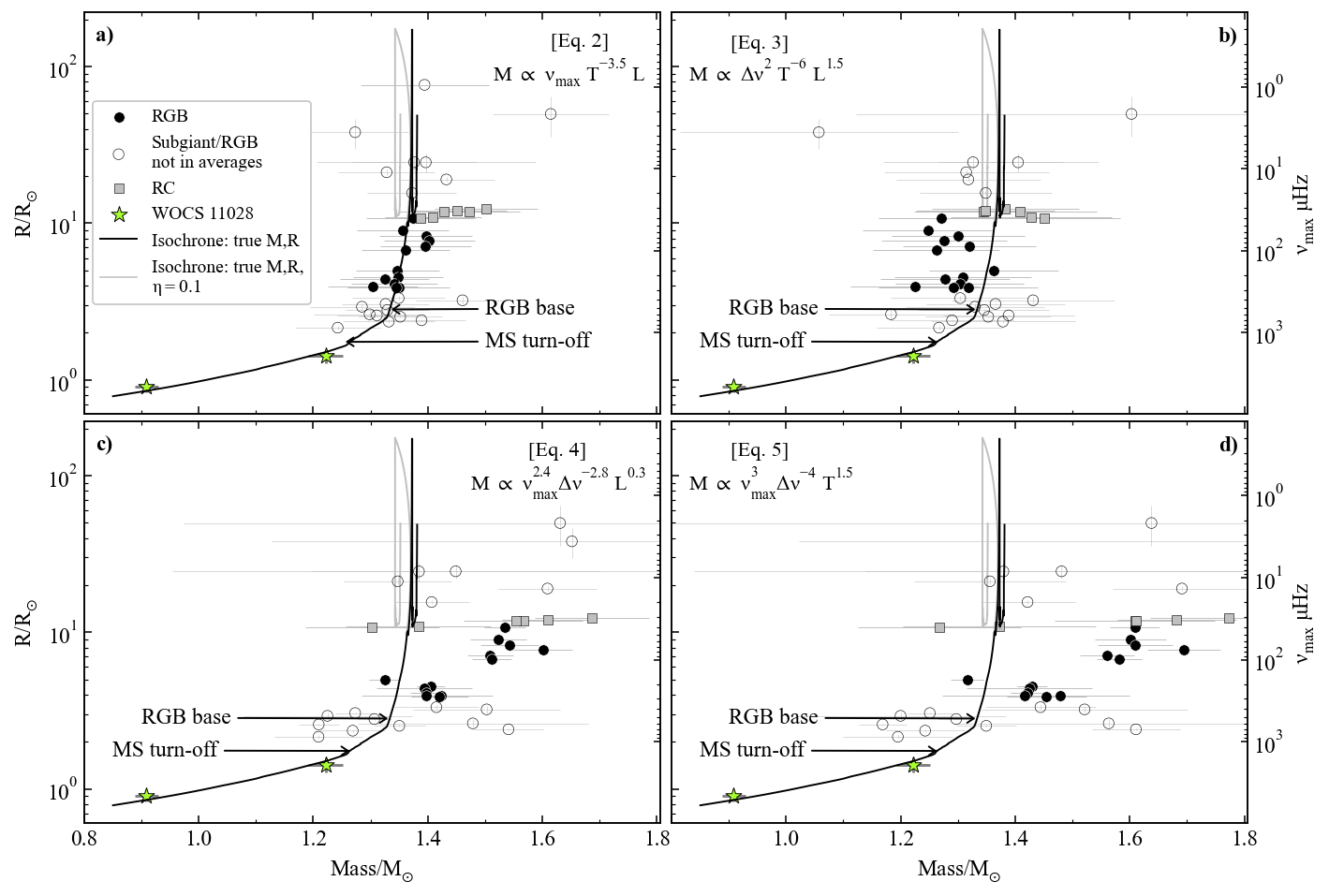}
    \caption{Mass-Radius diagrams of M67 presumed single stars calculated using equations \ref{eq:eq1} to \ref{eq:eq5}, with the mass equations as annotated on panels (a) to (d).
    Subgiants and red giants are shown as open circles, except those used to calculate a mean red giant branch mass, shown in filled black circles. Red clump stars are shown as grey squares. The black curve shows the true mass and radius of Isochrone A and, in light grey, a version including mass loss. 
    The main sequence turnoff and the base of the red giant branch are indicated with arrows following Figure~\ref{fig:CMDKIEL}.  For reference, we also show the eclipsing binary WOCS 11028 with star symbols, with masses and radii by \citet{2021AJ....161...59S}, where the error bars represent $5\sigma$.
    The legend in panel (a) applies to all panels. }
    \label{fig:mass_four}
\end{center}
\end{figure*}

Because Isochrone A is fitted to {\it Gaia} DR2 magnitudes, we calculate the luminosities of our M67 stars using the same photometric system and the formula $-2.5 \log L_{*}/L_{\odot} = G_{\mathrm{mag}} + BC_{G} - M_{\mathrm{bol},\odot}$, where $G_{\mathrm{mag}}$ is the absolute $G$ magnitude, and $BC_{G}$ is the bolometric correction. Bolometric corrections depend on \teff, $\log g$, and [Fe/H], and we obtain them through interpolation to tables from the MIST project\footnote{https://waps.cfa.harvard.edu/MIST/model\_grids.html}, which are based on the grid of stellar atmospheres and synthetic spectra described by \citet{2016ApJ...823..102C}. 
$M_{\mathrm{bol},\odot}$ is the absolute bolometric magnitude of the Sun that we take as 4.74 $\pm$ 0.01 magnitude following \citet{2015arXiv151006262M}.
To derive absolute $G$ magnitudes, we use the distance modulus ($9.614 \pm 0.049$) and the differential extinction corrections derived in R24. To address the magnitude-dependent trend in DR2 $G$ magnitudes seen by \citet{2018A&A...616A...4E} and \citet{2018MNRAS.479L.102C}, we set the $G_{\mathrm{mag}}$ uncertainties to 0.01 magnitudes, about 10 times the photometric precision estimated for DR2 sources at the brighter end ($G<13$). 
Solar reference values are taken as $\Delta\nu_{\odot}= 135.1 \mu \mathrm{Hz}$, $\nu_{\mathrm{max}, \odot} = 3090\, \mu \mathrm{Hz}$, and $T_{\mathrm{eff},\odot} = 5777 \mathrm{K}$ \citep{2011ApJ...743..143H}.

\begin{figure*}
\begin{center}
    \includegraphics[width=0.95\textwidth]{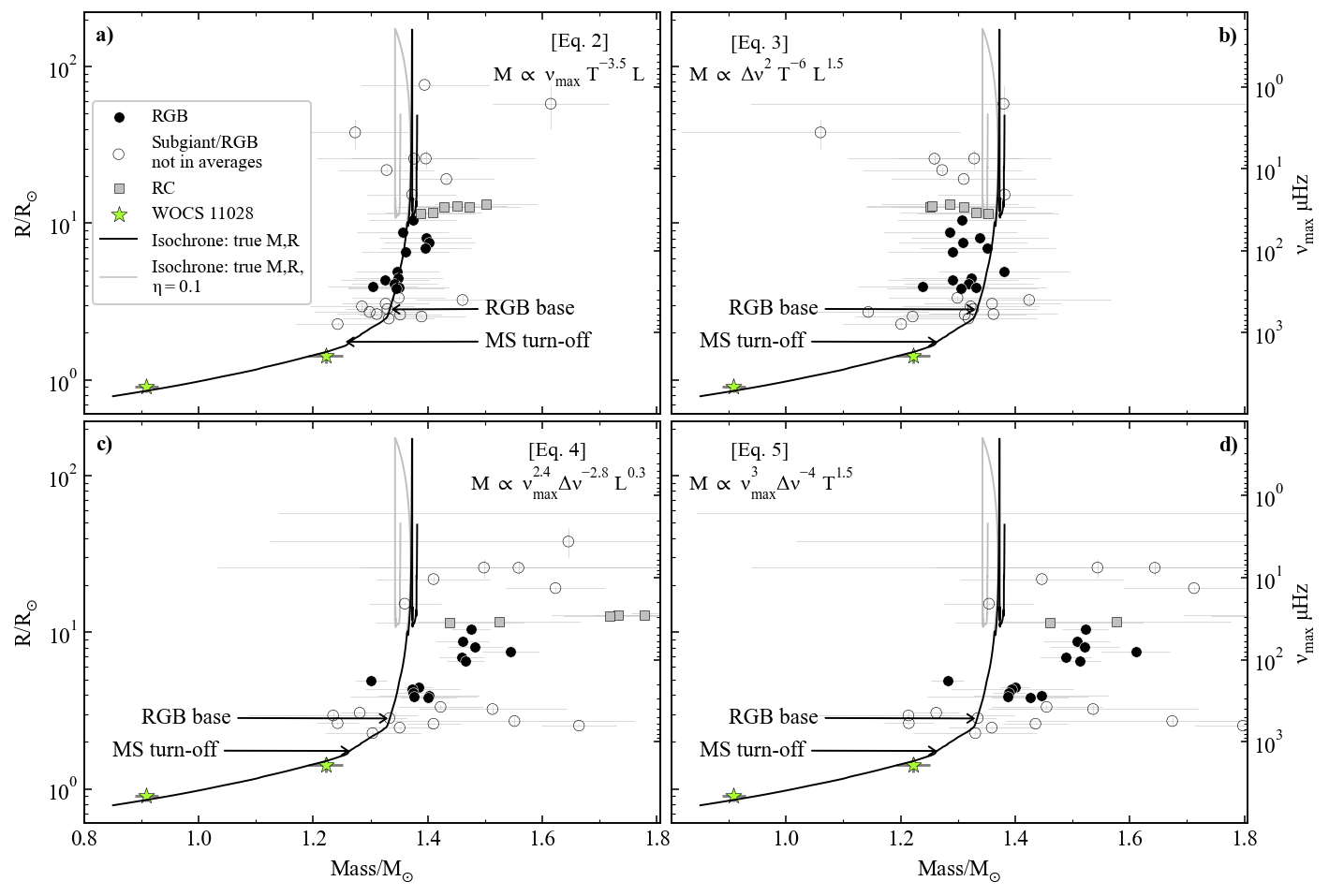}
    \caption{As in \autoref{fig:mass_four}, but after applying a correction factor $f_{\Delta\nu} = (\Delta\nu_{\mathrm{obs}} / \sqrt{\rho_{\mathrm{model}}})$. The corresponding model for each star was obtained from interpolating the observed \dnu\ into the isochrone's \dnu\ sequence prior to surface correction.} 
    \label{fig:mass_appendix_2}
\end{center}
\end{figure*}

\begin{figure*}
\begin{center}
    \includegraphics[width=0.95\textwidth]{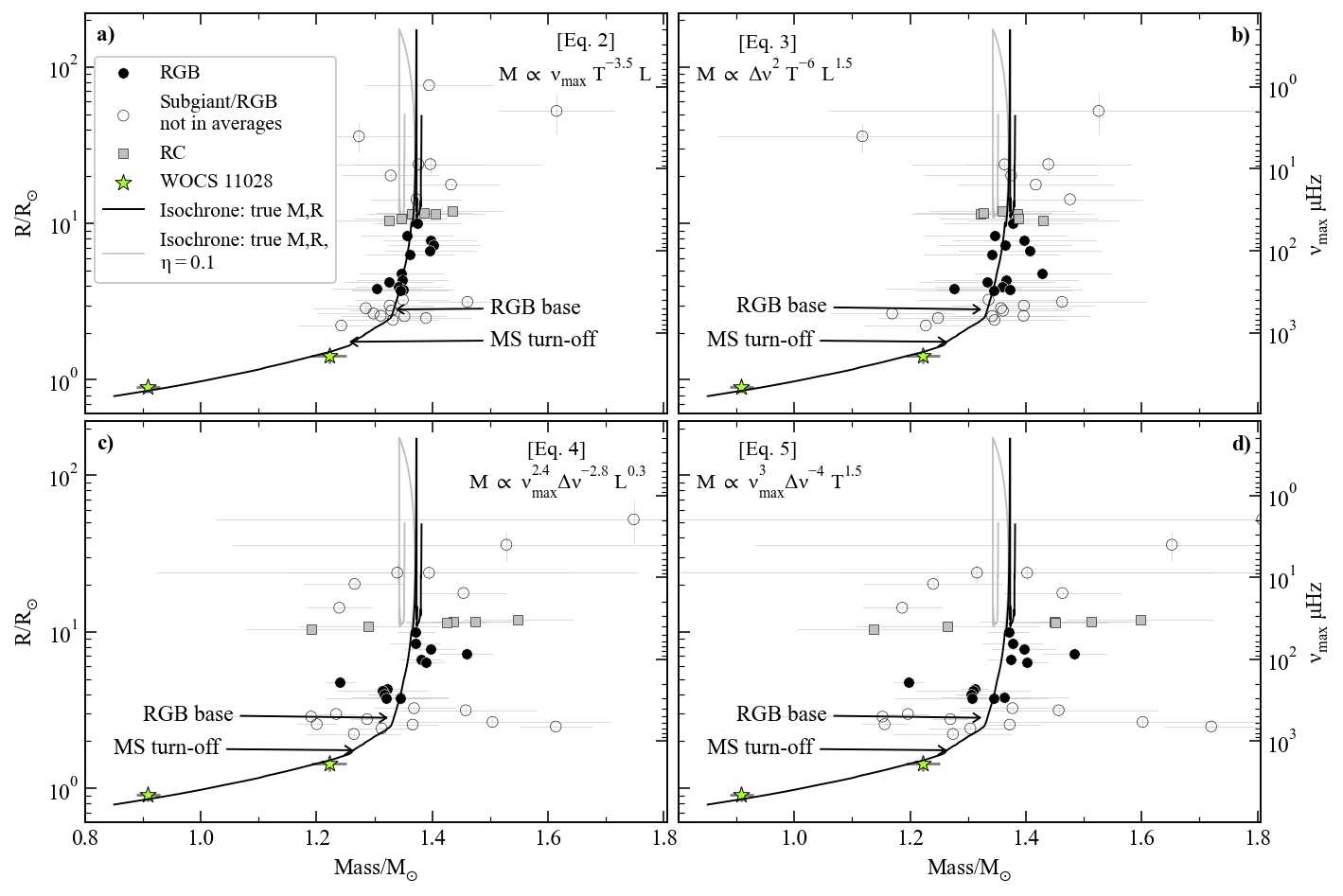}
    \caption{As in \autoref{fig:mass_four}, but applying the correction factor $f_{\Delta\nu} = (\Delta\nu_{\mathrm{obs}} / \sqrt{\rho_{\mathrm{surf}}})$, and $f_{\nu_{\mathrm{max}}}$. The corresponding model for each star was obtained from interpolating the observed \dnu\ into the isochrone's surface corrected \dnu\ sequence. Additionally, a red-clump-specific correction factor $f_{\nu_{\mathrm{max}}} = 1.0477$ has been applied. The mean red giant and mean red clump values from the four different equations show excellent agreement among them.}
    \label{fig:mass_corrected}
\end{center}
\end{figure*}

\begin{figure*}
\begin{center}
    \includegraphics[width=\textwidth]{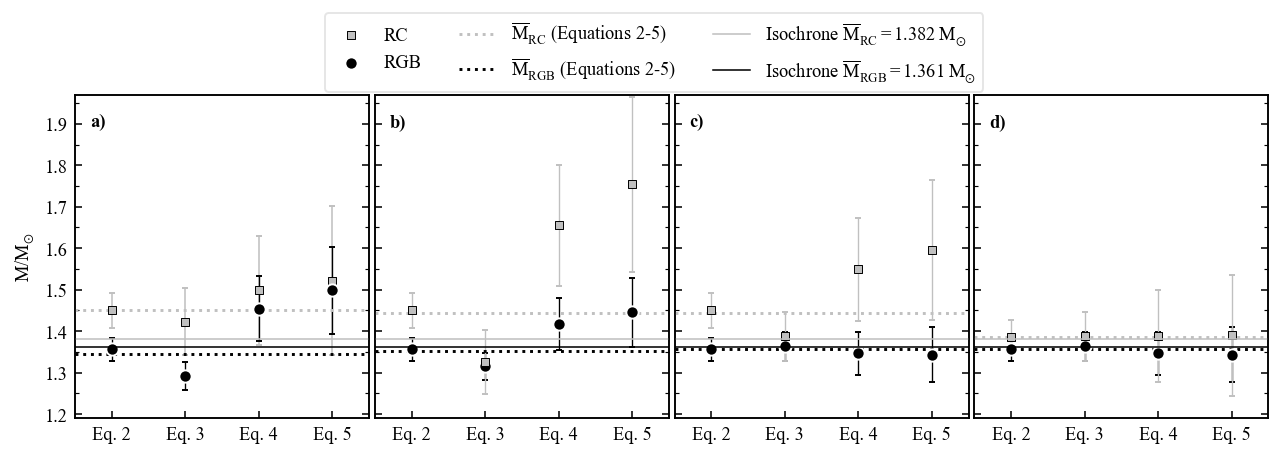}

\vspace{\floatsep}

    \begin{tabular}{cccccc}
         Panel & M/R figure & $f_{\Delta\nu}$ & $f_{\nu_{\mathrm{max}}}$  & $\overline{\mathrm{M}}_{\mathrm{RGB}}$ (Eqs. 2-5) & $\overline{\mathrm{M}}_{\mathrm{RC}}$ (Eqs. 2-5) \\
         \midrule
        \textbf{a} & \autoref{fig:mass_four} &1 & 1 & 1.345 $\pm$ 0.028 $\pm$ 0.081 & 1.452 $\pm$ 0.050 $\pm$ 0.039  \\
        \textbf{b} & \autoref{fig:mass_appendix_2}& $ (\Delta\nu_{\mathrm{obs}} / \sqrt{\rho_{\mathrm{model}}})^{*}$ & 1 & 1.352 $\pm$ 0.027 $\pm$ 0.051 & 1.444 $\pm$ 0.050 $\pm$ 0.168 \\
        \textbf{c} &   &$ (\Delta\nu_{\mathrm{obs}} / \sqrt{\rho_{\mathrm{surf}}})^{*}$ & 1 & 1.356 $\pm$ 0.027 $\pm$ 0.008 & 1.444 $\pm$ 0.046 $\pm$ 0.081 \\
        \textbf{d} & \autoref{fig:mass_corrected} & $ (\Delta\nu_{\mathrm{obs}} / \sqrt{\rho_{\mathrm{surf}}})^{*}$ & RC only, 1.0477 & 1.356 $\pm$ 0.027 $\pm$ 0.008 &  1.387 $\pm$ 0.044 $\pm$ 0.001 \\
    \end{tabular}
\vspace{\floatsep}
        \captionof{figure}{The scatter symbols represent M67 mean masses from the presumed single red giant stars (black circle) and red clump stars (grey square) within the range $25<\nu_{\mathrm{max}}<350$ \muhz, using solar-scaling relations from equations \ref{eq:eq1} to \ref{eq:eq4}. 
    The weighted averages of the masses across the four equations are shown as dotted lines: black for red giants, grey for red clump stars, and are listed in the table. The first annotated uncertainty arises from error propagation, while the second represents the standard deviation. The solid horizontal lines indicate the mean masses from the models along Isochrone A within the same \numax\ range. Panels a) to d) illustrate the progression of increasingly refined corrections for $f_{\Delta\nu}$ and $f_{\nu_{\mathrm{max}}}$, as provided in the table. ($^{*}$) indicates values are solar-scaled.}
    \label{fig:mass_study_vertical}

\end{center}
\end{figure*}
We calculated the masses and radii of the presumed single stars in our M67 sample, where the input \dnu\ and \numax\ was from our \textsc{pySYD} measurements, and \teff\ from \autoref{sec:temperatures} (\autoref{tab:pysyd_teff}), using equations \ref{eq:eq1} to \ref{eq:eq5}. First, we assumed $f_{\Delta\nu} = f_{\nu_{\mathrm{max}}}=1$. In \autoref{fig:mass_four} we compared these scaling masses and radii, represented by individual symbols, to the "true" model masses and radii along Isochrone A (solid black curve). Here, each panel represents
one of the four mass scaling relations. 
We found that the stars only align with the  isochrone for the \autoref{eq:eq1}-mass (the one without \dnu). 
In contrast, equations \ref{eq:eq2} to \ref{eq:eq4} show strong mass discrepancies that seem proportional to the power of \dnu\ in each equation, suggesting a non-unity \dcorr\ factor is needed. 
To accurately address the mass discrepancies, which clearly vary in magnitude along the isochrone, it is essential to incorporate a \dcorr\ factor that depends on stellar parameters.

One common approach to determining \dcorr\ is by interpolating within pre-computed model grids, where \dcorr\ is defined as the ratio between \dnu\ (measured from radial mode frequencies) and the square root of the mean stellar density. Examples include the grids by \citet{2017MNRAS.467.1433R} and \citet{2022RNAAS...6..168S}. However, these grids are typically calibrated to the Sun and often do not account for surface effects. In \autoref{fig:frac_sc}b, we previously demonstrated the impact of the surface term on \dnu. \autoref{fig:mass_appendix_2} demonstrates how ignoring the surface term will affect the mass estimates. The figure shows the scaling-based masses calculated using Equations \ref{eq:eq1} to \ref{eq:eq4}, similar to \autoref{fig:mass_four}, but now incorporating \dcorr\ = $\Delta\nu_{\mathrm{obs}}/\sqrt{\rho_{\mathrm{model}}}$ (solar-scaled). This approach is similar to using the aforementioned grids, but with the added benefit that our isochrone is calibrated to M67. For each star, we determined the model density $\rho_{\mathrm{model}}$ by interpolating the observed $\Delta\nu$ ($\Delta\nu_{\mathrm{obs}}$, derived from \textsc{pySYD} and listed in \autoref{tab:pysyd_teff}) over the modeled $\Delta\nu$ ($\Delta\nu_{\mathrm{model}}$), derived from radial frequencies along Isochrone A, \textit{prior} to applying surface corrections. This approach represents an improvement over the case where \dcorr=1, as the masses are now closer to agreeing amongst the four equations, and with the isochrone-predicted "truth" values. However, significant discrepancies remain, in particular regarding the most commonly used equation (\ref{eq:eq4}). \autoref{fig:mass_appendix_2}d shows mass discrepancies betweeen $\sim$ 3\% and 20\%  compared to the true isochrone masses. This agrees with the findings by \citet{2023MNRAS.523..916L}. It highlights the need for a refined $f_{\Delta\nu}$ approach, one that accounts for both the deviation of \dnu\ from the square root of stellar density, and the influence of the surface effect on \dnu. 

We achieved this by applying \dcorr = $\Delta\nu_{\mathrm{obs}}/\sqrt{\rho_{\mathrm{surf}}}$ (solar-scaled), where this time the density $\rho_{\mathrm{surf}}$ is determined by interpolating the observed $\Delta\nu$ over $\Delta\nu_{\mathrm{surf}}$ from radial frequencies along Isochrone A \textit{after} surface corrections. \autoref{fig:mass_corrected} shows that when using this more thorough correction we find a very good alignment amongst the red-giant-branch scaling masses from the four different equations\footnote{One could of course simply adopt the mass of our best-matched model as the corrected mass. However, that would not be a scaling-based mass, and hence it would not reflect the uncertainty from the other observables ($L$, \teff, \numax) that we use to obtain the corrected mass through \ref{eq:eq2}-\ref{eq:eq4}. However, for completeness, we list the best-matched model masses and radii in Table~\ref{tab:truemass}.}.

To quantify and summarise the improvement of these corrections, we first show, in \autoref{fig:mass_study_vertical}a (filled symbols), the average red giant branch mass across the black filled circles from \autoref{fig:mass_four}
for each mass equation before applying any corrections. We then show the results after applying the square-root-density correction that neglects the surface term in \autoref{fig:mass_study_vertical}b.
Comparing this with \autoref{fig:mass_study_vertical}c, which shows the averages after our last 
correction (square-root-density + surface term), clearly demonstrates that red giant branch scaling-based masses from each equation, on average, now agree with one another within uncertainties. Incidentally, they also align with Isochrone A's mean red-giant-branch mass (solid black line), although that was not necessarily the intent of this exercise.  
However, even after this correction, in \autoref{fig:mass_study_vertical}c the red clump masses (squares) still disagree amongst the different equations. Moreover, three of the averaged masses are so high that, if real, the stars would have evolved beyond the red clump phase by the age of this cluster. The exception this time is \autoref{eq:eq2}, the only one without \numax. We also see that the red clump mass disagreement seems to scale with the power of \numax. This suggests that this discrepancy can be removed using a \numax\ scaling factor. As previously stated, the need for a red clump specific \numax\ correction factor has been suggested before \citep{2012MNRAS.419.2077M, 2017ApJ...843...11V, 2018A&A...616A.104K}, and in our data, a factor of $f_{\nu_{\mathrm{max}}}=1.0477$ aligns all mean red clump masses to a consistent value of $\overline{\mathrm{M}}_{\mathrm{RC}} = 1.387 \pm 0.044 \mathrm{M}_{\odot}$, as shown in \autoref{fig:mass_study_vertical}d. A summary of the scaling masses described here is presented in the table below \autoref{fig:mass_study_vertical}.

The individual clump star masses that incorporate the $f_{\nu_{\mathrm{max}}}$ correction are shown in \autoref{fig:mass_corrected} with grey squared symbols. Although the mass spread observed in red clump stars is still greater than that of the red giant branch sample in black circles, it is consistent with the mass spread observed in other lower SNR stars like the subgiants in our sample.

\autoref{fig:mass_study_vertical}d shows that the averaged red-clump masses are compatible with the no-mass-loss scenario (solid grey lines in \autoref{fig:mass_study_vertical}) in agreement with the no-mass-loss result -- also from M67 asteroseismology -- by \citet{2016ApJ...832..133S}. 
Specifically, we find the mass difference between red giant branch and red clump stars to be $M_{\mathrm{RC}}-M_{\mathrm{RGB}} = 0.031 \pm 0.052\,$\msol, which should be compared to the no-mass-loss isochrone, which gives $M_{\mathrm{RC,iso}}-M_{\mathrm{RGB,iso}} = 0.021\,$\msol. Hence, the maximum mass loss consistent with our data within $2\sigma$, $(0.031-2\times 0.052)\,$\msol, would correspond to a Reimers $\eta\sim 0.23$.
 
Interestingly, \citet{2018ApJS..239...32P} also found a negligible difference in mass between the RGB and RC in the metal-rich open cluster NGC 6791: $0.02\pm 0.05\, \mathrm{M}_{\odot}$, using theoretically motivated corrections to the \dnu\ scaling relations, in agreement with the results from \citet{2012MNRAS.419.2077M, 2019ApJ...879...81A}, and \citet{2024arXiv241110520A}. These results suggest that mass loss rates considered normal in the past \citep{1975psae.book..229R, 1996A&A...305..849C}, may not be the norm in open clusters.

\begin{figure}
\begin{center}
    \includegraphics[width=\columnwidth]{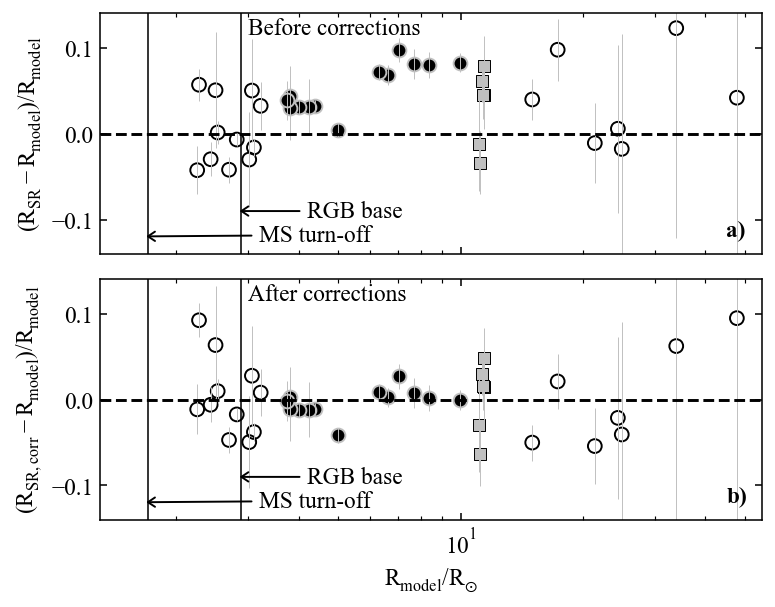}
    \caption{Fractional difference in scaling-based radius ($R_{\mathrm{SC}}$) and corresponding model radius ($R_{\mathrm{model}}$) where the model is found as the best match between the observed \textsc{pySYD}-\dnu\ and \dnusurf\ from Isochrone A. The symbols follow Figures~\ref{fig:mass_four} and \ref{fig:mass_corrected}. The RGB base and MS turn-off are indicated following \autoref{fig:CMDKIEL}. Panel a) shows the case without any corrections to the scaling relations, and panel b) shows the case after using the same correction as in Figures~\ref{fig:mass_corrected} and \ref{fig:mass_study_vertical}c and explained in the text.}
    \label{fig:frac_radius}
\end{center}
\end{figure}

With the square-root density + surface term $f_{\Delta{\nu}}$ corrections, and the red clump-specific $f_{\nu_{\mathrm{max}}}$ correction factor that we found by unifying the mass agreement with all four mass scaling-based equations, we now want to confirm that those same corrections also improve the scaling-based radii (\autoref{eq:eq5}). \autoref{fig:frac_radius}a shows the fractional difference between the uncorrected scaling-based radii relative to the true isochrone radii, which for each star is defined as the radius of the model whose \dnusurf\ matches the observed \textsc{pySYD}-\dnu. \autoref{fig:frac_radius}b shows that after applying our corrections, the agreement is indeed improved. Moreover, our best data, represented by the filled circles, show fractional differences remarkably close to zero. However, we see that some stars remain far from zero, despite having small uncertainties. This likely stems from underestimated \numax\ and \dnu\ \textsc{pySYD}-uncertainties that occur due to the stochastic nature of the oscillations, as stated in \autoref{sec:surface} and also seen in \autoref{fig:frac}.

To ensure that our mass results were not influenced by the models used to refine our calculations, we repeated our mass calculations for the best red giant branch data (filled black circles in Figures \ref{fig:mass_four} to \ref{fig:mass_corrected}). We do so using our observed \numax, \dnu, \teff, and $L$, but using the modifications to the mass equation (\ref{eq:eq4}) proposed by \citet{2022ApJ...927..167L}. 
This resulted in a mean red giant mass of $\overline{\mathrm{M}}_{\mathrm{RGB}} = 1.347 \pm 0.072\, \mathrm{M}_{\odot}$ with their APOGEE-based fit, and $\overline{\mathrm{M}}_{\mathrm{RGB}} = 1.383 \pm 0.076\, \mathrm{M}_{\odot}$ with their LAMOST fit. Our mean red giant mass estimate from the four equations $\overline{\mathrm{M}}_{\mathrm{RGB}} = 1.356 \pm 0.027 \pm 0.008\, \mathrm{M}_{\odot}$ is consistent with both fits within the margin of error but is closer to their APOGEE fit, and has significantly lower uncertainties.

\section{Model grid fitting}
\label{sec:model}

Having tested models from Isochrone A against M67 stars in the previous sections, we move on to perform model grid fitting using temperatures from \autoref{sec:temperatures}, individual frequencies from \autoref{sec:peak}, luminosities as described in \autoref{sec:mass}, and models across an age range spanning 2 Gyrs.

\begin{figure*}
\begin{center}
    \includegraphics[width=.7\textwidth]{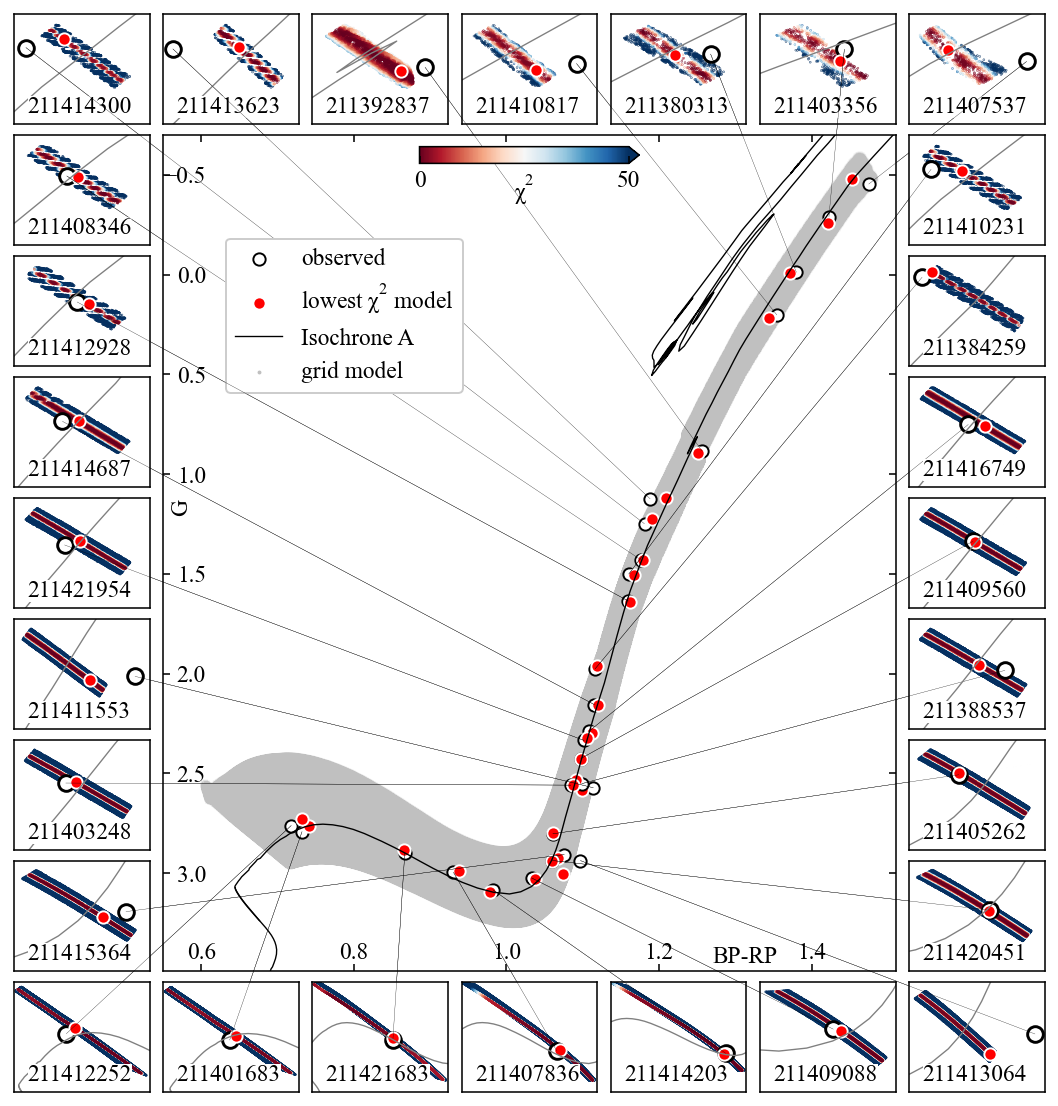}
    \caption{Isochrone A in the colour-magnitude diagram and single stars from \autoref{tab:modefreqs}
    shown in large white circles. Red circles show each star's best-fit model. Central panel: Individual models from our grid are shown in very small grey symbols. Due to the high grid density, they appear as a continuous grey area. 
    Small outer panels: the subgrid of the models selected by \dnusurf\ proximity are colour coded by combined $\chi^2$. EPIC IDs are annotated, and grey lines connect each star from the central panel with the corresponding subpanel. The scale of each subpanel is determined by the colour-magnitude range of each subgrid. Consequently, the subpanels do not necessarily share a common scaling.}
    \label{fig:mygrid}
\end{center}
\end{figure*}

To construct the model grid, we evolved tracks with initial masses between 1.23 and 1.44 \msol\ with mass step $10^{-4}$\msol\ using the same model parameters as Isochrone A and saved atmosphere-inclusive structure profiles within a stellar age range of  3.0-5.0 Gyr.
 We calculated radial frequencies as in \autoref{sec:surface} for every model in the grid with \numax < 950 \muhz\ and an age gap between models of at most 2 million years, resulting in over two million individual models. As in \autoref{sec:mass}, we used the power law derived in \autoref{sec:surface} to surface-correct the models' radial modes using the cubic method, thereby obtaining a \dnusurf\ for each model. 
For each presumed single star from \autoref{tab:modefreqs} we created a subgrid with models whose \dnusurf\ were within 1.5 times the average \textsc{pySYD}-\dnu\ uncertainty of the observed \dnu\ ($\sim$1\%, \autoref{tab:pysyd_teff})\footnote{This \dnusurf\ range for each subgrid was more than enough to span the best $\chi^{2}$ models and rapidly deteriorating $\chi^{2}$ models towards lower/higher frequencies, as will be shown in \autoref{fig:mygrid}.}. 
We then calculated a seismic $\chi^{2}$ from each star's individual radial mode frequencies (\autoref{tab:modefreqs}), and models from its subgrid as follows:
\begin{align*}
    \chi^{2}_{\mathrm{seismic}} &= \frac{1}{N} \sum_{n=0}^{N} \frac{(\nu_{\mathrm{obs},n} - \nu_{\mathrm{model},n})^{2}}{\sigma_{\mathrm{obs},n}^{2}}
\end{align*}
\noindent where $N$ is the number of observed radial modes for each star. We also derived a classical $\chi^{2}$ based on luminosity and effective temperature, assigning equal weights to both components:

\begin{align*}
    \chi^{2}_{\mathrm{classical}} &= \frac{1}{2}\left (\frac{(L_{\mathrm{obs}} - L_{\mathrm{model}})^{2}}{\sigma_{L_{\mathrm{obs}}}^{2}}\right)+ \frac{1}{2} \left (\frac{(T_{\mathrm{eff, obs}} - T_{\mathrm{eff, model}})^{2}}{\sigma_{T_{\mathrm{eff, obs}}}^{2}}\right)
\end{align*}
\noindent Following \citet{2018ApJ...864...99J}, and to avoid introducing bias through arbitrary weighting choices, we combined seismic and classical "goodness of fit" metrics into one $\chi^{2}$, as follows:
\begin{align*}
 \quad   \chi^{2} &= \;  \frac{1}{2} \;   \chi^{2}_{\mathrm{seismic}} \; +\;  \frac{1}{2} \;  \chi^{2}_{\mathrm{classical}}.
\end{align*}

It is important to note that seismic and classical $\chi^2$ statistics are not independent, as both luminosities and temperatures are informed by the seismology. While we assign equal weights to these statistics, $\chi^{2}_{\mathrm{seismic}}$ imposes strong constraints, with only a narrow "slice" of good-fit models along lines of equal density in the grid. Beyond this slice, the fit deteriorates rapidly as model frequencies deviate from observed values. In contrast, $\chi^{2}_{\mathrm{classical}}$ declines more gradually, introducing a gradient to the final $\chi^2$ statistic. This results in the best-fit models being located within an eliptical region of the grid. 

We selected a best-match result for each star based on the minimisation of $\chi^2$. 
If the models in our grid accurately represent the seismic properties (stellar interiors) and classical observables of the cluster, the best-match models should align closely with the observations in the colour-magnitude diagram. This expectation is largely met in Figure~\ref{fig:mygrid} (main panel), where Isochrone A is shown in black, and the full model grid is represented by dense grey dots, forming a broad grey band. While some individual matches are not perfect, importantly, there is no systematic offset (bias) where all the best-match models (red circles) are consistently brighter, fainter, redder, or bluer than the observed stars (white circles). This absence of directional bias suggests no overall density discrepancy between the matched models and the stars, a crucial observation for the discussion at the end of this section.
This finding is further supported by the $\chi^2$ distribution along the grid for each star. The close-ups shown along the perimeter of Figure~\ref{fig:mygrid} show each star's subgrid, chosen based on the \dnusurf\ proximity to the observed \dnu. The models from each subgrid are colour-coded according to their $\chi^2$ values, and the diagonal stripes seen in most subgrids correspond to lines of equal density, with the darkest red stripe indicating models that are good matches to the observed mode frequencies and to \teff\ and $L$ simultaneously. 

\begin{figure*}
\begin{center}
    \includegraphics[width=.65\textwidth]{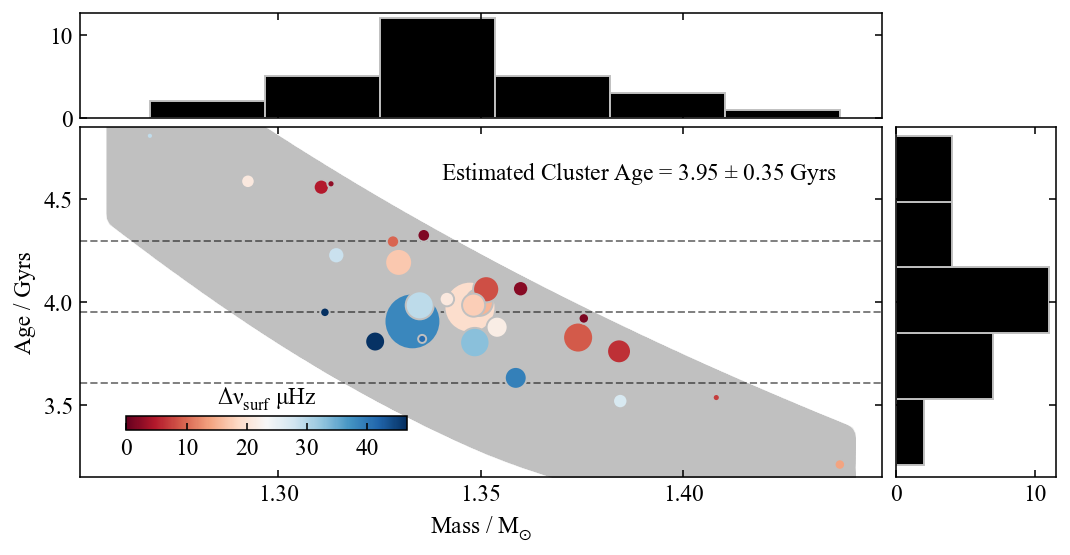}
    \caption{Ages and masses of the best-fit models to the final M67 sample. The grey band corresponds to the  masses and ages of individual models forming the full grid, and appears as a continuous area due to the high model density. Symbols are colour-coded by model \dnusurf, and their sizes are scaled inversely to their combined $\chi^{2}$ values. Histograms show number of stars in each mass or age bin.}
    \label{fig:fit}
\end{center}
\end{figure*}

The parameters of the best-fit models are presented in \autoref{tab:fitresults}. In Figures \ref{fig:echelles1} to \ref{fig:echelles4}, we show the observed radial modes (small cyan circle) and best-fit model (open orange circles) in echelle diagrams for each star. These figures include collapsed echelles to highlight the radial ridge positions.

According to $\chi^{2}$, the worse-fitted stars are those of EPIC IDs 211410817 ($\chi^{2}$=1.1096), 211414300 ($\chi^{2}$=1.1322), and 211413064 ($\chi^{2}$=1.4938). Despite not being flagged as multiple star systems or having poor photometry, these stars deviate from the cluster sequence in the colour-magnitude diagram. This could indicate underestimated uncertainties, potential issues with their membership classification, unrecognized multiplicity, or other astrophysical phenomena affecting their photometry. Indeed, there is doubt about the membership classifications of 211410817 and 211413064. According to R24, these two stars were classified as cluster members by \citet{2015AJ....150...97G}, but could not be confirmed with {\it Gaia} DR3 astrometry because the radial velocities of 211410817 are not available, and the proper motions of 211413064 seem to contradict the membership status by \citet{2015AJ....150...97G}.

To get an overview of the fitting results for the sample as an ensemble, we show in Figure~\ref{fig:fit} the ages and masses of the best-fit model for each star, where the colour-coding indicates the model's \dnusurf. As expected, for a fixed age, the least evolved stars (dark blue) are of lower mass than the most evolved stars (dark red). The diagonal trend at fixed \dnusurf\ (colour) reflects the intrinsic age-mass relation of the model grid, arising from age being the only parameter varied (because the other parameters were already fit to the cluster in R24). For clarity, we also show in this mass-age diagram all models in the grid, appearing as a continuous grey band. The histograms along both axis show the mass and age distributions of the $\chi^2$ fit.

Lastly, we estimated an average cluster age from the best-fit models using the inverse of their combined $\chi^2$ as weights, represented by the circle size in Figure~\ref{fig:fit}. This resulted in a weighted average cluster age of $3.95 \pm 0.35\, \mathrm{Gyrs}$\footnote{A non-weighted average of the ages results in $4.01 \pm 0.35$ Gyrs.}, where the uncertainty is derived from the standard deviation of the age values in the best-fit models, and is indicated by horizontal dashed lines around the average. This age aligns with that obtained via classical isochrone fitting to the cluster's main sequence and red-giant branch stars by R24, though we obtain a larger uncertainty here. The fact that we obtain an age similar to R24 is likely because all models along the darkest red stripes in the subgrids of \autoref{fig:mygrid}, which indicate low combined $\chi^2$, are also good seismic matches. There, younger models would appear on the left side of the stripe and older ones on the right). Thus, age is predominantly determined by $\chi^{2}_{\mathrm{classical}}$ along the stripe, closely approximating classical isochrone fitting.  Our larger age uncertainty here may be attributed to our smaller sample size of 28 stars compared to the nearly 200 stars in the R24 sample, which also included main-sequence and turnoff stars.
Because we find the same age as R24, like them, we conclude our age estimate is in agreement with the estimates from \citet{2004PASP..116..997V} ($4.0 \pm 0.4\, \mathrm{Gyrs}$ from $BV$ photometry), \citet{2009ApJ...698.1872S} ($3.5-4.0\, \mathrm{Gyrs}$ from 2MASS photometry), \citet{2010A&A...513A..50B} ($3.9 \pm 0.1\, \mathrm{Gyrs}$ from the magnitude of the white-dwarf cooling sequence), \citet{2016ApJ...823...16B} ($4.2 \pm 0.2\, \mathrm{Gyrs}$ from gyrochronology) and \citet{2022A&A...665A.126N} ($3.9\, \mathrm{Gyrs}$ from {\it Gaia} DR2 photometry using PARSEC isochrones v2.0).

The fact that our averaged age agrees with the age of Isochorne A 
means that Isochrone A agrees not only with the classical observables of all the cluster stars (as per R24) but also with the radial mode frequencies of the subgiants and red giants (this work).
This has implications for the unresolved issue of the eclipsing binary WOCS 11028. Notably, \citet{2021AJ....161...59S} reported precise masses for the eclipsing binary WOCS 11028 near the cluster's TAMS, but the primary component, WOCS 11028a, does not align with any existing isochrone for the cluster \citep{2021AJ....161...59S, 2022A&A...665A.126N}. There are discrepancies of at least 0.17 magnitudes between the deconvoluted primary photometry and the predicted luminosity for the primary's mass ($1.222 \pm 0.006$ \msol), with the discrepancy with Isochrone A being 0.2 magnitudes. To further investigate this discrepancy, we applied our $\chi^2$ fitting approach as described above, but instead, we used models from an alternative isochrone made to fit the primary component and the rest of the cluster's morphology in the colour-magnitude diagram (Isochrone EB from R24). In that case, the fit revealed a one-directional offset for all stars in our sample, where the best seismic fits on the red giant branch were at least 0.1 magnitudes fainter than the observed values. This further confirms that the mass of WOCS 11028a cannot be reconciled with the stars in our seismic sample in a scenario where they all have evolved independently and were born from the same primordial molecular cloud. This therefore leaves us with two possibilities. Either the inconsistency stems from WOCS 11028a, or more concerning, it could suggest our current models may be flawed for masses (or evolutionary phases) close to the ones of WOCS 11028a. 
Given that (1) the cluster membership of WOCS 11028 is undisputed, (2) there are indications the binary components have not interacted in the past \citep{2021AJ....161...59S}, and (3) their fundamental parameters are fairly well established, our inability to align WOCS 11028a with the rest of the stars supports the need for further studies into the issue.

\section{Summary, conclusions, and future work}

In this work, we used Kepler/K2 data to obtain seismic parameters of subgiants and red giants of the Open Cluster M67. Specifically:

\begin{enumerate}
    \item We obtained and evaluated $\log g$ against the modelled isochrone from R24 in the Kiel diagram and found remarkably good agreement.
    \item We measured the ratio of cluster stars with possible internal magnetic fields in their cores 
    using the mode suppression method and found that the suppression rate in M67 exceeds that found in comparable field red giant branch stars. By looking at the suppression rate of red clump stars, we provide evidence that magnetic cores do not normally survive the helium flash.
    \item We measured individual mode frequencies, which allowed us to develop an expression for correcting the surface term that changes smoothly with stellar evolution. 
    \item For the first time, we mapped the evolution of the phase term $\epsilon$ in the transition between main sequence turn-off and red giant branch stars.
    \item We successfully calibrated the masses and radii from scaling relations against our models and showed how reliable the masses derived from scaling relations truly are. 
    \item  We performed $\chi^2$ fitting over a grid of models using the radial modes of 28 stars, including giants and subgiants. This approach yielded the most precise subgiant and red giant branch star masses reported to date for the cluster. Importantly, our seismic fitting results are consistent with classical isochrone fitting methods. Still, our age estimates remain subject to potential age systematics stemming from our choices of input physics (see Figures 42 and 43 in \cite{2014EAS....65...99L}).
Howerver, we note that the corrections applied in this work, whether for \teff\ offsets, the surface term, or $f_{\nu_{\mathrm{max}}}$ and $f_{\Delta\nu}$ correction factors agree with literature values.
In this context it is also worth noting that our models agree with so-called small frequency separations, which probe the deep interiors of stars (Reyes 2025, in preparation).
\end{enumerate}

The main avenues for future work include a more detailed investigation of the red clump stars, such as identifying their individual frequencies, conducting in-depth comparisons with models, and assessing whether the surface correction differs from that of the red giant branch stars. Furthermore, one could incorporate non-radial mixed modes in the analysis for all the stars. One could potentially obtain improvements by using SED fitting to derive effective temperatures and bolometric fluxes. Finally, the characterisation of the yellow straggler (Appendix~\ref{sec:blue}) would benefit from a comprehensive modeling approach that includes uncertainty estimations of the fundamental stellar properties.

\section*{Software}

This work uses the following software packages:
\begin{itemize}
    \item \textsc{FAMED} version 0.0.1 \citet{2020A&A...640A.130C}.
    \item \textsc{MESA} version 23.05.1 \citet{2011ApJS..192....3P, 2013ApJS..208....4P,  2015ApJS..220...15P, 2018ApJS..234...34P, 2019ApJS..243...10P, 2023ApJS..265...15J}.
    \item \textsc{GYRE} version 6.0.1 \citet{2013MNRAS.435.3406T}.
\end{itemize}

\section*{Acknowledgements}

D.S. is supported by the Australian Research Council (DP190100666).

R.D.M. acknowledges the support of NASA grants NNX15AW69G, NSSC18K0449, and NSSC19K0105.

The authors thank Joel Ong, Guy Davies, and Jim Fuller for useful comments on the manuscript.

This paper includes data collected by the Kepler mission and obtained from the MAST data archive at the Space Telescope Science Institute (STScI). Funding for the Kepler mission is provided by the NASA Science Mission Directorate. STScI is operated by the Association of Universities for Research in Astronomy, Inc., under NASA contract NAS 5–26555.

This work has made use of data from the European Space Agency (ESA) mission
{\it Gaia} (\url{https://www.cosmos.esa.int/Gaia}), processed by the {\it Gaia}
Data Processing and Analysis Consortium (DPAC,
\url{https://www.cosmos.esa.int/web/Gaia/dpac/consortium}). Funding for the DPAC
has been provided by national institutions, in particular the institutions
participating in the {\it Gaia} Multilateral Agreement.

This research includes computations using the computational cluster Katana supported by Research Technology Services at UNSW Sydney, \url{https://doi.org/10.26190/669x-a286}.

\section*{Data Availability}
 \label{sec:availability}
Isochrone A is available for download at \href{https://doi.org/10.5281/zenodo.12616440}{DOI: 10.5281/zenodo.12616440}. 

\noindent The grid of models presented in this work can be found at \href{https://doi.org/10.5281/zenodo.14927665}{DOI: 10.5281/zenodo.14927665}.


\bibliographystyle{mnras}

\begin{thebibliography}{}
\makeatletter
\relax
\def\mn@urlcharsother{\let\do\@makeother \do\$\do\&\do\#\do\^\do\_\do\%\do\~}
\def\mn@doi{\begingroup\mn@urlcharsother \@ifnextchar [ {\mn@doi@} {\mn@doi@[]}}
\def\mn@doi@[#1]#2{\def\@tempa{#1}\ifx\@tempa\@empty \href {http://dx.doi.org/#2} {doi:#2}\else \href {http://dx.doi.org/#2} {#1}\fi \endgroup}
\def\mn@eprint#1#2{\mn@eprint@#1:#2::\@nil}
\def\mn@eprint@arXiv#1{\href {http://arxiv.org/abs/#1} {{\tt arXiv:#1}}}
\def\mn@eprint@dblp#1{\href {http://dblp.uni-trier.de/rec/bibtex/#1.xml} {dblp:#1}}
\def\mn@eprint@#1:#2:#3:#4\@nil{\def\@tempa {#1}\def\@tempb {#2}\def\@tempc {#3}\ifx \@tempc \@empty \let \@tempc \@tempb \let \@tempb \@tempa \fi \ifx \@tempb \@empty \def\@tempb {arXiv}\fi \@ifundefined {mn@eprint@\@tempb}{\@tempb:\@tempc}{\expandafter \expandafter \csname mn@eprint@\@tempb\endcsname \expandafter{\@tempc}}}

\bibitem[\protect\citeauthoryear{{Aerts}}{{Aerts}}{2021}]{2021RvMP...93a5001A}
{Aerts} C.,  2021, \mn@doi [Reviews of Modern Physics] {10.1103/RevModPhys.93.015001}, \href {https://ui.adsabs.harvard.edu/abs/2021RvMP...93a5001A} {93, 015001}

\bibitem[\protect\citeauthoryear{{Aguirre B{\o}rsen-Koch} et~al.,}{{Aguirre B{\o}rsen-Koch} et~al.}{2022}]{2022MNRAS.509.4344A}
{Aguirre B{\o}rsen-Koch} V.,  et~al., 2022, \mn@doi [\mnras] {10.1093/mnras/stab2911}, \href {https://ui.adsabs.harvard.edu/abs/2022MNRAS.509.4344A} {509, 4344}

\bibitem[\protect\citeauthoryear{{An}, {Pinsonneault}, {Terndrup}  \& {Chung}}{{An} et~al.}{2019}]{2019ApJ...879...81A}
{An} D.,  {Pinsonneault} M.~H.,  {Terndrup} D.~M.,   {Chung} C.,  2019, \mn@doi [\apj] {10.3847/1538-4357/ab23ed}, \href {https://ui.adsabs.harvard.edu/abs/2019ApJ...879...81A} {879, 81}

\bibitem[\protect\citeauthoryear{{Ash}, {Pinsonneault}, {Vrard}  \& {Zinn}}{{Ash} et~al.}{2024}]{2024arXiv241110520A}
{Ash} A.~L.,  {Pinsonneault} M.~H.,  {Vrard} M.,   {Zinn} J.,  2024, \mn@doi [arXiv e-prints] {10.48550/arXiv.2411.10520}, \href {https://ui.adsabs.harvard.edu/abs/2024arXiv241110520A} {p. arXiv:2411.10520}

\bibitem[\protect\citeauthoryear{{Ball}}{{Ball}}{2021}]{2021RNAAS...5....7B}
{Ball} W.~H.,  2021, \mn@doi [Research Notes of the American Astronomical Society] {10.3847/2515-5172/abd9cb}, \href {https://ui.adsabs.harvard.edu/abs/2021RNAAS...5....7B} {5, 7}

\bibitem[\protect\citeauthoryear{{Ball} \& {Gizon}}{{Ball} \& {Gizon}}{2014}]{2014A&A...568A.123B}
{Ball} W.~H.,  {Gizon} L.,  2014, \mn@doi [\aap] {10.1051/0004-6361/201424325}, \href {https://ui.adsabs.harvard.edu/abs/2014A&A...568A.123B} {568, A123}

\bibitem[\protect\citeauthoryear{{Barnes}, {Weingrill}, {Fritzewski}, {Strassmeier}  \& {Platais}}{{Barnes} et~al.}{2016}]{2016ApJ...823...16B}
{Barnes} S.~A.,  {Weingrill} J.,  {Fritzewski} D.,  {Strassmeier} K.~G.,   {Platais} I.,  2016, \mn@doi [\apj] {10.3847/0004-637X/823/1/16}, \href {https://ui.adsabs.harvard.edu/abs/2016ApJ...823...16B} {823, 16}

\bibitem[\protect\citeauthoryear{{Belkacem}, {Samadi}, {Mosser}, {Goupil}  \& {Ludwig}}{{Belkacem} et~al.}{2013}]{2013ASPC..479...61B}
{Belkacem} K.,  {Samadi} R.,  {Mosser} B.,  {Goupil} M.~J.,   {Ludwig} H.~G.,  2013, in {Shibahashi} H.,  {Lynas-Gray} A.~E.,  eds,  Astronomical Society of the Pacific Conference Series Vol. 479, Progress in Physics of the Sun and Stars: A New Era in Helio- and Asteroseismology. p.~61 (\mn@eprint {arXiv} {1307.3132}), \mn@doi{10.48550/arXiv.1307.3132}

\bibitem[\protect\citeauthoryear{{Bellini} et~al.,}{{Bellini} et~al.}{2010}]{2010A&A...513A..50B}
{Bellini} A.,  et~al., 2010, \mn@doi [\aap] {10.1051/0004-6361/200913721}, \href {https://ui.adsabs.harvard.edu/abs/2010A&A...513A..50B} {513, A50}

\bibitem[\protect\citeauthoryear{{Bertelli Motta}, {Pasquali}, {Caffau}  \& {Grebel}}{{Bertelli Motta} et~al.}{2018}]{2018MNRAS.480.4314B}
{Bertelli Motta} C.,  {Pasquali} A.,  {Caffau} E.,   {Grebel} E.~K.,  2018, \mn@doi [\mnras] {10.1093/mnras/sty2147}, \href {https://ui.adsabs.harvard.edu/abs/2018MNRAS.480.4314B} {480, 4314}

\bibitem[\protect\citeauthoryear{{Brogaard} et~al.,}{{Brogaard} et~al.}{2018}]{2018MNRAS.476.3729B}
{Brogaard} K.,  et~al., 2018, \mn@doi [\mnras] {10.1093/mnras/sty268}, \href {https://ui.adsabs.harvard.edu/abs/2018MNRAS.476.3729B} {476, 3729}

\bibitem[\protect\citeauthoryear{{Brown}, {Gilliland}, {Noyes}  \& {Ramsey}}{{Brown} et~al.}{1991}]{1991ApJ...368..599B}
{Brown} T.~M.,  {Gilliland} R.~L.,  {Noyes} R.~W.,   {Ramsey} L.~W.,  1991, \mn@doi [\apj] {10.1086/169725}, \href {https://ui.adsabs.harvard.edu/abs/1991ApJ...368..599B} {368, 599}

\bibitem[\protect\citeauthoryear{{Buder} et~al.,}{{Buder} et~al.}{2021}]{2021MNRAS.506..150B}
{Buder} S.,  et~al., 2021, \mn@doi [\mnras] {10.1093/mnras/stab1242}, \href {https://ui.adsabs.harvard.edu/abs/2021MNRAS.506..150B} {506, 150}

\bibitem[\protect\citeauthoryear{{Cantiello}, {Fuller}  \& {Bildsten}}{{Cantiello} et~al.}{2016}]{2016ApJ...824...14C}
{Cantiello} M.,  {Fuller} J.,   {Bildsten} L.,  2016, \mn@doi [\apj] {10.3847/0004-637X/824/1/14}, \href {https://ui.adsabs.harvard.edu/abs/2016ApJ...824...14C} {824, 14}

\bibitem[\protect\citeauthoryear{{Carraro}, {Girardi}, {Bressan}  \& {Chiosi}}{{Carraro} et~al.}{1996}]{1996A&A...305..849C}
{Carraro} G.,  {Girardi} L.,  {Bressan} A.,   {Chiosi} C.,  1996, \aap, \href {https://ui.adsabs.harvard.edu/abs/1996A&A...305..849C} {305, 849}

\bibitem[\protect\citeauthoryear{{Casagrande} \& {VandenBerg}}{{Casagrande} \& {VandenBerg}}{2018}]{2018MNRAS.479L.102C}
{Casagrande} L.,  {VandenBerg} D.~A.,  2018, \mn@doi [\mnras] {10.1093/mnrasl/sly104}, \href {https://ui.adsabs.harvard.edu/abs/2018MNRAS.479L.102C} {479, L102}

\bibitem[\protect\citeauthoryear{{Casagrande} et~al.,}{{Casagrande} et~al.}{2016}]{2016MNRAS.455..987C}
{Casagrande} L.,  et~al., 2016, \mn@doi [\mnras] {10.1093/mnras/stv2320}, \href {https://ui.adsabs.harvard.edu/abs/2016MNRAS.455..987C} {455, 987}

\bibitem[\protect\citeauthoryear{{Casagrande} et~al.,}{{Casagrande} et~al.}{2021}]{2021MNRAS.507.2684C}
{Casagrande} L.,  et~al., 2021, \mn@doi [\mnras] {10.1093/mnras/stab2304}, \href {https://ui.adsabs.harvard.edu/abs/2021MNRAS.507.2684C} {507, 2684}

\bibitem[\protect\citeauthoryear{{Chaplin} \& {Miglio}}{{Chaplin} \& {Miglio}}{2013}]{2013ARA&A..51..353C}
{Chaplin} W.~J.,  {Miglio} A.,  2013, \mn@doi [\araa] {10.1146/annurev-astro-082812-140938}, \href {https://ui.adsabs.harvard.edu/abs/2013ARA&A..51..353C} {51, 353}

\bibitem[\protect\citeauthoryear{{Choi}, {Dotter}, {Conroy}, {Cantiello}, {Paxton}  \& {Johnson}}{{Choi} et~al.}{2016}]{2016ApJ...823..102C}
{Choi} J.,  {Dotter} A.,  {Conroy} C.,  {Cantiello} M.,  {Paxton} B.,   {Johnson} B.~D.,  2016, \mn@doi [\apj] {10.3847/0004-637X/823/2/102}, \href {https://ui.adsabs.harvard.edu/abs/2016ApJ...823..102C} {823, 102}

\bibitem[\protect\citeauthoryear{{Chontos}, {Huber}, {Sayeed}  \& {Yamsiri}}{{Chontos} et~al.}{2021}]{2021arXiv210800582C}
{Chontos} A.,  {Huber} D.,  {Sayeed} M.,   {Yamsiri} P.,  2021, arXiv e-prints, \href {https://ui.adsabs.harvard.edu/abs/2021arXiv210800582C} {p. arXiv:2108.00582}

\bibitem[\protect\citeauthoryear{{Christensen-Dalsgaard}, {Dappen}  \& {Lebreton}}{{Christensen-Dalsgaard} et~al.}{1988}]{1988Natur.336..634C}
{Christensen-Dalsgaard} J.,  {Dappen} W.,   {Lebreton} Y.,  1988, \mn@doi [\nat] {10.1038/336634a0}, \href {https://ui.adsabs.harvard.edu/abs/1988Natur.336..634C} {336, 634}

\bibitem[\protect\citeauthoryear{{Christensen-Dalsgaard} et~al.,}{{Christensen-Dalsgaard} et~al.}{1996}]{1996Sci...272.1286C}
{Christensen-Dalsgaard} J.,  et~al., 1996, \mn@doi [Science] {10.1126/science.272.5266.1286}, \href {https://ui.adsabs.harvard.edu/abs/1996Sci...272.1286C} {272, 1286}

\bibitem[\protect\citeauthoryear{{Corsaro} \& {De Ridder}}{{Corsaro} \& {De Ridder}}{2014}]{2014A&A...571A..71C}
{Corsaro} E.,  {De Ridder} J.,  2014, \mn@doi [\aap] {10.1051/0004-6361/201424181}, \href {https://ui.adsabs.harvard.edu/abs/2014A&A...571A..71C} {571, A71}

\bibitem[\protect\citeauthoryear{{Corsaro} et~al.,}{{Corsaro} et~al.}{2012}]{2012ApJ...757..190C}
{Corsaro} E.,  et~al., 2012, \mn@doi [\apj] {10.1088/0004-637X/757/2/190}, \href {https://ui.adsabs.harvard.edu/abs/2012ApJ...757..190C} {757, 190}

\bibitem[\protect\citeauthoryear{{Corsaro}, {McKeever}  \& {Kuszlewicz}}{{Corsaro} et~al.}{2020}]{2020A&A...640A.130C}
{Corsaro} E.,  {McKeever} J.~M.,   {Kuszlewicz} J.~S.,  2020, \mn@doi [\aap] {10.1051/0004-6361/202037930}, \href {https://ui.adsabs.harvard.edu/abs/2020A&A...640A.130C} {640, A130}

\bibitem[\protect\citeauthoryear{{Cox}}{{Cox}}{2000}]{2000asqu.book.....C}
{Cox} A.~N.,  2000, {Allen's astrophysical quantities}

\bibitem[\protect\citeauthoryear{{Evans} et~al.,}{{Evans} et~al.}{2018}]{2018A&A...616A...4E}
{Evans} D.~W.,  et~al., 2018, \mn@doi [\aap] {10.1051/0004-6361/201832756}, \href {https://ui.adsabs.harvard.edu/abs/2018A&A...616A...4E} {616, A4}

\bibitem[\protect\citeauthoryear{{Frandsen}, {Jones}, {Kjeldsen}, {Viskum}, {Hjorth}, {Andersen}  \& {Thomsen}}{{Frandsen} et~al.}{1995}]{1995A&A...301..123F}
{Frandsen} S.,  {Jones} A.,  {Kjeldsen} H.,  {Viskum} M.,  {Hjorth} J.,  {Andersen} N.~H.,   {Thomsen} B.,  1995, \aap, \href {https://ui.adsabs.harvard.edu/abs/1995A&A...301..123F} {301, 123}

\bibitem[\protect\citeauthoryear{{Fuller}, {Cantiello}, {Stello}, {Garcia}  \& {Bildsten}}{{Fuller} et~al.}{2015}]{2015Sci...350..423F}
{Fuller} J.,  {Cantiello} M.,  {Stello} D.,  {Garcia} R.~A.,   {Bildsten} L.,  2015, \mn@doi [Science] {10.1126/science.aac6933}, \href {https://ui.adsabs.harvard.edu/abs/2015Sci...350..423F} {350, 423}

\bibitem[\protect\citeauthoryear{{Gaia Collaboration} et~al.,}{{Gaia Collaboration} et~al.}{2018}]{2018A&A...616A...1G}
{Gaia Collaboration} et~al., 2018, \mn@doi [\aap] {10.1051/0004-6361/201833051}, \href {https://ui.adsabs.harvard.edu/abs/2018A&A...616A...1G} {616, A1}

\bibitem[\protect\citeauthoryear{{Gaia Collaboration} et~al.,}{{Gaia Collaboration} et~al.}{2022}]{2022arXiv220605595G}
{Gaia Collaboration} et~al., 2022, \mn@doi [arXiv e-prints] {10.48550/arXiv.2206.05595}, \href {https://ui.adsabs.harvard.edu/abs/2022arXiv220605595G} {p. arXiv:2206.05595}

\bibitem[\protect\citeauthoryear{{Gaia Collaboration} et~al.,}{{Gaia Collaboration} et~al.}{2023}]{2023A&A...674A...1G}
{Gaia Collaboration} et~al., 2023, \mn@doi [\aap] {10.1051/0004-6361/202243940}, \href {https://ui.adsabs.harvard.edu/abs/2023A&A...674A...1G} {674, A1}

\bibitem[\protect\citeauthoryear{{Geller}, {Latham}  \& {Mathieu}}{{Geller} et~al.}{2015}]{2015AJ....150...97G}
{Geller} A.~M.,  {Latham} D.~W.,   {Mathieu} R.~D.,  2015, \mn@doi [\aj] {10.1088/0004-6256/150/3/97}, \href {https://ui.adsabs.harvard.edu/abs/2015AJ....150...97G} {150, 97}

\bibitem[\protect\citeauthoryear{{Gilliland} et~al.,}{{Gilliland} et~al.}{1993}]{1993AJ....106.2441G}
{Gilliland} R.~L.,  et~al., 1993, \mn@doi [\aj] {10.1086/116814}, \href {https://ui.adsabs.harvard.edu/abs/1993AJ....106.2441G} {106, 2441}

\bibitem[\protect\citeauthoryear{{Handberg}}{{Handberg}}{2013}]{2013PhDT.......408H}
{Handberg} R.,  2013, PhD thesis, Aarhus University, Denmark

\bibitem[\protect\citeauthoryear{{Handberg}, {Brogaard}, {Miglio}, {Bossini}, {Elsworth}, {Slumstrup}, {Davies}  \& {Chaplin}}{{Handberg} et~al.}{2017}]{2017MNRAS.472..979H}
{Handberg} R.,  {Brogaard} K.,  {Miglio} A.,  {Bossini} D.,  {Elsworth} Y.,  {Slumstrup} D.,  {Davies} G.~R.,   {Chaplin} W.~J.,  2017, \mn@doi [\mnras] {10.1093/mnras/stx1929}, \href {https://ui.adsabs.harvard.edu/abs/2017MNRAS.472..979H} {472, 979}

\bibitem[\protect\citeauthoryear{{Harvey}}{{Harvey}}{1985}]{1985ESASP.235..199H}
{Harvey} J.,  1985, in {Rolfe} E.,  {Battrick} B.,  eds,  ESA Special Publication Vol. 235, Future Missions in Solar, Heliospheric \& Space Plasma Physics. p.~199

\bibitem[\protect\citeauthoryear{{Hekker}}{{Hekker}}{2020}]{2020FrASS...7....3H}
{Hekker} S.,  2020, \mn@doi [Frontiers in Astronomy and Space Sciences] {10.3389/fspas.2020.00003}, \href {https://ui.adsabs.harvard.edu/abs/2020FrASS...7....3H} {7, 3}

\bibitem[\protect\citeauthoryear{{Houdek} \& {Dupret}}{{Houdek} \& {Dupret}}{2015}]{2015LRSP...12....8H}
{Houdek} G.,  {Dupret} M.-A.,  2015, \mn@doi [Living Reviews in Solar Physics] {10.1007/lrsp-2015-8}, \href {https://ui.adsabs.harvard.edu/abs/2015LRSP...12....8H} {12, 8}

\bibitem[\protect\citeauthoryear{Howell et~al.,}{Howell et~al.}{2014}]{Howell_2014}
Howell S.~B.,  et~al., 2014, \mn@doi [Publications of the Astronomical Society of the Pacific] {10.1086/676406}, 126, 398

\bibitem[\protect\citeauthoryear{{Howell}, {Campbell}, {Stello}  \& {De Silva}}{{Howell} et~al.}{2022}]{2022MNRAS.515.3184H}
{Howell} M.,  {Campbell} S.~W.,  {Stello} D.,   {De Silva} G.~M.,  2022, \mn@doi [\mnras] {10.1093/mnras/stac1918}, \href {https://ui.adsabs.harvard.edu/abs/2022MNRAS.515.3184H} {515, 3184}

\bibitem[\protect\citeauthoryear{{Huber}, {Stello}, {Bedding}, {Chaplin}, {Arentoft}, {Quirion}  \& {Kjeldsen}}{{Huber} et~al.}{2009}]{2009CoAst.160...74H}
{Huber} D.,  {Stello} D.,  {Bedding} T.~R.,  {Chaplin} W.~J.,  {Arentoft} T.,  {Quirion} P.~O.,   {Kjeldsen} H.,  2009, \mn@doi [Communications in Asteroseismology] {10.48550/arXiv.0910.2764}, \href {https://ui.adsabs.harvard.edu/abs/2009CoAst.160...74H} {160, 74}

\bibitem[\protect\citeauthoryear{{Huber} et~al.,}{{Huber} et~al.}{2011}]{2011ApJ...743..143H}
{Huber} D.,  et~al., 2011, \mn@doi [\apj] {10.1088/0004-637X/743/2/143}, \href {https://ui.adsabs.harvard.edu/abs/2011ApJ...743..143H} {743, 143}

\bibitem[\protect\citeauthoryear{{Huber}, {Bryson}  \& {et al.}}{{Huber} et~al.}{2017}]{2017yCat.4034....0H}
{Huber} D.,  {Bryson} S.~T.,   {et al.} 2017, VizieR Online Data Catalog, \href {https://ui.adsabs.harvard.edu/abs/2017yCat.4034....0H} {p. IV/34}

\bibitem[\protect\citeauthoryear{{Jermyn} et~al.,}{{Jermyn} et~al.}{2023}]{2023ApJS..265...15J}
{Jermyn} A.~S.,  et~al., 2023, \mn@doi [\apjs] {10.3847/1538-4365/acae8d}, \href {https://ui.adsabs.harvard.edu/abs/2023ApJS..265...15J} {265, 15}

\bibitem[\protect\citeauthoryear{{Joyce} \& {Chaboyer}}{{Joyce} \& {Chaboyer}}{2018}]{2018ApJ...864...99J}
{Joyce} M.,  {Chaboyer} B.,  2018, \mn@doi [\apj] {10.3847/1538-4357/aad464}, \href {https://ui.adsabs.harvard.edu/abs/2018ApJ...864...99J} {864, 99}

\bibitem[\protect\citeauthoryear{{Kallinger} et~al.,}{{Kallinger} et~al.}{2010}]{2010A&A...509A..77K}
{Kallinger} T.,  et~al., 2010, \mn@doi [\aap] {10.1051/0004-6361/200811437}, \href {https://ui.adsabs.harvard.edu/abs/2010A&A...509A..77K} {509, A77}

\bibitem[\protect\citeauthoryear{{Kallinger}, {Beck}, {Stello}  \& {Garcia}}{{Kallinger} et~al.}{2018}]{2018A&A...616A.104K}
{Kallinger} T.,  {Beck} P.~G.,  {Stello} D.,   {Garcia} R.~A.,  2018, \mn@doi [\aap] {10.1051/0004-6361/201832831}, \href {https://ui.adsabs.harvard.edu/abs/2018A&A...616A.104K} {616, A104}

\bibitem[\protect\citeauthoryear{{Khan} et~al.,}{{Khan} et~al.}{2019}]{2019A&A...628A..35K}
{Khan} S.,  et~al., 2019, \mn@doi [\aap] {10.1051/0004-6361/201935304}, \href {https://ui.adsabs.harvard.edu/abs/2019A&A...628A..35K} {628, A35}

\bibitem[\protect\citeauthoryear{{Kjeldsen} \& {Bedding}}{{Kjeldsen} \& {Bedding}}{1995}]{1995A&A...293...87K}
{Kjeldsen} H.,  {Bedding} T.~R.,  1995, \mn@doi [\aap] {10.48550/arXiv.astro-ph/9403015}, \href {https://ui.adsabs.harvard.edu/abs/1995A&A...293...87K} {293, 87}

\bibitem[\protect\citeauthoryear{{Kjeldsen}, {Bedding}  \& {Christensen-Dalsgaard}}{{Kjeldsen} et~al.}{2008}]{2008ApJ...683L.175K}
{Kjeldsen} H.,  {Bedding} T.~R.,   {Christensen-Dalsgaard} J.,  2008, \mn@doi [\apjl] {10.1086/591667}, \href {https://ui.adsabs.harvard.edu/abs/2008ApJ...683L.175K} {683, L175}

\bibitem[\protect\citeauthoryear{{Lebreton}, {Goupil}  \& {Montalb{\'a}n}}{{Lebreton} et~al.}{2014}]{2014EAS....65...99L}
{Lebreton} Y.,  {Goupil} M.~J.,   {Montalb{\'a}n} J.,  2014, in {Lebreton} Y.,  {Valls-Gabaud} D.,   {Charbonnel} C.,  eds,  EAS Publications Series Vol. 65, EAS Publications Series. pp 99--176 (\mn@eprint {arXiv} {1410.5336}), \mn@doi{10.1051/eas/1465004}

\bibitem[\protect\citeauthoryear{{Li}, {Bedding}, {Christensen-Dalsgaard}, {Stello}, {Li}  \& {Keen}}{{Li} et~al.}{2020}]{2020MNRAS.495.3431L}
{Li} T.,  {Bedding} T.~R.,  {Christensen-Dalsgaard} J.,  {Stello} D.,  {Li} Y.,   {Keen} M.~A.,  2020, \mn@doi [\mnras] {10.1093/mnras/staa1350}, \href {https://ui.adsabs.harvard.edu/abs/2020MNRAS.495.3431L} {495, 3431}

\bibitem[\protect\citeauthoryear{{Li}, {Li}, {Bi}, {Bedding}, {Davies}  \& {Du}}{{Li} et~al.}{2022}]{2022ApJ...927..167L}
{Li} T.,  {Li} Y.,  {Bi} S.,  {Bedding} T.~R.,  {Davies} G.,   {Du} M.,  2022, \mn@doi [\apj] {10.3847/1538-4357/ac4fbf}, \href {https://ui.adsabs.harvard.edu/abs/2022ApJ...927..167L} {927, 167}

\bibitem[\protect\citeauthoryear{{Li} et~al.,}{{Li} et~al.}{2023}]{2023MNRAS.523..916L}
{Li} Y.,  et~al., 2023, \mn@doi [\mnras] {10.1093/mnras/stad1445}, \href {https://ui.adsabs.harvard.edu/abs/2023MNRAS.523..916L} {523, 916}

\bibitem[\protect\citeauthoryear{{Li}, {Bi}, {Davies}, {Bedding}, {Li}, {Stello}  \& {Reyes}}{{Li} et~al.}{2024}]{2024MNRAS.530.2810L}
{Li} T.,  {Bi} S.,  {Davies} G.~R.,  {Bedding} T.~R.,  {Li} Y.,  {Stello} D.,   {Reyes} C.,  2024, \mn@doi [\mnras] {10.1093/mnras/stae1026}, \href {https://ui.adsabs.harvard.edu/abs/2024MNRAS.530.2810L} {530, 2810}

\bibitem[\protect\citeauthoryear{{Lindegren} et~al.,}{{Lindegren} et~al.}{2021}]{2021A&A...649A...4L}
{Lindegren} L.,  et~al., 2021, \mn@doi [\aap] {10.1051/0004-6361/202039653}, \href {https://ui.adsabs.harvard.edu/abs/2021A&A...649A...4L} {649, A4}

\bibitem[\protect\citeauthoryear{{Liu}, {Deng}, {Ch{\'a}vez}, {Bertone}, {Davo}  \& {Mata-Ch{\'a}vez}}{{Liu} et~al.}{2008}]{2008MNRAS.390..665L}
{Liu} G.~Q.,  {Deng} L.,  {Ch{\'a}vez} M.,  {Bertone} E.,  {Davo} A.~H.,   {Mata-Ch{\'a}vez} M.~D.,  2008, \mn@doi [\mnras] {10.1111/j.1365-2966.2008.13741.x}, \href {https://ui.adsabs.harvard.edu/abs/2008MNRAS.390..665L} {390, 665}

\bibitem[\protect\citeauthoryear{{Lund} et~al.,}{{Lund} et~al.}{2017}]{2017ApJ...835..172L}
{Lund} M.~N.,  et~al., 2017, \mn@doi [\apj] {10.3847/1538-4357/835/2/172}, \href {https://ui.adsabs.harvard.edu/abs/2017ApJ...835..172L} {835, 172}

\bibitem[\protect\citeauthoryear{{Mamajek} et~al.,}{{Mamajek} et~al.}{2015}]{2015arXiv151006262M}
{Mamajek} E.~E.,  et~al., 2015, \mn@doi [arXiv e-prints] {10.48550/arXiv.1510.06262}, \href {https://ui.adsabs.harvard.edu/abs/2015arXiv151006262M} {p. arXiv:1510.06262}

\bibitem[\protect\citeauthoryear{{Mathieu} \& {Latham}}{{Mathieu} \& {Latham}}{1986}]{1986AJ.....92.1364M}
{Mathieu} R.~D.,  {Latham} D.~W.,  1986, \mn@doi [\aj] {10.1086/114269}, \href {https://ui.adsabs.harvard.edu/abs/1986AJ.....92.1364M} {92, 1364}

\bibitem[\protect\citeauthoryear{{Miglio} et~al.,}{{Miglio} et~al.}{2012}]{2012MNRAS.419.2077M}
{Miglio} A.,  et~al., 2012, \mn@doi [\mnras] {10.1111/j.1365-2966.2011.19859.x}, \href {https://ui.adsabs.harvard.edu/abs/2012MNRAS.419.2077M} {419, 2077}

\bibitem[\protect\citeauthoryear{{Miglio} et~al.,}{{Miglio} et~al.}{2013}]{2013EPJWC..4303004M}
{Miglio} A.,  et~al., 2013, in European Physical Journal Web of Conferences. p. 03004 (\mn@eprint {arXiv} {1301.1515}), \mn@doi{10.1051/epjconf/20134303004}

\bibitem[\protect\citeauthoryear{{Miglio} et~al.,}{{Miglio} et~al.}{2016}]{2016MNRAS.461..760M}
{Miglio} A.,  et~al., 2016, \mn@doi [\mnras] {10.1093/mnras/stw1555}, \href {https://ui.adsabs.harvard.edu/abs/2016MNRAS.461..760M} {461, 760}

\bibitem[\protect\citeauthoryear{{Miglio} et~al.,}{{Miglio} et~al.}{2021}]{2021A&A...645A..85M}
{Miglio} A.,  et~al., 2021, \mn@doi [\aap] {10.1051/0004-6361/202038307}, \href {https://ui.adsabs.harvard.edu/abs/2021A&A...645A..85M} {645, A85}

\bibitem[\protect\citeauthoryear{{Montalb{\'a}n} et~al.,}{{Montalb{\'a}n} et~al.}{2021}]{2021NatAs...5..640M}
{Montalb{\'a}n} J.,  et~al., 2021, \mn@doi [Nature Astronomy] {10.1038/s41550-021-01347-7}, \href {https://ui.adsabs.harvard.edu/abs/2021NatAs...5..640M} {5, 640}

\bibitem[\protect\citeauthoryear{{Mosser} et~al.,}{{Mosser} et~al.}{2013}]{2013A&A...550A.126M}
{Mosser} B.,  et~al., 2013, \mn@doi [\aap] {10.1051/0004-6361/201220435}, \href {https://ui.adsabs.harvard.edu/abs/2013A&A...550A.126M} {550, A126}

\bibitem[\protect\citeauthoryear{{Nguyen} et~al.,}{{Nguyen} et~al.}{2022}]{2022A&A...665A.126N}
{Nguyen} C.~T.,  et~al., 2022, \mn@doi [\aap] {10.1051/0004-6361/202244166}, \href {https://ui.adsabs.harvard.edu/abs/2022A&A...665A.126N} {665, A126}

\bibitem[\protect\citeauthoryear{{Paxton}, {Bildsten}, {Dotter}, {Herwig}, {Lesaffre}  \& {Timmes}}{{Paxton} et~al.}{2011}]{2011ApJS..192....3P}
{Paxton} B.,  {Bildsten} L.,  {Dotter} A.,  {Herwig} F.,  {Lesaffre} P.,   {Timmes} F.,  2011, \mn@doi [\apjs] {10.1088/0067-0049/192/1/3}, \href {https://ui.adsabs.harvard.edu/abs/2011ApJS..192....3P} {192, 3}

\bibitem[\protect\citeauthoryear{{Paxton} et~al.,}{{Paxton} et~al.}{2013}]{2013ApJS..208....4P}
{Paxton} B.,  et~al., 2013, \mn@doi [\apjs] {10.1088/0067-0049/208/1/4}, \href {https://ui.adsabs.harvard.edu/abs/2013ApJS..208....4P} {208, 4}

\bibitem[\protect\citeauthoryear{{Paxton} et~al.,}{{Paxton} et~al.}{2015}]{2015ApJS..220...15P}
{Paxton} B.,  et~al., 2015, \mn@doi [\apjs] {10.1088/0067-0049/220/1/15}, \href {https://ui.adsabs.harvard.edu/abs/2015ApJS..220...15P} {220, 15}

\bibitem[\protect\citeauthoryear{{Paxton} et~al.,}{{Paxton} et~al.}{2018}]{2018ApJS..234...34P}
{Paxton} B.,  et~al., 2018, \mn@doi [\apjs] {10.3847/1538-4365/aaa5a8}, \href {https://ui.adsabs.harvard.edu/abs/2018ApJS..234...34P} {234, 34}

\bibitem[\protect\citeauthoryear{{Paxton} et~al.,}{{Paxton} et~al.}{2019}]{2019ApJS..243...10P}
{Paxton} B.,  et~al., 2019, \mn@doi [\apjs] {10.3847/1538-4365/ab2241}, \href {https://ui.adsabs.harvard.edu/abs/2019ApJS..243...10P} {243, 10}

\bibitem[\protect\citeauthoryear{{Perets} \& {Fabrycky}}{{Perets} \& {Fabrycky}}{2009}]{2009ApJ...697.1048P}
{Perets} H.~B.,  {Fabrycky} D.~C.,  2009, \mn@doi [\apj] {10.1088/0004-637X/697/2/1048}, \href {https://ui.adsabs.harvard.edu/abs/2009ApJ...697.1048P} {697, 1048}

\bibitem[\protect\citeauthoryear{{Pinsonneault} et~al.,}{{Pinsonneault} et~al.}{2018}]{2018ApJS..239...32P}
{Pinsonneault} M.~H.,  et~al., 2018, \mn@doi [\apjs] {10.3847/1538-4365/aaebfd}, \href {https://ui.adsabs.harvard.edu/abs/2018ApJS..239...32P} {239, 32}

\bibitem[\protect\citeauthoryear{{Planck Collaboration} et~al.,}{{Planck Collaboration} et~al.}{2016}]{2016A&A...594A..13P}
{Planck Collaboration} et~al., 2016, \mn@doi [\aap] {10.1051/0004-6361/201525830}, \href {https://ui.adsabs.harvard.edu/abs/2016A&A...594A..13P} {594, A13}

\bibitem[\protect\citeauthoryear{{Reimers}}{{Reimers}}{1975}]{1975psae.book..229R}
{Reimers} D.,  1975, in , Problems in stellar atmospheres and envelopes..
Springer-Verlag, pp 229--256

\bibitem[\protect\citeauthoryear{{Reyes}, {Stello}, {Hon}, {Trampedach}, {Sandquist}  \& {Pinsonneault}}{{Reyes} et~al.}{2024}]{2024MNRAS.tmp.1617R}
{Reyes} C.,  {Stello} D.,  {Hon} M.,  {Trampedach} R.,  {Sandquist} E.,   {Pinsonneault} M.~H.,  2024, \mn@doi [\mnras] {10.1093/mnras/stae1650}, \href {https://ui.adsabs.harvard.edu/abs/2024MNRAS.tmp.1617R} {}

\bibitem[\protect\citeauthoryear{{Rodrigues} et~al.,}{{Rodrigues} et~al.}{2017}]{2017MNRAS.467.1433R}
{Rodrigues} T.~S.,  et~al., 2017, \mn@doi [\mnras] {10.1093/mnras/stx120}, \href {https://ui.adsabs.harvard.edu/abs/2017MNRAS.467.1433R} {467, 1433}

\bibitem[\protect\citeauthoryear{{Sanders}}{{Sanders}}{1977}]{1977A&AS...27...89S}
{Sanders} W.~L.,  1977, \aaps, \href {https://ui.adsabs.harvard.edu/abs/1977A&AS...27...89S} {27, 89}

\bibitem[\protect\citeauthoryear{{Sandquist} et~al.,}{{Sandquist} et~al.}{2021}]{2021AJ....161...59S}
{Sandquist} E.~L.,  et~al., 2021, \mn@doi [\aj] {10.3847/1538-3881/abca8d}, \href {https://ui.adsabs.harvard.edu/abs/2021AJ....161...59S} {161, 59}

\bibitem[\protect\citeauthoryear{{Sarajedini}, {Dotter}  \& {Kirkpatrick}}{{Sarajedini} et~al.}{2009}]{2009ApJ...698.1872S}
{Sarajedini} A.,  {Dotter} A.,   {Kirkpatrick} A.,  2009, \mn@doi [\apj] {10.1088/0004-637X/698/2/1872}, \href {https://ui.adsabs.harvard.edu/abs/2009ApJ...698.1872S} {698, 1872}

\bibitem[\protect\citeauthoryear{{Schwarz}}{{Schwarz}}{1978}]{1978AnSta...6..461S}
{Schwarz} G.,  1978, Annals of Statistics, \href {https://ui.adsabs.harvard.edu/abs/1978AnSta...6..461S} {6, 461}

\bibitem[\protect\citeauthoryear{{Serenelli} et~al.,}{{Serenelli} et~al.}{2017}]{2017ApJS..233...23S}
{Serenelli} A.,  et~al., 2017, \mn@doi [\apjs] {10.3847/1538-4365/aa97df}, \href {https://ui.adsabs.harvard.edu/abs/2017ApJS..233...23S} {233, 23}

\bibitem[\protect\citeauthoryear{{Sharma}, {Stello}, {Bland-Hawthorn}, {Huber}  \& {Bedding}}{{Sharma} et~al.}{2016}]{2016ApJ...822...15S}
{Sharma} S.,  {Stello} D.,  {Bland-Hawthorn} J.,  {Huber} D.,   {Bedding} T.~R.,  2016, \mn@doi [\apj] {10.3847/0004-637X/822/1/15}, \href {https://ui.adsabs.harvard.edu/abs/2016ApJ...822...15S} {822, 15}

\bibitem[\protect\citeauthoryear{{Skrutskie} et~al.,}{{Skrutskie} et~al.}{2006}]{2006AJ....131.1163S}
{Skrutskie} M.~F.,  et~al., 2006, \mn@doi [\aj] {10.1086/498708}, \href {https://ui.adsabs.harvard.edu/abs/2006AJ....131.1163S} {131, 1163}

\bibitem[\protect\citeauthoryear{{Stello} \& {Sharma}}{{Stello} \& {Sharma}}{2022}]{2022RNAAS...6..168S}
{Stello} D.,  {Sharma} S.,  2022, \mn@doi [Research Notes of the American Astronomical Society] {10.3847/2515-5172/ac8b12}, \href {https://ui.adsabs.harvard.edu/abs/2022RNAAS...6..168S} {6, 168}

\bibitem[\protect\citeauthoryear{{Stello} et~al.,}{{Stello} et~al.}{2006}]{2006MNRAS.373.1141S}
{Stello} D.,  et~al., 2006, \mn@doi [\mnras] {10.1111/j.1365-2966.2006.11060.x}, \href {https://ui.adsabs.harvard.edu/abs/2006MNRAS.373.1141S} {373, 1141}

\bibitem[\protect\citeauthoryear{{Stello}, {Bruntt}, {Preston}  \& {Buzasi}}{{Stello} et~al.}{2008}]{2008ApJ...674L..53S}
{Stello} D.,  {Bruntt} H.,  {Preston} H.,   {Buzasi} D.,  2008, \mn@doi [\apjl] {10.1086/528936}, \href {https://ui.adsabs.harvard.edu/abs/2008ApJ...674L..53S} {674, L53}

\bibitem[\protect\citeauthoryear{{Stello}, {Chaplin}, {Basu}, {Elsworth}  \& {Bedding}}{{Stello} et~al.}{2009}]{2009MNRAS.400L..80S}
{Stello} D.,  {Chaplin} W.~J.,  {Basu} S.,  {Elsworth} Y.,   {Bedding} T.~R.,  2009, \mn@doi [\mnras] {10.1111/j.1745-3933.2009.00767.x}, \href {https://ui.adsabs.harvard.edu/abs/2009MNRAS.400L..80S} {400, L80}

\bibitem[\protect\citeauthoryear{{Stello}, {Cantiello}, {Fuller}, {Garcia}  \& {Huber}}{{Stello} et~al.}{2016a}]{2016PASA...33...11S}
{Stello} D.,  {Cantiello} M.,  {Fuller} J.,  {Garcia} R.~A.,   {Huber} D.,  2016a, \mn@doi [\pasa] {10.1017/pasa.2016.9}, \href {https://ui.adsabs.harvard.edu/abs/2016PASA...33...11S} {33, e011}

\bibitem[\protect\citeauthoryear{{Stello}, {Cantiello}, {Fuller}, {Huber}, {Garc{\'\i}a}, {Bedding}, {Bildsten}  \& {Silva Aguirre}}{{Stello} et~al.}{2016b}]{2016Natur.529..364S}
{Stello} D.,  {Cantiello} M.,  {Fuller} J.,  {Huber} D.,  {Garc{\'\i}a} R.~A.,  {Bedding} T.~R.,  {Bildsten} L.,   {Silva Aguirre} V.,  2016b, \mn@doi [\nat] {10.1038/nature16171}, \href {https://ui.adsabs.harvard.edu/abs/2016Natur.529..364S} {529, 364}

\bibitem[\protect\citeauthoryear{{Stello} et~al.,}{{Stello} et~al.}{2016c}]{2016ApJ...832..133S}
{Stello} D.,  et~al., 2016c, \mn@doi [\apj] {10.3847/0004-637X/832/2/133}, \href {https://ui.adsabs.harvard.edu/abs/2016ApJ...832..133S} {832, 133}

\bibitem[\protect\citeauthoryear{{Stokholm}, {Aguirre B{\o}rsen-Koch}, {Stello}, {Hon}  \& {Reyes}}{{Stokholm} et~al.}{2023}]{2023MNRAS.524.1634S}
{Stokholm} A.,  {Aguirre B{\o}rsen-Koch} V.,  {Stello} D.,  {Hon} M.,   {Reyes} C.,  2023, \mn@doi [\mnras] {10.1093/mnras/stad1912}, \href {https://ui.adsabs.harvard.edu/abs/2023MNRAS.524.1634S} {524, 1634}

\bibitem[\protect\citeauthoryear{{Townsend} \& {Teitler}}{{Townsend} \& {Teitler}}{2013}]{2013MNRAS.435.3406T}
{Townsend} R.~H.~D.,  {Teitler} S.~A.,  2013, \mn@doi [\mnras] {10.1093/mnras/stt1533}, \href {https://ui.adsabs.harvard.edu/abs/2013MNRAS.435.3406T} {435, 3406}

\bibitem[\protect\citeauthoryear{{Trampedach}, {Stein}, {Christensen-Dalsgaard}, {Nordlund}  \& {Asplund}}{{Trampedach} et~al.}{2014}]{2014MNRAS.442..805T}
{Trampedach} R.,  {Stein} R.~F.,  {Christensen-Dalsgaard} J.,  {Nordlund} {\r{A}}.,   {Asplund} M.,  2014, \mn@doi [\mnras] {10.1093/mnras/stu889}, \href {https://ui.adsabs.harvard.edu/abs/2014MNRAS.442..805T} {442, 805}

\bibitem[\protect\citeauthoryear{{VandenBerg} \& {Stetson}}{{VandenBerg} \& {Stetson}}{2004}]{2004PASP..116..997V}
{VandenBerg} D.~A.,  {Stetson} P.~B.,  2004, \mn@doi [\pasp] {10.1086/426340}, \href {https://ui.adsabs.harvard.edu/abs/2004PASP..116..997V} {116, 997}

\bibitem[\protect\citeauthoryear{{Vanderburg} \& {Johnson}}{{Vanderburg} \& {Johnson}}{2014}]{2014PASP..126..948V}
{Vanderburg} A.,  {Johnson} J.~A.,  2014, \mn@doi [\pasp] {10.1086/678764}, \href {https://ui.adsabs.harvard.edu/abs/2014PASP..126..948V} {126, 948}

\bibitem[\protect\citeauthoryear{{Vanderburg} et~al.,}{{Vanderburg} et~al.}{2016}]{2016ApJS..222...14V}
{Vanderburg} A.,  et~al., 2016, \mn@doi [\apjs] {10.3847/0067-0049/222/1/14}, \href {https://ui.adsabs.harvard.edu/abs/2016ApJS..222...14V} {222, 14}

\bibitem[\protect\citeauthoryear{{Viani}, {Basu}, {Chaplin}, {Davies}  \& {Elsworth}}{{Viani} et~al.}{2017}]{2017ApJ...843...11V}
{Viani} L.~S.,  {Basu} S.,  {Chaplin} W.~J.,  {Davies} G.~R.,   {Elsworth} Y.,  2017, \mn@doi [\apj] {10.3847/1538-4357/aa729c}, \href {https://ui.adsabs.harvard.edu/abs/2017ApJ...843...11V} {843, 11}

\bibitem[\protect\citeauthoryear{{White}, {Bedding}, {Stello}, {Christensen-Dalsgaard}, {Huber}  \& {Kjeldsen}}{{White} et~al.}{2011}]{2011ApJ...743..161W}
{White} T.~R.,  {Bedding} T.~R.,  {Stello} D.,  {Christensen-Dalsgaard} J.,  {Huber} D.,   {Kjeldsen} H.,  2011, \mn@doi [\apj] {10.1088/0004-637X/743/2/161}, \href {https://ui.adsabs.harvard.edu/abs/2011ApJ...743..161W} {743, 161}

\makeatother
\end{thebibliography}
\input{M67.bbl}



\appendix

\section{Yellow straggler EPIC 211415650}
\label{sec:blue}
\begin{figure*}
\begin{minipage}[!htb]{\columnwidth}
\centering
\includegraphics[width=\columnwidth]{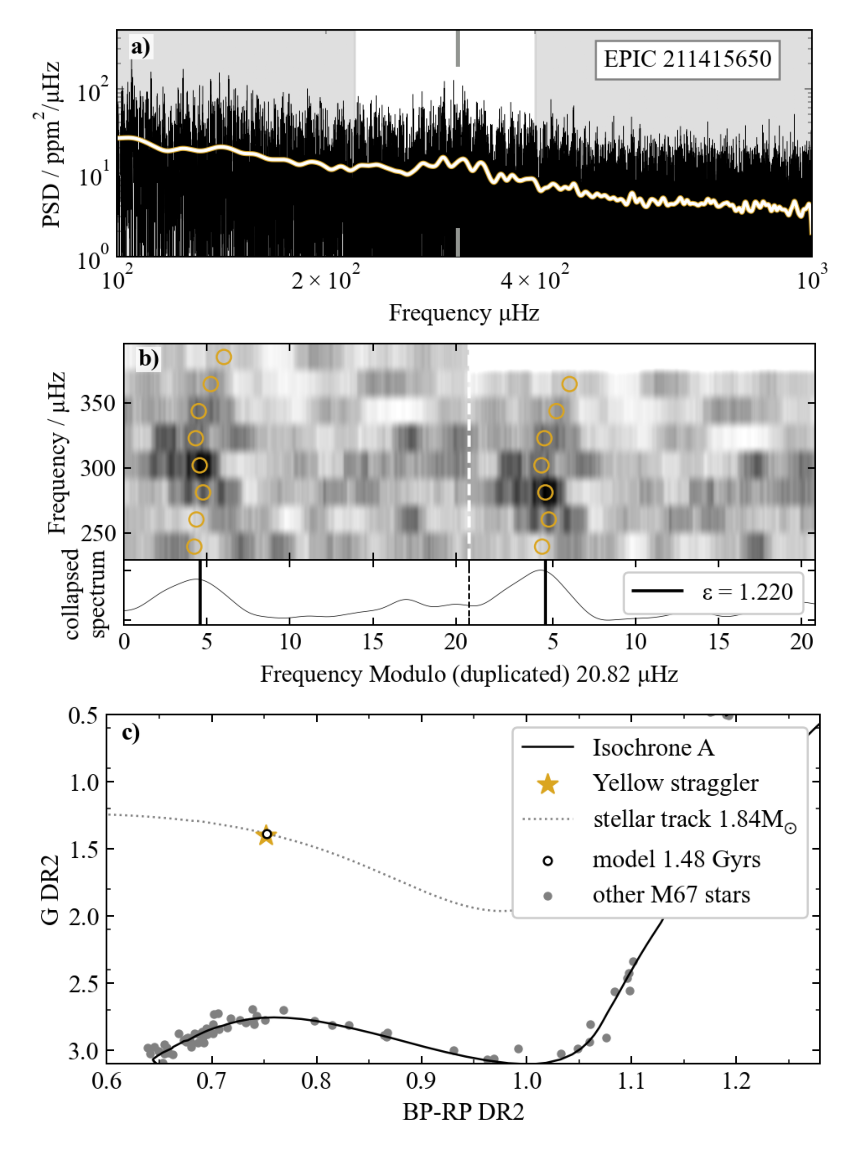}
\end{minipage}%
\begin{minipage}[!htb]{\columnwidth}
\centering
\renewcommand{\arraystretch}{1.08}
\begin{tabular}{lc}   
    \toprule
         \multicolumn{2}{c}{Other Names} \\
         \midrule 
         Sanders ID& 1072\\
         WOCS ID& 2008\\
         EPIC ID& 211415650\\
         \midrule
          \multicolumn{2}{c}{Observed seismic parameters} \\
          \midrule
        \numax\ \textsc{pySYD}  & 312.79 $\pm$ 4.69 \muhz\ \\ 
        \dnu\ \textsc{pySYD} & 20.67 $\pm$ 1.16 \muhz\ \\
         \midrule
           \multicolumn{2}{c}{Seismically derived parameters} \\
         \midrule
         Seismic $\log g$ & 3.448 $\pm$ 0.008 cm/$s^2$\\
         \teff\ IRFM & 5897 $\pm$ 52 K\\
         Luminosity & 19.955 $\pm$ 0.973 \lsol\ \\
         Mass \autoref{eq:eq1} & 1.880 $\pm$ 0.112 \msol\ \\ Mass \autoref{eq:eq2} &  1.922 $\pm$ 0.273 \msol\ \\Mass \autoref{eq:eq3} & 1.822 $\pm$ 0.289 \msol\ \\Mass \autoref{eq:eq4} & 1.798 $\pm$ 0.404 \msol\ \\
         Mean mass & 1.874 $\pm$ 0.134 \msol\ \\
         Radius \autoref{eq:eq5} & 4.193 $\pm$ 0.466 \rsol\ \\
         \midrule
           \multicolumn{2}{c}{Model parameters} \\
         \midrule 
         \numax\  & 310.52 \muhz \\
         \dnusurf\  & 20.89 \muhz \\ 
         $\log g$ & 3.44 cm/$s^2$ \\
         \teff\ & 5910 K \\
         Luminosity & 19.957 \lsol\ \\
         Mass & 1.8402 \msol\ \\
         Radius & 4.279 \rsol\ \\
         Age  & 1.4838  Gyrs \\

         \bottomrule
\end{tabular}
\end{minipage}%
\caption{Yellow Straggler EPIC 211415650. Left: a) The star's power spectrum (in black) and smoothed power spectrum (light yellow curve). The region of excess power is highlighted with a white background, and the frequency of maximum oscillation power \numax\ is indicated with short vertical grey lines. b) duplicated echelle diagram of the yellow straggler and modelled surface-corrected radial modes (open orange circles). c) HR diagram showing the absolute yellow straggler photometry and the modelled star from (b). Right: Table detailing seismically derived stellar parameters of the straggler and specifications of the model we show in the left-side image. All seismically derived parameters use our corrected scaling relations as described in the main text, and the mean mass corresponds to the weighted average over the four equations using their given uncertainties as weights. }
\label{fig:yellow}
\end{figure*}

We report the first seismic analysis of the yellow straggler star EPIC 211415650. This star had short cadence data from campaigns 16 and 18 only, the last data received from \textit{K2}. We treated the data as described in the main text. We calculated all seismically derived parameters in the same way as for the main sample of this work, applying the described corrections to the scaling relations. A 1.84\msol, 1.48 gigayears old model calculated similarly to the models from Isochrone A is a good fit to the observed colour and magnitude of the yellow straggler and simultaneously to its observed \numax\ and \dnu. See \autoref{fig:yellow}.

\section{Tables}

\begin{table*}
    \centering
    \caption{M67 Seismic sample. Identifiers are from the K2 Ecliptic Plane Input Catalog \citep{2017yCat.4034....0H}. \numax\ and \dnu\ are results from \textsc{pySYD} (\autoref{sec:seismicdata}). \teff\ and seismic $\log g$ were obtained using the iterative infrared flux method (\autoref{sec:temperatures}). }
    \begin{tabular}{lcccc}
EPIC ID & \numax\ \muhz & \dnu\ \muhz  & $T_{\mathrm{eff}}$ K & $\log g$ $\mathrm{cm/s^{2}}$\\
\hline
211376143$^{1}$ & 0.99 $\pm$ 0.03 & - & 3646 $\pm$ 56 & 0.84 $\pm$ 0.03 \\
211414329 & 2.5 $\pm$ 0.06 & 0.49 $\pm$ 0.07 & 3895 $\pm$ 36 & 1.26 $\pm$ 0.03 \\
211443624 & 4.68 $\pm$ 0.22 & 0.78 $\pm$ 0.08 & 4073 $\pm$ 33 & 1.55 $\pm$ 0.10 \\
211414351 & 8.29 $\pm$ 1.04 & 1.31 $\pm$ 0.03 & 4252 $\pm$ 30 & 1.76 $\pm$ 0.04 \\
211407537 & 8.91 $\pm$ 0.50 & 1.35 $\pm$ 0.05 & 4217 $\pm$ 34 & 1.83 $\pm$ 0.07 \\
211409660$^{4}$ & 9.14 $\pm$ 0.42 & 1.38 $\pm$ 0.04 & 4246 $\pm$ 43 & 1.84 $\pm$ 0.06 \\
211403356 & 10.9 $\pm$ 0.11 & 1.62 $\pm$ 0.04 & 4279 $\pm$ 48 & 1.92 $\pm$ 0.01 \\
211380313 & 16.63 $\pm$ 0.20 & 2.12 $\pm$ 0.03 & 4353 $\pm$ 39 & 2.11 $\pm$ 0.02 \\
211410817 & 20.66 $\pm$ 0.33 & 2.61 $\pm$ 0.02 & 4404 $\pm$ 30 & 2.20 $\pm$ 0.02 \\
211406541$^{4}$ & 36.60 $\pm$ 2.15 & 4.40 $\pm$ 0.07 & 4628 $\pm$ 33 & 2.46 $\pm$ 0.08 \\
211418433$^{3}$ & 37.41 $\pm$ 1.36 & 4.30 $\pm$ 0.02 & 4698 $\pm$ 42 & 2.48 $\pm$ 0.05 \\
211410523$^{3}$ & 38.88 $\pm$ 0.77 & 4.17 $\pm$ 0.07 & 4699 $\pm$ 40 & 2.49 $\pm$ 0.03 \\
211415732$^{3}$ & 39.00 $\pm$ 0.94 & 4.35 $\pm$ 0.11 & 4701 $\pm$ 34 & 2.49 $\pm$ 0.03 \\
211420284$^{3}$ & 39.22 $\pm$ 0.50 & 4.19 $\pm$ 0.05 & 4690 $\pm$ 37 & 2.50 $\pm$ 0.02 \\
211413402$^{3}$ & 39.73 $\pm$ 0.35 & 4.42 $\pm$ 0.09 & 4645 $\pm$ 55 & 2.50 $\pm$ 0.01 \\
211406540$^{3}$ & 39.82 $\pm$ 0.74 & 4.15 $\pm$ 0.05 & 4711 $\pm$ 40 & 2.50 $\pm$ 0.02 \\
211417056$^{3}$ & 39.92 $\pm$ 0.73 & 4.21 $\pm$ 0.03 & 4715 $\pm$ 40 & 2.51 $\pm$ 0.02 \\
211392837 & 47.8 $\pm$ 0.28 & 4.82 $\pm$ 0.02 & 4584 $\pm$ 35 & 2.58 $\pm$ 0.01 \\
211413623 & 67.0 $\pm$ 0.74 & 6.29 $\pm$ 0.02 & 4707 $\pm$ 42 & 2.73 $\pm$ 0.02 \\
211396385$^{4}$ & 78.42 $\pm$ 0.45 & 7.03 $\pm$ 0.02 & 4704 $\pm$ 44 & 2.80 $\pm$ 0.01 \\
211414300 & 79.3 $\pm$ 0.80 & 7.13 $\pm$ 0.04 & 4728 $\pm$ 39 & 2.80 $\pm$ 0.01 \\
211408346 & 96.6 $\pm$ 1.11 & 8.18 $\pm$ 0.02 & 4753 $\pm$ 35 & 2.89 $\pm$ 0.02 \\
211410231 & 105.5 $\pm$ 0.93 & 8.94 $\pm$ 0.02 & 4782 $\pm$ 38 & 2.93 $\pm$ 0.01 \\
211412928 & 118.0 $\pm$ 0.82 & 9.71 $\pm$ 0.02 & 4795 $\pm$ 38 & 2.98 $\pm$ 0.01 \\
211384259 & 175.8 $\pm$ 0.99 & 13.81 $\pm$ 0.04 & 4894 $\pm$ 32 & 3.15 $\pm$ 0.01 \\
211411629$^{2, 4}$ & 189.6 $\pm$ 0.00 & 14.38 $\pm$ 0.02 & 4847 $\pm$ 53 & 3.19 $\pm$ 0.01 \\
211406144$^{2, 4}$ & 193.1 $\pm$ 1.21 & 14.44 $\pm$ 0.03 & 4858 $\pm$ 34 & 3.20 $\pm$ 0.01 \\
211414687 & 203.7 $\pm$ 0.90 & 15.23 $\pm$ 0.03 & 4883 $\pm$ 38 & 3.22 $\pm$ 0.01 \\
211416749 & 234.8 $\pm$ 0.65 & 16.80 $\pm$ 0.04 & 4895 $\pm$ 44 & 3.28 $\pm$ 0.01 \\
211421954 & 247.9 $\pm$ 5.00 & 17.56 $\pm$ 0.20 & 4926 $\pm$ 36 & 3.31 $\pm$ 0.03 \\
211409560 & 277.3 $\pm$ 1.45 & 19.13 $\pm$ 0.04 & 4943 $\pm$ 37 & 3.36 $\pm$ 0.01 \\
211411553 & 309.6 $\pm$ 9.71 & 20.75 $\pm$ 0.19 & 4909 $\pm$ 43 & 3.40 $\pm$ 0.04 \\
211388537 & 312.2 $\pm$ 6.57 & 20.76 $\pm$ 0.26 & 4975 $\pm$ 41 & 3.41 $\pm$ 0.03 \\
211403248 & 320.6 $\pm$ 1.66 & 21.24 $\pm$ 0.22 & 4965 $\pm$ 55 & 3.42 $\pm$ 0.01 \\
211405262 & 431.0 $\pm$ 7.15 & 26.72 $\pm$ 0.29 & 5040 $\pm$ 44 & 3.55 $\pm$ 0.02 \\
211415364 & 446.1 $\pm$ 4.24 & 28.32 $\pm$ 0.02 & 4989 $\pm$ 41 & 3.57 $\pm$ 0.01 \\
211420451 & 462.1 $\pm$ 21.06 & 29.48 $\pm$ 0.50 & 5035 $\pm$ 35 & 3.58 $\pm$ 0.06 \\
211413064 & 488.8 $\pm$ 6.20 & 28.80 $\pm$ 0.81 & 4953 $\pm$ 32 & 3.60 $\pm$ 0.02 \\
211409088 & 545.5 $\pm$ 3.91 & 32.89 $\pm$ 0.02 & 5095 $\pm$ 36 & 3.66 $\pm$ 0.01 \\
211411922$^{5}$ & 570.2 $\pm$ 3.96 & 35.19 $\pm$ 0.24 & 5209 $\pm$ 30 & 3.68 $\pm$ 0.01 \\
211414203 & 644.2 $\pm$ 6.85 & 38.60 $\pm$ 0.27 & 5245 $\pm$ 33 & 3.74 $\pm$ 0.01 \\
211417812$^{2}$ & 651.2 $\pm$ 11.25 & 39.77 $\pm$ 1.63 & 5365 $\pm$ 52 & 3.75 $\pm$ 0.02 \\
211409644$^{2}$ & 668.8 $\pm$ 7.60 & 37.70 $\pm$ 1.43 & 5391 $\pm$ 37 & 3.76 $\pm$ 0.02 \\
211407836 & 675.1 $\pm$ 8.16 & 38.96 $\pm$ 0.12 & 5365 $\pm$ 31 & 3.76 $\pm$ 0.02 \\
211414081$^{2, 4}$ & 693.9 $\pm$ 12.48 & 42.24 $\pm$ 0.59 & 5816 $\pm$ 41 & 3.79 $\pm$ 0.02 \\
211421683 & 704.7 $\pm$ 11.54 & 41.62 $\pm$ 0.23 & 5559 $\pm$ 50 & 3.79 $\pm$ 0.02 \\
211435003 & 705.6 $\pm$ 36.57 & 39.74 $\pm$ 0.75 & 5712 $\pm$ 39 & 3.79 $\pm$ 0.07 \\
211403555$^{4}$ & 743.6 $\pm$ 14.17 & 41.36 $\pm$ 0.33 & 5512 $\pm$ 42 & 3.81 $\pm$ 0.03 \\
211401683 & 782.0 $\pm$ 18.93 & 46.67 $\pm$ 0.37 & 5971 $\pm$ 43 & 3.85 $\pm$ 0.03 \\
211412252 & 847.1 $\pm$ 11.49 & 46.00 $\pm$ 0.26 & 5973 $\pm$ 37 & 3.88 $\pm$ 0.02 \\
 \hline
 \end{tabular}
 \begin{center}
\footnotesize
\item[] $^{1}$ Unable to measure \dnu; $^{2}$ RUWE > 1.2; $^{3}$ Red clump star; 
$^{4}$ Non-single star; $^{5}$ Other photometric variability (see end of \autoref{sec:temperatures})\\
 \end{center}
    \label{tab:pysyd_teff}
\end{table*}

{\renewcommand{\arraystretch}{1.05} 
\begin{table*}
    \caption{M67 sample described in \autoref{sec:peak}: observed $\ell$=0 mode frequencies, uncertainties and signal-to-noise-ratio.}
    \begin{tabular}{p{0.09\textwidth}| p{0.28\textwidth}| p{0.25\textwidth}|p{0.26\textwidth}}
EPIC ID & $\ell$=0 observed \muhz\ & $\sigma_{\ell=0}\, \mu\mathrm{Hz}$ & SNR \\
\hline
211407537 & [7.73, 9.06, 10.42, 11.76] & [0.04, 0.03, 0.03, 0.02]  & [6.03, 9.62, 20.69, 7.54] \\
211409660$^1$ & [6.51, 7.79, 9.18, 10.55] & [0.04, 0.08, 0.06, 0.05]  & [1.85, 2.86, 17.04, 8.27] \\
211403356 & [9.45, 11.06, 12.67] & [0.02, 0.02, 0.02]  & [4.94, 28.35, 15.66] \\
211380313 & [16.76, 18.86] & [0.02, 0.03]  & [30.07, 19.97] \\
211410817 & [15.37, 17.71, 20.28, 22.89, 25.55] & [0.04, 0.03, 0.03, 0.04, 0.04]  & [10.99, 3.61, 9.92, 9.31, 5.00] \\
211406541$^1$ & [30.32, 34.33, 38.72, 43.13, 47.55] & [0.22, 0.35, 0.16, 0.24, 0.20]  & [4.19, 19.06, 11.32, 8.39, 8.06] \\
211392837 & [34.65, 39.13, 44.00, 48.79, 53.71] & [0.11, 0.12, 0.11, 0.13, 0.11]  & [5.30, 7.01, 28.10, 43.39, 17.16] \\
211413623 & [51.61, 57.60, 64.07, 70.34, 76.70] & [0.12, 0.10, 0.07, 0.10, 0.11]  & [3.57, 9.80, 53.80, 73.68, 18.01] \\
211396385$^1$ & [71.30, 78.34, 85.34] & [0.08, 0.10, 0.10]  & [31.61, 51.49, 28.19] \\
211414300 & [58.32, 65.25, 72.20, 79.30, 86.44, 93.65] & [0.04, 0.04, 0.04, 0.09, 0.05, 0.09]  & [7.03, 4.38, 8.10, 19.90, 7.33, 18.70] \\
211408346 & [83.70, 91.86, 100.05] & [0.08, 0.12, 0.09]  & [11.05, 17.90, 19.55] \\
211410231 & [82.39, 91.06, 100.02, 108.95, 117.91] & [0.07, 0.07, 0.07, 0.09, 0.12]  & [16.12, 14.21, 26.71, 42.56, 11.27] \\
211412928 & [90.02, 99.59, 109.21, 118.95, 128.63] & [0.06, 0.08, 0.06, 0.08, 0.08]  & [7.79, 11.05, 20.82, 58.19, 33.49] \\
211384259 & [142.59, 156.18, 169.89, 183.65, 197.53, 211.26] & [0.14, 0.14, 0.11, 0.09, 0.07, 0.07]  & [6.17, 7.87, 31.79, 24.75, 15.40, 8.17] \\
211411629$^1$ & [134.11, 147.98, 161.88, 176.24, 190.64, 204.83, 219.50] & [0.37, 0.29, 0.06, 0.11, 0.06, 0.07, 0.37]  & [6.37, 4.27, 7.71, 26.45, 13.15, 19.62, 4.29] \\
211406144$^1$ & [149.78, 163.82, 178.32, 192.74, 207.10, 221.71, 236.42] & [0.33, 0.07, 0.07, 0.12, 0.22, 0.19, 0.20]  & [4.93, 13.82, 32.70, 20.29, 9.45, 8.43, 4.91] \\
211414687 & [172.96, 188.17, 203.37, 218.49, 234.00] & [0.13, 0.41, 0.23, 0.22, 0.27]  & [4.75, 7.20, 20.58, 18.06, 3.06] \\
211416749 & [226.22, 242.98, 260.07, 277.20] & [0.23, 0.22, 0.26, 0.48]  & [61.26, 29.41, 4.12, 4.73] \\
211421954 & [183.43, 200.35, 217.68, 235.41, 252.87, 270.45, 288.35] & [0.15, 0.15, 0.11, 0.15, 0.30, 0.14, 0.26]  & [6.52, 3.07, 9.23, 35.94, 14.07, 22.23, 3.70] \\
211409560 & [218.95, 237.83, 256.94, 276.03, 295.15, 314.43, 334.03] & [0.18, 0.38, 0.25, 0.18, 0.13, 0.29, 0.29]  & [8.11, 6.35, 16.31, 60.28, 11.96, 8.32, 7.01] \\
211411553 & [259.93, 280.60, 301.29, 321.99, 343.11, 364.35] & [0.26, 0.19, 0.48, 0.28, 0.34, 0.21]  & [6.42, 9.49, 36.58, 20.21, 16.87, 8.64] \\
211388537 & [238.86, 259.20, 279.88, 300.58, 321.63, 342.46, 363.09] & [0.15, 0.15, 0.21, 0.30, 0.27, 0.29, 0.30]  & [6.24, 8.23, 8.85, 6.36, 7.45, 13.33, 4.88] \\
211403248 & [288.54, 309.51, 330.88, 352.39] & [0.39, 0.15, 0.39, 0.15]  & [4.82, 16.25, 41.15, 15.23] \\
211405262 & [364.05, 390.76, 417.38, 444.47, 471.71] & [0.37, 0.40, 0.19, 0.37, 0.33]  & [5.92, 1.21, 11.15, 6.52, 4.64] \\
211415364 & [410.88, 439.35, 467.67, 495.91] & [0.30, 0.19, 0.19, 0.50]  & [7.59, 15.96, 16.01, 4.47] \\
211420451 & [395.46, 425.24, 454.41, 483.55] & [0.53, 0.49, 0.29, 0.51]  & [6.41, 4.77, 10.43, 7.29] \\
211413064 & [421.18, 449.92, 479.21, 509.02, 537.66, 566.50] & [0.28, 0.17, 0.26, 0.49, 0.28, 0.49]  & [7.98, 4.53, 14.34, 5.92, 2.33, 2.85] \\
211409088 & [443.29, 476.06, 508.79, 541.69, 574.64, 607.88, 640.94] & [0.41, 0.64, 0.25, 0.48, 0.50, 0.22, 0.59]  & [6.46, 8.36, 4.41, 18.49, 9.64, 6.93, 3.89] \\
211411922$^1$ & [524.91, 558.66, 592.36] & [0.51, 0.29, 0.38]  & [6.46, 6.71, 6.86] \\
211414203$^*$ & [593.83, 632.20, 670.52] & [0.67, 0.33, 0.33]  & [2.10, 4.43, 3.84] \\
211417812$^1$ & [608.08, 647.72, 687.35, 727.39, 766.63] & [0.46, 0.19, 0.47, 0.67, 0.67]  & [3.12, 4.85, 4.99, 3.51, 3.03] \\
211409644$^1$ & [605.23, 644.41, 683.90] & [0.25, 0.25, 0.44]  & [2.98, 3.03, 4.23] \\
211407836 & [562.53, 601.31, 640.41, 679.13, 718.65] & [0.35, 0.23, 0.34, 0.37, 0.34]  & [3.09, 4.62, 8.24, 5.81, 3.84] \\
211414081$^{*,1}$ & [608.98, 650.80, 693.93, 736.46, 779.04] & [0.25, 0.50, 0.35, 0.33, 0.50]  & [4.41, 2.85, 3.21, 4.26, 2.26] \\
211421683$^*$ & [631.10, 672.46, 713.97, 756.25] & [0.35, 0.25, 0.40, 0.52]  & [3.64, 3.06, 7.82, 3.10] \\
211403555$^{*,1}$ & [712.78, 754.13, 794.75, 836.54] & [0.19, 0.45, 0.25, 0.25]  & [4.73, 3.86, 2.84, 3.14] \\
211401683$^*$ & [798.29, 844.56, 891.83] & [0.30, 0.20, 0.45]  & [6.06, 2.40, 6.82] \\
211412252$^*$ & [837.82, 883.92] & [0.22, 0.40]  & [5.55, 3.81] \\
\hline
\end{tabular}
\label{tab:modefreqs}
    \begin{center}
    \footnotesize
    \item[] ($^{1}$) Stars excluded from the final fit of \autoref{sec:model} because of binarity, high RUWE, or other photometric variability, ($^{*}$) Low SNR stars.\\
     \end{center}
\end{table*}
}

\begin{table*}
\caption{Scaling-based Mass and Radius for presumed single stars in our sample using Equations \ref{eq:eq1}-\ref{eq:eq5}, after applying the final corrections \dcorr\ and $f_{\nu_{\mathrm{max}}}$ (See \autoref{sec:mass}). $M_{\mathrm{surf}}$ and $R_{\mathrm{surf}}$ refer to the masses and radii from isochrone models found by interpolating $\Delta\nu_{\mathrm{obs}}$ onto \dnusurf, following the notation from the Table under Figure~\ref{fig:mass_study_vertical}. All masses are in units of \msol\ and radii are in units of \rsol. }
    \centering
    \begin{tabular}{cccccccccc}
EPIC ID  & Mass Eq.(2) & Mass Eq.(3) & Mass Eq.(4) &  Mass Eq.(5) & Radius Eq.(6) & & $M_{\mathrm{surf}}$ & $R_{\mathrm{surf}}$\\
\hline   
211412252 &   1.389 $\pm$ 0.076 & 1.248 $\pm$ 0.103 & 1.613 $\pm$ 0.063 & 1.720 $\pm$ 0.082 & 2.484 $\pm$ 0.045 & &1.3200  & 2.274 \\
211401683 &   1.243 $\pm$ 0.075 & 1.227 $\pm$ 0.106 & 1.265 $\pm$ 0.081 & 1.274 $\pm$ 0.102 & 2.225 $\pm$ 0.065 & &1.3186  & 2.251 \\
211403555 &   1.441 $\pm$ 0.085 & 1.392 $\pm$ 0.122 & 1.513 $\pm$ 0.081 & 1.545 $\pm$ 0.103 & 2.564 $\pm$ 0.065 & &1.3292  & 2.438 \\
211435003 &   1.298 $\pm$ 0.097 & 1.169 $\pm$ 0.107 & 1.504 $\pm$ 0.204 & 1.602 $\pm$ 0.277 & 2.656 $\pm$ 0.171 & &1.3320  & 2.498 \\
211421683 &   1.332 $\pm$ 0.080 & 1.346 $\pm$ 0.123 & 1.312 $\pm$ 0.059 & 1.304 $\pm$ 0.073 & 2.414 $\pm$ 0.049 & &1.3283  & 2.429 \\
211414081 &   1.410 $\pm$ 0.081 & 1.488 $\pm$ 0.132 & 1.308 $\pm$ 0.079 & 1.266 $\pm$ 0.100 & 2.370 $\pm$ 0.080 & &1.3272  & 2.408 \\
211407836 &   1.352 $\pm$ 0.073 & 1.342 $\pm$ 0.109 & 1.366 $\pm$ 0.046 & 1.372 $\pm$ 0.054 & 2.553 $\pm$ 0.035 & &1.3325  & 2.528 \\
211409644 &   1.403 $\pm$ 0.078 & 1.350 $\pm$ 0.153 & 1.481 $\pm$ 0.164 & 1.515 $\pm$ 0.236 & 2.692 $\pm$ 0.207 & &1.3340  & 2.580 \\
211417812 &   1.491 $\pm$ 0.092 & 1.703 $\pm$ 0.212 & 1.238 $\pm$ 0.153 & 1.143 $\pm$ 0.198 & 2.372 $\pm$ 0.200 & &1.3320  & 2.496 \\
211414203 &   1.305 $\pm$ 0.071 & 1.366 $\pm$ 0.114 & 1.224 $\pm$ 0.043 & 1.191 $\pm$ 0.051 & 2.449 $\pm$ 0.043 & &1.3332  & 2.543 \\
211411922 &   1.311 $\pm$ 0.070 & 1.396 $\pm$ 0.114 & 1.201 $\pm$ 0.035 & 1.156 $\pm$ 0.041 & 2.569 $\pm$ 0.040 & &1.3360  & 2.695 \\
211409088 &   1.329 $\pm$ 0.073 & 1.360 $\pm$ 0.115 & 1.287 $\pm$ 0.029 & 1.270 $\pm$ 0.031 & 2.768 $\pm$ 0.022 & &1.3380  & 2.816 \\
211413064 &   1.461 $\pm$ 0.081 & 1.463 $\pm$ 0.146 & 1.458 $\pm$ 0.123 & 1.457 $\pm$ 0.171 & 3.154 $\pm$ 0.179 & &1.3420  & 3.069 \\
211420451 &   1.285 $\pm$ 0.091 & 1.357 $\pm$ 0.123 & 1.191 $\pm$ 0.143 & 1.152 $\pm$ 0.176 & 2.873 $\pm$ 0.163 & &1.3420  & 3.023 \\
211415364 &   1.326 $\pm$ 0.076 & 1.397 $\pm$ 0.123 & 1.234 $\pm$ 0.034 & 1.196 $\pm$ 0.037 & 2.986 $\pm$ 0.031 & &1.3426  & 3.103 \\
211405262 &   1.349 $\pm$ 0.081 & 1.336 $\pm$ 0.123 & 1.368 $\pm$ 0.071 & 1.377 $\pm$ 0.092 & 3.251 $\pm$ 0.088 & &1.3440  & 3.225 \\
211403248 &   1.345 $\pm$ 0.084 & 1.345 $\pm$ 0.136 & 1.345 $\pm$ 0.046 & 1.345 $\pm$ 0.063 & 3.739 $\pm$ 0.082 & &1.3500  & 3.744 \\
211388537 &   1.304 $\pm$ 0.079 & 1.276 $\pm$ 0.117 & 1.345 $\pm$ 0.085 & 1.363 $\pm$ 0.111 & 3.813 $\pm$ 0.125 & &1.3509  & 3.802 \\
211411553 &   1.350 $\pm$ 0.089 & 1.372 $\pm$ 0.126 & 1.320 $\pm$ 0.106 & 1.307 $\pm$ 0.132 & 3.761 $\pm$ 0.136 & &1.3510  & 3.804 \\
211409560 &   1.342 $\pm$ 0.075 & 1.360 $\pm$ 0.117 & 1.316 $\pm$ 0.026 & 1.306 $\pm$ 0.027 & 3.966 $\pm$ 0.030 & &1.3527  & 4.014 \\
211421954 &   1.325 $\pm$ 0.078 & 1.333 $\pm$ 0.118 & 1.313 $\pm$ 0.078 & 1.308 $\pm$ 0.100 & 4.202 $\pm$ 0.129 & &1.3544  & 4.252 \\
211416749 &   1.348 $\pm$ 0.078 & 1.366 $\pm$ 0.124 & 1.323 $\pm$ 0.023 & 1.312 $\pm$ 0.024 & 4.331 $\pm$ 0.030 & &1.3560  & 4.379 \\
211414687 &   1.338 $\pm$ 0.075 & 1.379 $\pm$ 0.120 & 1.283 $\pm$ 0.024 & 1.260 $\pm$ 0.025 & 4.560 $\pm$ 0.033 & &1.3577  & 4.675 \\
211406144 &   1.362 $\pm$ 0.075 & 1.386 $\pm$ 0.117 & 1.330 $\pm$ 0.029 & 1.316 $\pm$ 0.030 & 4.793 $\pm$ 0.039 & &1.3580  & 4.843 \\
211411629 &   1.361 $\pm$ 0.084 & 1.412 $\pm$ 0.139 & 1.291 $\pm$ 0.020 & 1.263 $\pm$ 0.022 & 4.740 $\pm$ 0.030 & &1.3580  & 4.856 \\
211384259 &   1.347 $\pm$ 0.073 & 1.428 $\pm$ 0.119 & 1.241 $\pm$ 0.026 & 1.198 $\pm$ 0.026 & 4.783 $\pm$ 0.040 & &1.3591  & 4.990 \\
211412928 &   1.361 $\pm$ 0.077 & 1.341 $\pm$ 0.117 & 1.390 $\pm$ 0.031 & 1.402 $\pm$ 0.035 & 6.348 $\pm$ 0.055 & &1.3640  & 6.290 \\
211410231 &   1.396 $\pm$ 0.079 & 1.407 $\pm$ 0.123 & 1.380 $\pm$ 0.036 & 1.374 $\pm$ 0.042 & 6.651 $\pm$ 0.070 & &1.3640  & 6.635 \\
211408346 &   1.402 $\pm$ 0.079 & 1.364 $\pm$ 0.117 & 1.459 $\pm$ 0.047 & 1.484 $\pm$ 0.055 & 7.234 $\pm$ 0.093 & &1.3654  & 7.044 \\
211414300 &   1.397 $\pm$ 0.080 & 1.397 $\pm$ 0.124 & 1.396 $\pm$ 0.046 & 1.396 $\pm$ 0.056 & 7.757 $\pm$ 0.123 & &1.3660  & 7.700 \\
211396385 &   1.401 $\pm$ 0.083 & 1.392 $\pm$ 0.128 & 1.414 $\pm$ 0.030 & 1.420 $\pm$ 0.035 & 7.874 $\pm$ 0.070 & &1.3661  & 7.775 \\
211413623 &   1.356 $\pm$ 0.080 & 1.346 $\pm$ 0.122 & 1.370 $\pm$ 0.044 & 1.377 $\pm$ 0.053 & 8.384 $\pm$ 0.116 & &1.3677  & 8.373 \\
211392837 &   1.375 $\pm$ 0.077 & 1.377 $\pm$ 0.119 & 1.372 $\pm$ 0.032 & 1.370 $\pm$ 0.036 & 9.975 $\pm$ 0.104 & &1.3697  & 9.981 \\
211406541 &   1.129 $\pm$ 0.091 & 1.264 $\pm$ 0.114 & 0.963 $\pm$ 0.142 & 0.900 $\pm$ 0.168 & 9.213 $\pm$ 0.609 & &1.3714  & 10.643 \\
211410817 &   1.373 $\pm$ 0.078 & 1.476 $\pm$ 0.126 & 1.239 $\pm$ 0.057 & 1.186 $\pm$ 0.068 & 14.255 $\pm$ 0.314 & &1.3738  & 15.007 \\
211380313 &   1.432 $\pm$ 0.085 & 1.417 $\pm$ 0.135 & 1.454 $\pm$ 0.076 & 1.463 $\pm$ 0.102 & 17.699 $\pm$ 0.561 & &1.3740  & 17.334 \\
211403356 &   1.328 $\pm$ 0.084 & 1.375 $\pm$ 0.150 & 1.266 $\pm$ 0.087 & 1.240 $\pm$ 0.119 & 20.233 $\pm$ 0.936 & &1.3746  & 21.391 \\
211409660 &   1.387 $\pm$ 0.105 & 1.380 $\pm$ 0.157 & 1.397 $\pm$ 0.197 & 1.401 $\pm$ 0.261 & 23.502 $\pm$ 1.820 & &1.3752  & 24.031 \\
211407537 &   1.376 $\pm$ 0.109 & 1.363 $\pm$ 0.160 & 1.394 $\pm$ 0.242 & 1.402 $\pm$ 0.322 & 23.854 $\pm$ 2.297 & &1.3752  & 24.374 \\
211414351 &   1.397 $\pm$ 0.191 & 1.439 $\pm$ 0.143 & 1.339 $\pm$ 0.416 & 1.316 $\pm$ 0.514 & 23.902 $\pm$ 3.251 & &1.3753  & 24.915 \\
211443624 &   1.273 $\pm$ 0.093 & 1.118 $\pm$ 0.251 & 1.528 $\pm$ 0.473 & 1.652 $\pm$ 0.719 & 36.030 $\pm$ 7.615 & &1.3760  & 33.920 \\
211414329 &   1.615 $\pm$ 0.102 & 1.526 $\pm$ 0.467 & 1.748 $\pm$ 0.722 & 1.808 $\pm$ 1.065 & 52.345 $\pm$ 15.348 & &1.3760  & 47.817 \\
211376143$^*$  & 1.394 $\pm$ 0.112 & & & & & & 1.3780 & 75.205\\
211418433$^{**}$  &  1.325 $\pm$ 0.092 & 1.430 $\pm$ 0.131 & 1.191 $\pm$ 0.112 & 1.138 $\pm$ 0.133 & 10.445 $\pm$ 0.418 & &1.3840  & 11.151\\
211410523$^{**}$  &  1.406 $\pm$ 0.085 & 1.385 $\pm$ 0.131 & 1.437 $\pm$ 0.098 & 1.450 $\pm$ 0.130 & 11.567 $\pm$ 0.440 & &1.3860  & 11.395\\
211415732$^{**}$  &  1.346 $\pm$ 0.081 & 1.388 $\pm$ 0.137 & 1.289 $\pm$ 0.121 & 1.265 $\pm$ 0.160 & 10.788 $\pm$ 0.609 & &1.3840  & 11.116\\
211420284$^{**}$  &  1.364 $\pm$ 0.079 & 1.322 $\pm$ 0.119 & 1.424 $\pm$ 0.069 & 1.451 $\pm$ 0.092 & 11.528 $\pm$ 0.317 & &1.3860  & 11.353\\
211413402$^{**}$  &  1.438 $\pm$ 0.093 & 1.505 $\pm$ 0.165 & 1.350 $\pm$ 0.085 & 1.314 $\pm$ 0.116 & 10.924 $\pm$ 0.462 & &1.3840  & 11.116\\
211406540$^{**}$  &  1.435 $\pm$ 0.087 & 1.359 $\pm$ 0.126 & 1.547 $\pm$ 0.094 & 1.598 $\pm$ 0.127 & 11.991 $\pm$ 0.392 & &1.3860  & 11.436\\
211417056$^{**}$  &  1.386 $\pm$ 0.083 & 1.327 $\pm$ 0.119 & 1.474 $\pm$ 0.076 & 1.513 $\pm$ 0.097 & 11.651 $\pm$ 0.272 & &1.3860  & 11.315\\
\hline
    \end{tabular}
        \begin{center}
    \footnotesize
    \item[] ($^*$) denotes the star for which we only have \numax. Hence, $M_{\mathrm{surf}}$, and $R_{\mathrm{surf}}$ are obtained based on \numax\ only, and only mass from Equation (2) is presented. 
     \item[] ($^{**}$) indicates red clump stars.\\
     \end{center}
    \label{tab:truemass}
\end{table*}

{\renewcommand{\arraystretch}{1.15} 
\begin{table*}
    \caption{Parameters from $\chi^2$ best-fit models to M67 presumed single stars, as defined in \autoref{sec:model}.} 
    \begin{tabular}{cccccccccc}
EPIC ID & Age Gyrs & Mass \msol\ & \numax\ \muhz\ & \dnusurf\ \muhz\ & \amlt\ & $f_{\mathrm{ov}}$ & $\chi^{2}_{\mathrm{seismic}}$ & $\chi^{2}_{\mathrm{classical}}$ & $\chi^{2}_{\mathrm{combined}}$\\
\hline
211407537 & 3.9197  & 1.375 & 9.17 & 1.33 & 1.907  & 0.0054 & 0.5582   & 0.4808 & 0.5195 \\
211403356 & 4.3238  & 1.336 & 11.67 & 1.62 & 1.908  & 0.0041 & 0.3214   & 0.3935 & 0.3575 \\
211380313 & 4.0640  & 1.360 & 16.63 & 2.14 & 1.907  & 0.0048 & 0.0253   & 0.4577 & 0.2415 \\
211410817 & 4.5743  & 1.313 & 21.00 & 2.55 & 1.908  & 0.0034 & 1.5171   & 0.7021 & 1.1096 \\
211392837 & 4.5578  & 1.311 & 47.83 & 4.85 & 1.911  & 0.0033 & 0.0817   & 0.4120 & 0.2469 \\
211413623 & 3.7596  & 1.384 & 68.19 & 6.31 & 1.913  & 0.0057 & 0.1801   & 0.0189 & 0.0995 \\
211414300 & 3.5348  & 1.408 & 79.69 & 7.11 & 1.915  & 0.0065 & 2.1828   & 0.0817 & 1.1322 \\
211408346 & 4.0614  & 1.351 & 94.82 & 8.21 & 1.920  & 0.0046 & 0.1246   & 0.0414 & 0.0830 \\
211410231 & 3.8266  & 1.374 & 105.98 & 8.92 & 1.922  & 0.0053 & 0.0948   & 0.0313 & 0.0630 \\
211412928 & 4.2934  & 1.328 & 117.30 & 9.73 & 1.925  & 0.0038 & 0.7254   & 0.0190 & 0.3722 \\
211384259 & 3.2090  & 1.439 & 187.95 & 13.81 & 1.916  & 0.0076 & 0.9900   & 0.0448 & 0.5174 \\
211414687 & 3.9979  & 1.350 & 210.21 & 15.27 & 1.920  & 0.0045 & 0.0707   & 0.0823 & 0.0765 \\
211416749 & 4.1920  & 1.330 & 239.41 & 16.94 & 1.927  & 0.0039 & 0.0251   & 0.1313 & 0.0782 \\
211421954 & 3.9842  & 1.348 & 252.70 & 17.60 & 1.929  & 0.0045 & 0.1989   & 0.0141 & 0.1065 \\
211409560 & 3.9743  & 1.347 & 282.42 & 19.16 & 1.934  & 0.0044 & 0.0342   & 0.0094 & 0.0218 \\
211411553 & 4.5865  & 1.292 & 312.25 & 20.88 & 1.938  & 0.0028 & 0.1903   & 0.4301 & 0.3102 \\
211388537 & 4.0135  & 1.342 & 314.56 & 20.84 & 1.939  & 0.0043 & 0.4274   & 0.1143 & 0.2709 \\
211403248 & 3.8768  & 1.354 & 327.18 & 21.46 & 1.941  & 0.0047 & 0.2462   & 0.0097 & 0.1279 \\
211405262 & 3.5177  & 1.384 & 441.64 & 27.05 & 1.949  & 0.0057 & 0.5382   & 0.0099 & 0.2741 \\
211415364 & 4.2268  & 1.314 & 464.69 & 28.45 & 1.949  & 0.0034 & 0.0903   & 0.3386 & 0.2145 \\
211420451 & 3.9833  & 1.335 & 486.65 & 29.41 & 1.950  & 0.0040 & 0.1368   & 0.0074 & 0.0721 \\
211413064 & 4.8076  & 1.268 & 474.66 & 29.14 & 1.949  & 0.0022 & 0.9324   & 2.0552 & 1.4938 \\
211409088 & 3.8022  & 1.349 & 561.36 & 32.98 & 1.952  & 0.0045 & 0.1202   & 0.0112 & 0.0657 \\
211414203$^*$  & 3.9053  & 1.333 & 670.13 & 38.52 & 1.952  & 0.0040 & 0.0141   & 0.0222 & 0.0182 \\
211407836 & 3.6305  & 1.359 & 676.93 & 39.07 & 1.951  & 0.0048 & 0.2090   & 0.0280 & 0.1185 \\
211421683$^*$  & 3.8200  & 1.335 & 703.27 & 41.21 & 1.939  & 0.0041 & 1.7198   & 0.0391 & 0.8795 \\
211401683$^*$ & 3.9493  & 1.312 & 792.49 & 46.52 & 1.892  & 0.0033 & 0.8109   & 0.4498 & 0.6304 \\
211412252$^*$ & 3.8071  & 1.324 & 785.80 & 46.18 & 1.886  & 0.0037 & 0.0933   & 0.1941 & 0.1437 \\
\hline
\end{tabular}
    \begin{center}
    \footnotesize
    \item[] $^{*}$ Low SNR stars\\
     \end{center}
\label{tab:fitresults}
\end{table*}
}

\section{Visibility measurements}

\begin{figure*}
\begin{center}
    \includegraphics[width=.83\textwidth]{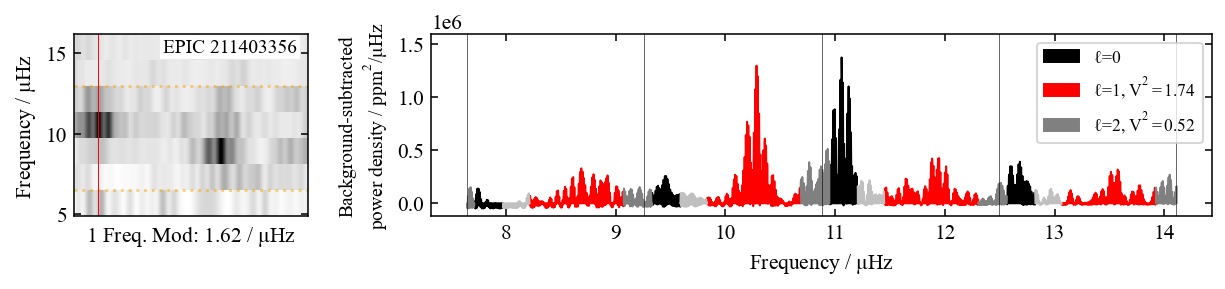}
    \includegraphics[width=.83\textwidth]{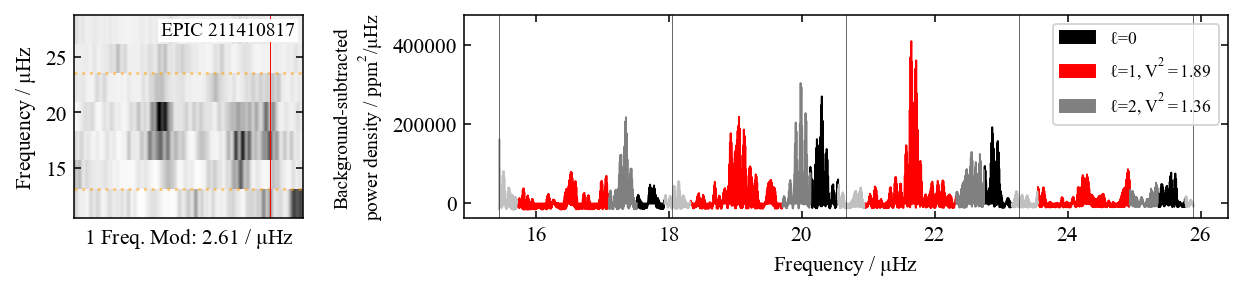}
    \includegraphics[width=.83\textwidth]{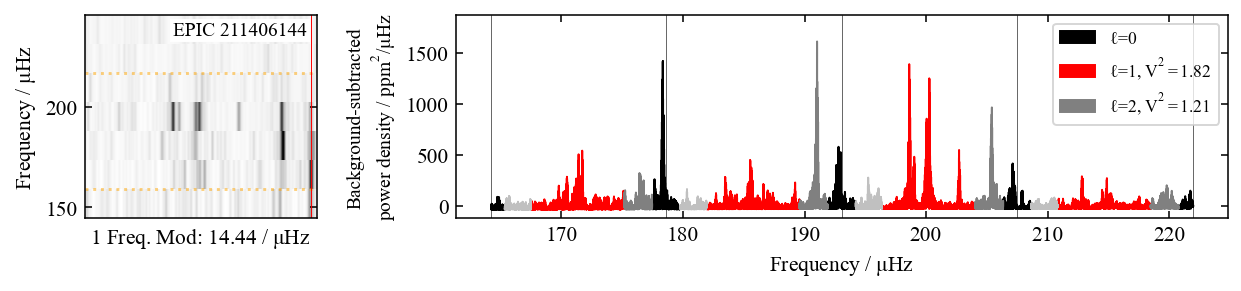}
    \includegraphics[width=.83\textwidth]{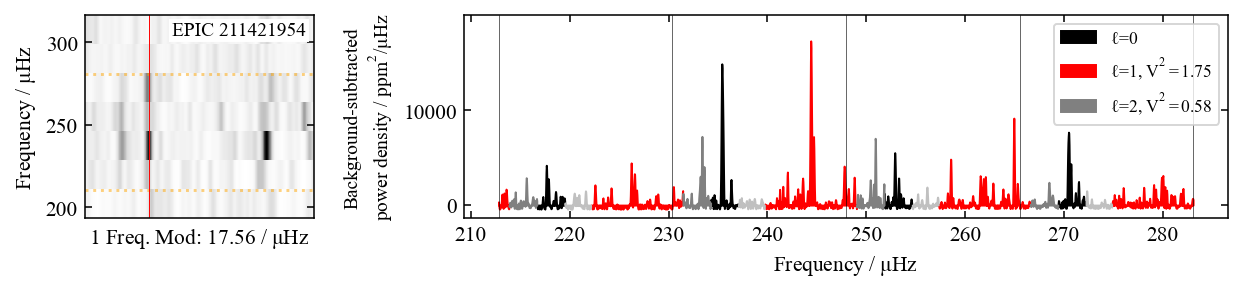}
    \includegraphics[width=.83\textwidth]{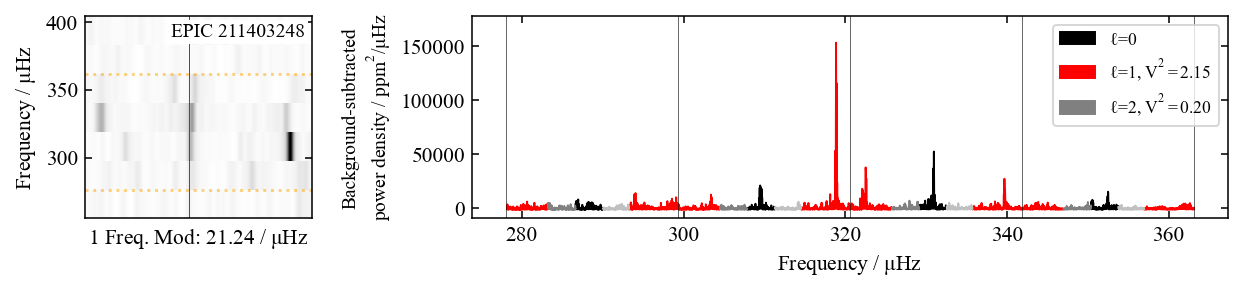}
    \includegraphics[width=.83\textwidth]{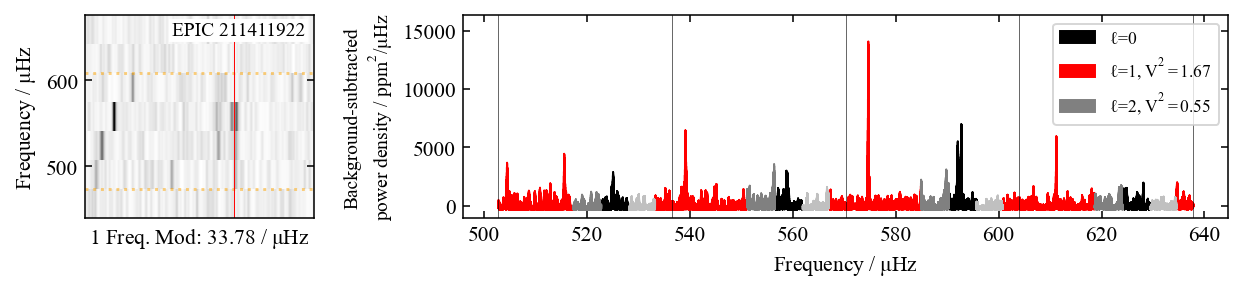}
    
    \caption{Echelle diagrams and power spectra centred around \numax\ of red-giant-branch stars with "normal" $\ell$=1 visibilities or having $\ell$=1 above the fiducial line shown in Figure \ref{fig:vis}. In the echelle diagram $\epsilon$ is indicated with a thin red line. In the power spectra the y-axis corresponds to background-subtracted power density, and frequency ranges corresponding to each mode degree $\ell$ are colour coded as per the legend, while the ranges where no power is expected ($\ell$=3) are shown in silver.}
    \label{fig:suppression1}
\end{center}
\end{figure*}

\begin{figure*}
\begin{center}
    \includegraphics[width=.83\textwidth]{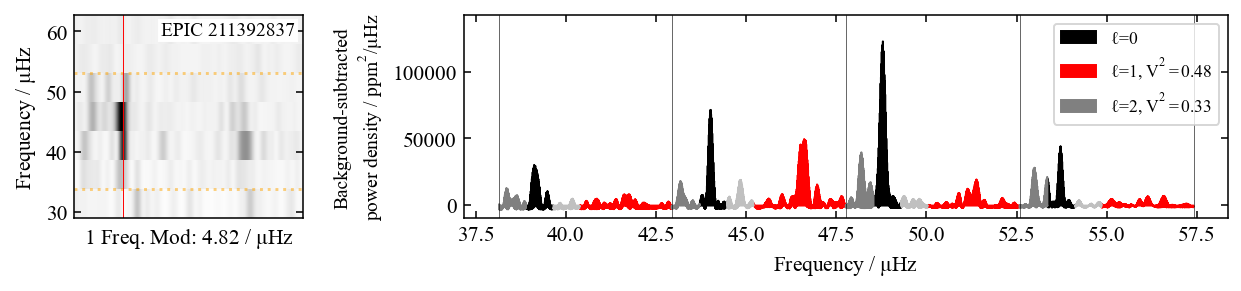}
    \includegraphics[width=.83\textwidth]{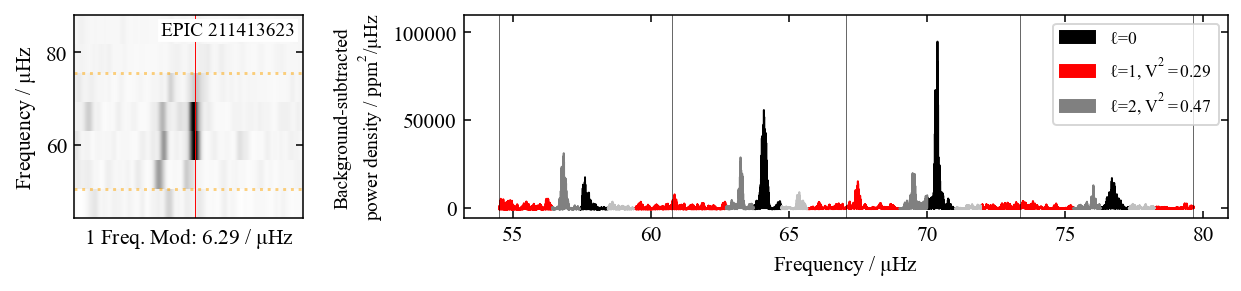}
    \includegraphics[width=.83\textwidth]{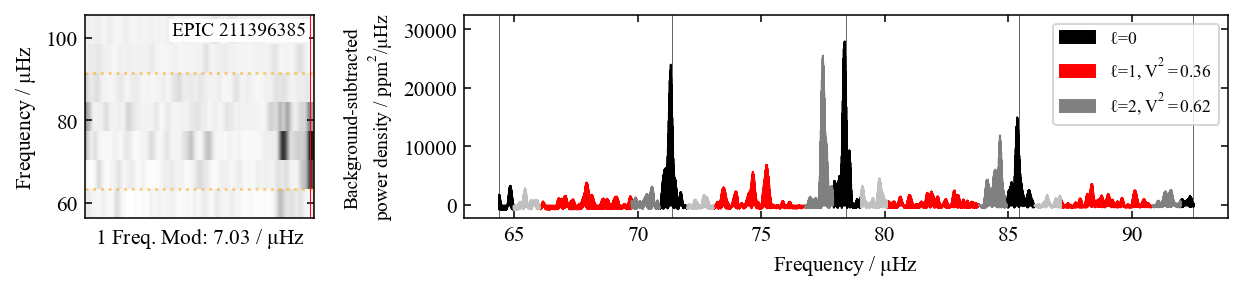}
    \includegraphics[width=.83\textwidth]{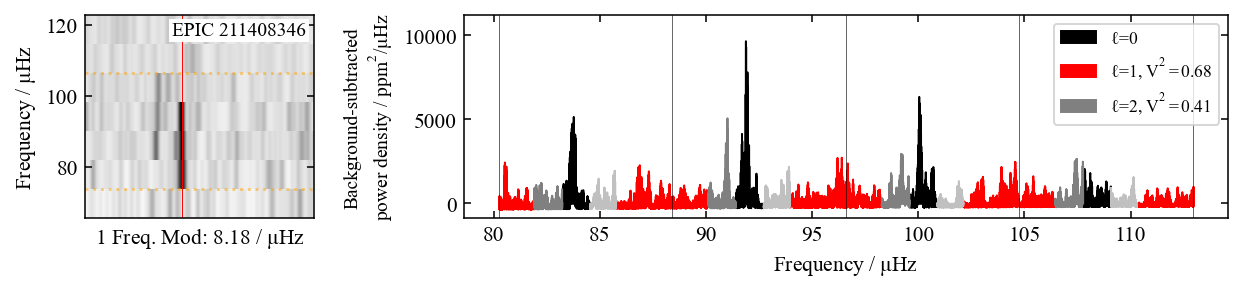}
    \includegraphics[width=.83\textwidth]{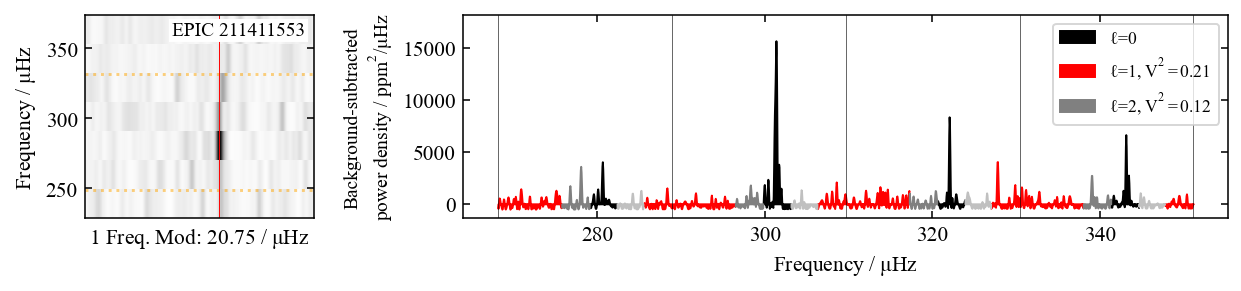}
    \includegraphics[width=.83\textwidth]{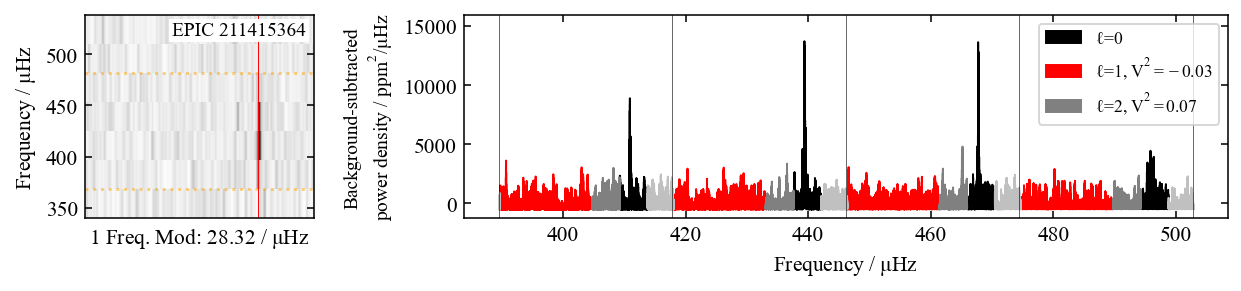}
    \caption{As Figure \ref{fig:suppression1}, but showing red-giant-branch stars with suppressed $\ell$=1 modes, or having $\ell$=1 visibility below the fiducial line shown in Figure \ref{fig:vis}.
    }
\label{fig:suppression2}
\end{center}
\end{figure*}

We present examples of visibility measurements presented in Figure \ref{fig:vis}. Figure \ref{fig:suppression1} shows red-giant-branch stars of normal $\ell$=1 visibility, and Figure \ref{fig:suppression2} shows examples of $\ell$=1-suppressed red-giant-branch stars.

\section{Stars with large phase-term diferences: cubic vs. inverse-cubic}
\begin{figure*}
\begin{center}
    \includegraphics[width=.9\textwidth]{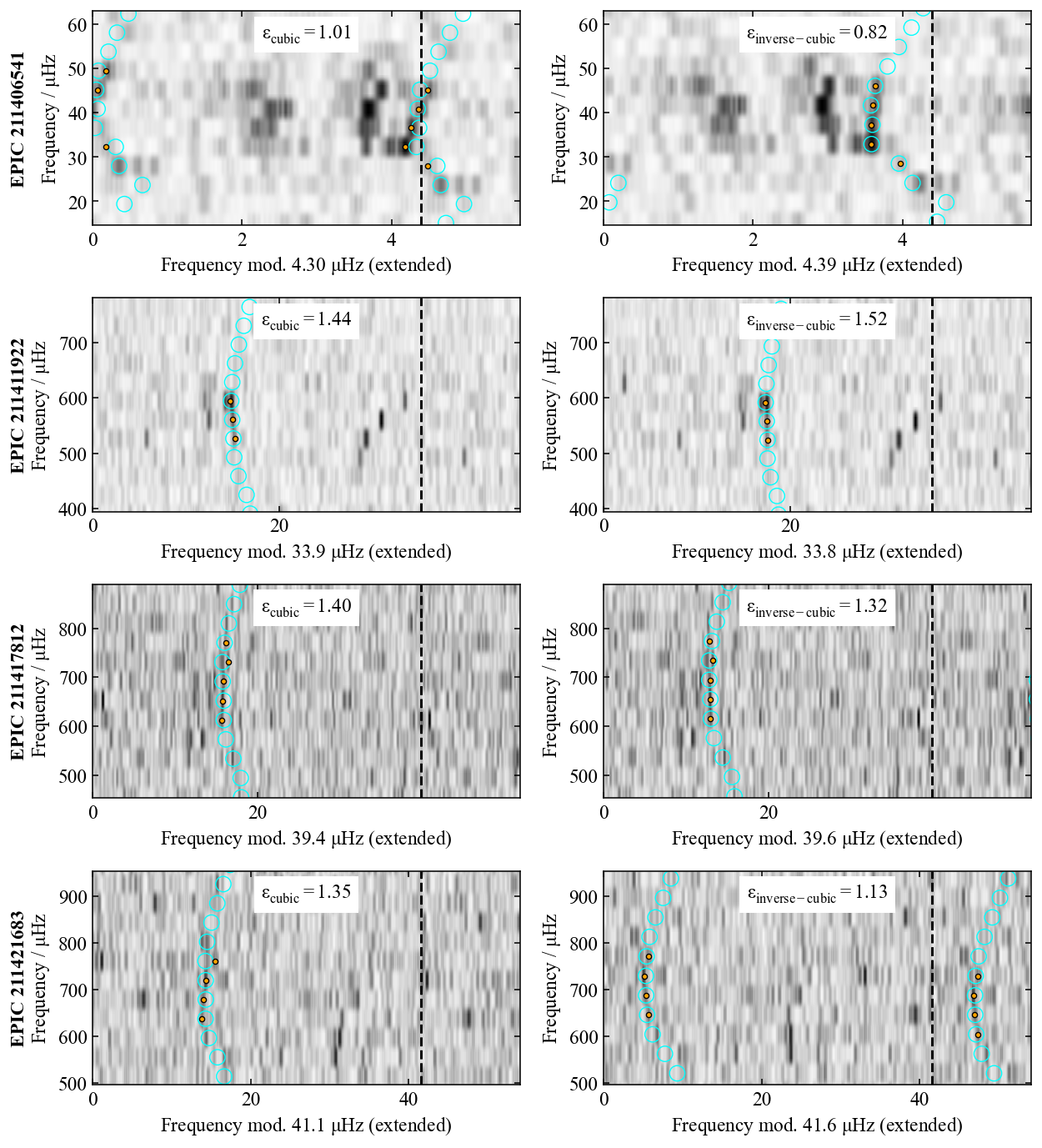}
    \caption{The figure shows the extended echelle diagrams for the four stars with the largest offsets in $\epsilon$ between the pure cubic (left) and inverse-cubic (right) corrected models. Observed radial mode frequencies are highlighted in orange, and surface-corrected model frequencies are marked with cyan circles. For each star, the diagrams use surface-corrected $\Delta\nu$ from the same models, applying either the pure cubic or inverse-cubic correction as annotated.} 
    \label{fig:cubic_vs_invcubic}
\end{center}
\end{figure*}

In \autoref{fig:cubic_vs_invcubic}, we present echelle diagrams for four stars, selected because they exhibit the largest differences in $\epsilon$ between their cubic and inverse-cubic surface-corrected models within the sample. Each row corresponds to the same star, with the echelle diagrams constructed using either the cubic or inverse-cubic surface-corrected \dnu\ and displaying the corresponding surface-corrected modeled frequencies, as indicated.

For star 211406541, neither the cubic nor the inverse-cubic correction appears optimal. This mainly arises because the number of observed modes is too low to describe the strong curvature in the radial ridge of the model with high confidence.  As a result, the cubic fit (left) does not align well with the observed modes (in yellow), particularly around 30-40\muhz. Although the inverse-cubic fit (right) does align with the observed modes, the reliability of the fit is questionable because the resulting $\epsilon$ value falls far from the smooth cluster sequence in \autoref{fig:phaseterm}.

For star 211411922, it is also challenging to determine which is the better correction, this time because both models fit the observed frequencies reasonably well. Here, only three radial modes were identified and used for the fits, resulting in larger uncertainties and hence different $\epsilon$ values.

For stars 211417812 and 211421683, the higher-order radial modes in the inverse-cubic corrected models tend to fall in regions with no power beyond the fitted modes. For the cubic-corrected models, those modes coincide with potential low-amplitude modes (not highlighted in yellow).
These interpretations are speculative, because the signals under consideration are statistically insignificant.

\section{Echelle diagrams: stars and their best-fit grid models}

\begin{figure*}
\begin{center}
    \includegraphics[width=\textwidth]{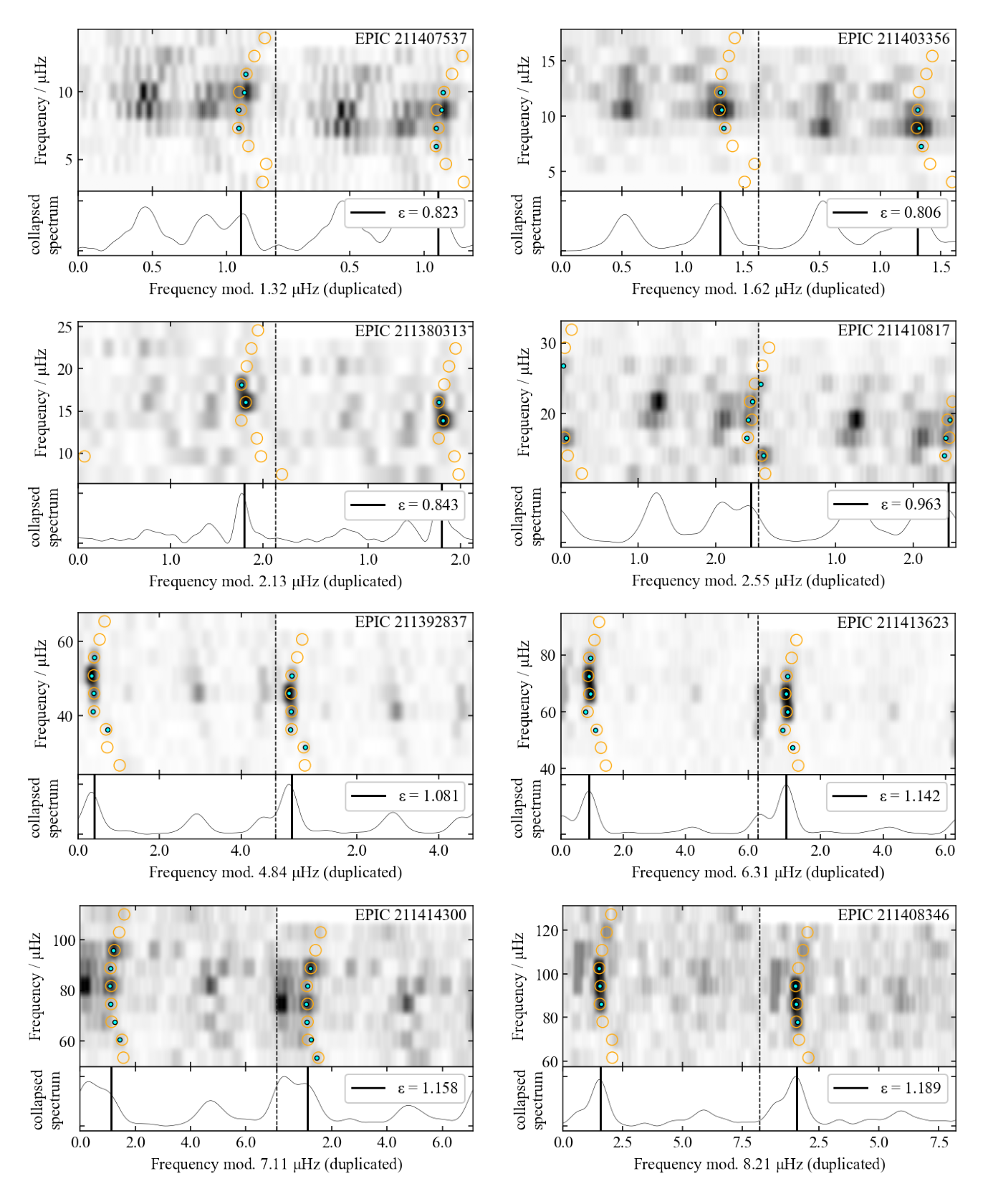}
    \caption{Duplicated echelle diagrams from our M67 sample used in the final grid fit (stars 1-8 of 28). Modelled radial modes from  each star's best-fit model ( \autoref{tab:fitresults}) are represented with open circles. Smaller cyan symbols represent the peakbagged radial modes (\autoref{tab:modefreqs}). A vertically collapsed version of each echelle diagram is shown at the bottom, and a vertical black line indicates the phase term $\epsilon$ calculated from the surface corrected model. The power spectra have been smoothed down for clarity of the image.}
    \label{fig:echelles1}
\end{center}
\end{figure*}

\begin{figure*}
\begin{center}
    \includegraphics[width=\textwidth]{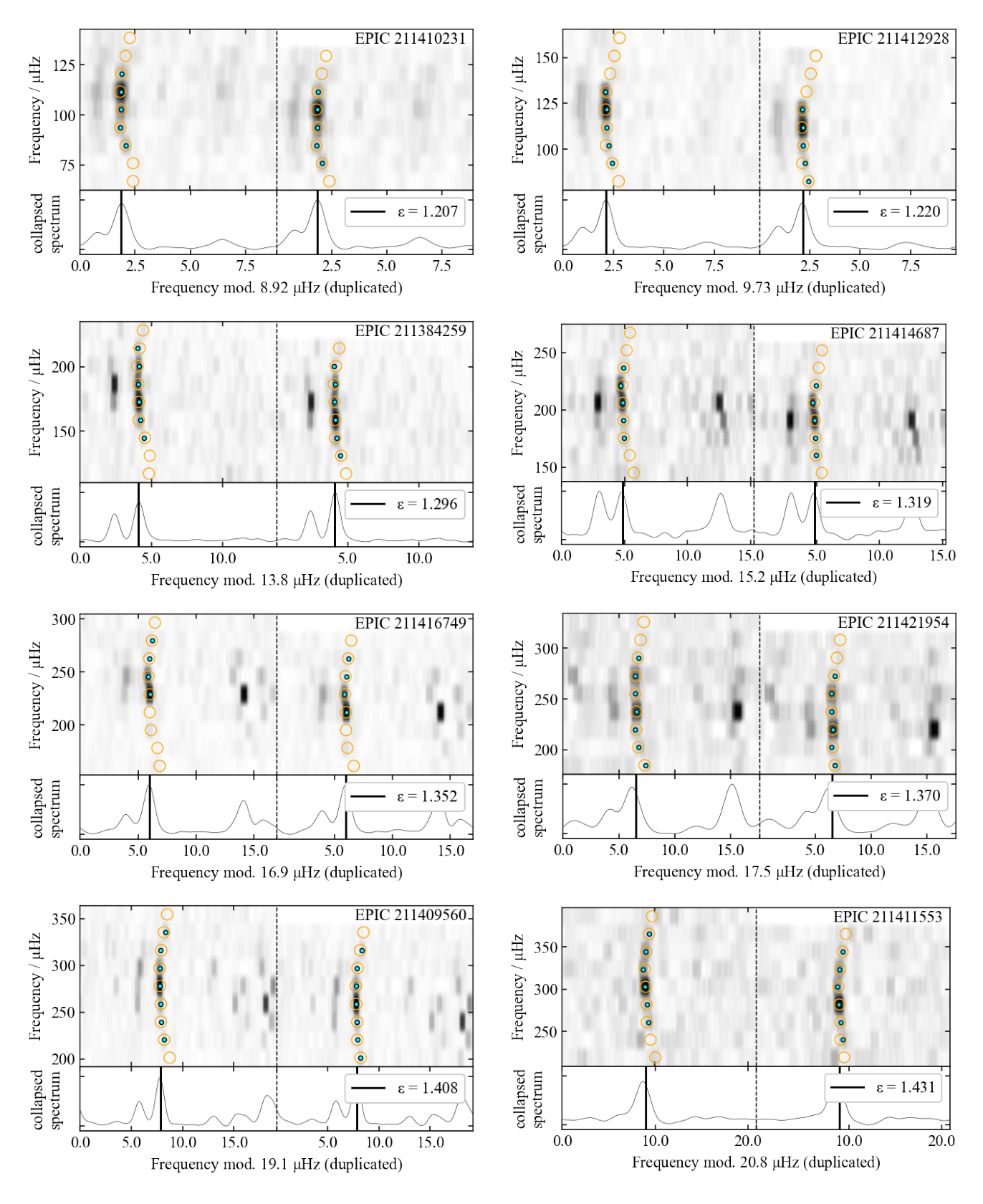}
    \caption{Duplicated echelle diagrams from our M67 sample used in the final grid fit (stars 9-16 of 28). Modelled radial modes from  each star's best-fit model ( \autoref{tab:fitresults}) are represented with open circles. Smaller cyan symbols represent the peakbagged radial modes (\autoref{tab:modefreqs}). A vertically collapsed version of each echelle diagram is shown at the bottom, and a vertical black line indicates the phase term $\epsilon$ calculated from the surface corrected model. The power spectra have been smoothed down for clarity of the image.}
    \label{fig:echelles2}
\end{center}
\end{figure*}

\begin{figure*}
\begin{center}
    \includegraphics[width=\textwidth]{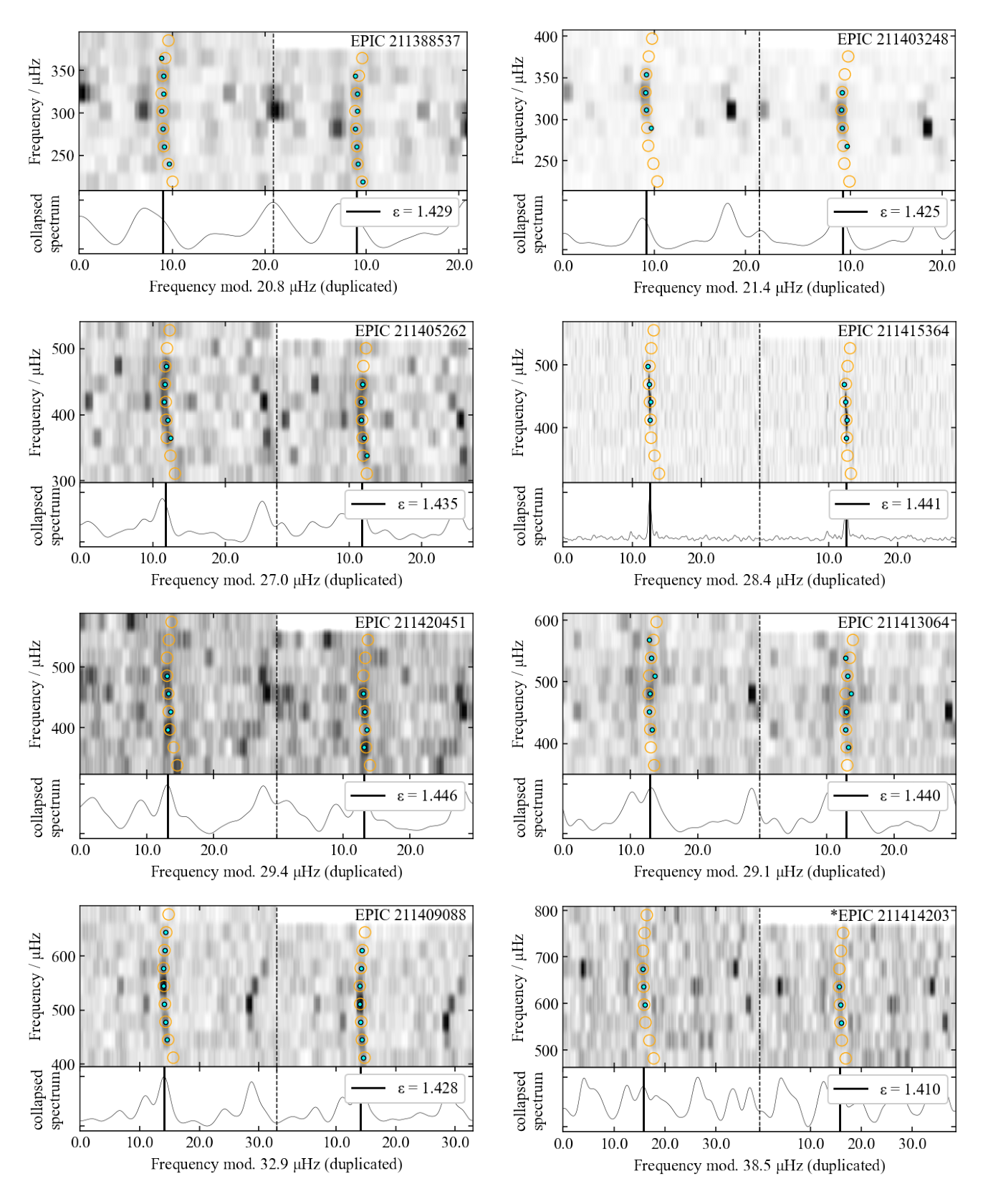}
    \caption{Duplicated echelle diagrams from our M67 sample used in the final grid fit (stars 17-24 of 28). Modelled radial modes from  each star's best-fit model ( \autoref{tab:fitresults}) are represented with open circles. Smaller cyan symbols represent the peakbagged radial modes (\autoref{tab:modefreqs}). A vertically collapsed version of each echelle diagram is shown at the bottom, and a vertical black line indicates the phase term $\epsilon$ calculated from the surface corrected model. The power spectra have been smoothed down for clarity of the image.}
    \label{fig:echelles3}
\end{center}
\end{figure*}

\begin{figure*}
\begin{center}
    \includegraphics[width=\textwidth]{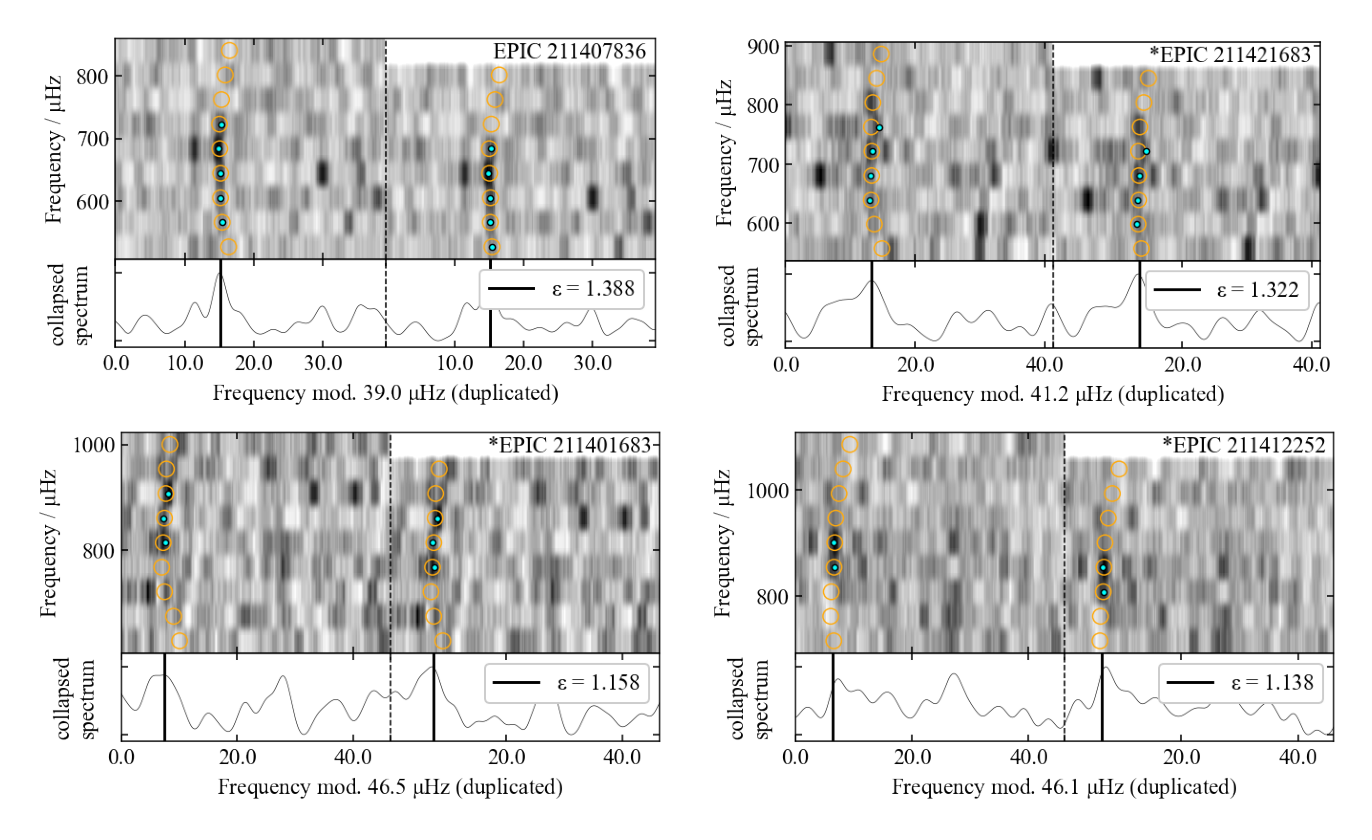}
    \caption{Duplicated echelle diagrams from our M67 sample used in the final grid fit (stars 25-28 of 28). Modelled radial modes from  each star's best-fit model ( \autoref{tab:fitresults}) are represented with open circles. Smaller cyan symbols represent the peakbagged radial modes (\autoref{tab:modefreqs}). A vertically collapsed version of each echelle diagram is shown at the bottom, and a vertical black line indicates the phase term $\epsilon$ calculated from the surface corrected model. The power spectra have been smoothed down for clarity of the image.}
    \label{fig:echelles4}
\end{center}
\end{figure*}

\section{Red clump stars}
\label{sec:redclump}

\begin{figure*}
\begin{center}
    \includegraphics[width=.83\textwidth]{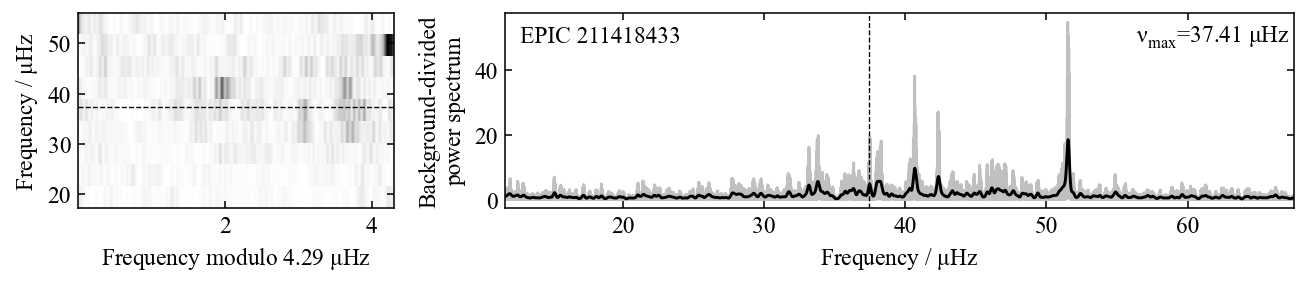}
    \includegraphics[width=.83\textwidth]{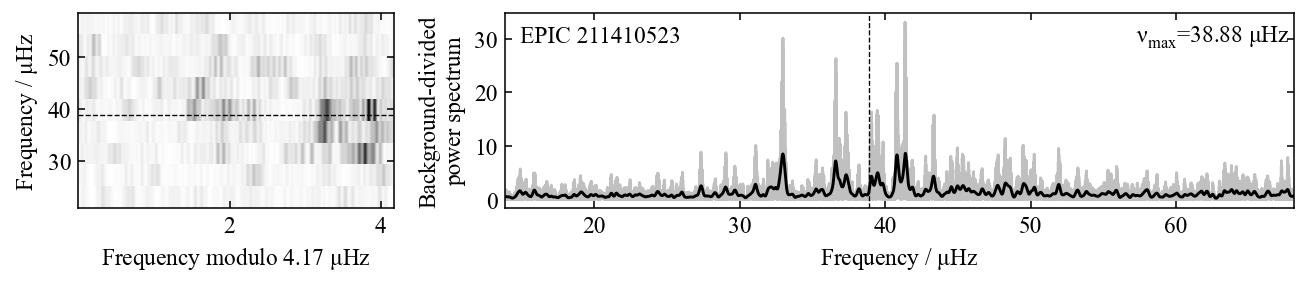}
    \includegraphics[width=.83\textwidth]{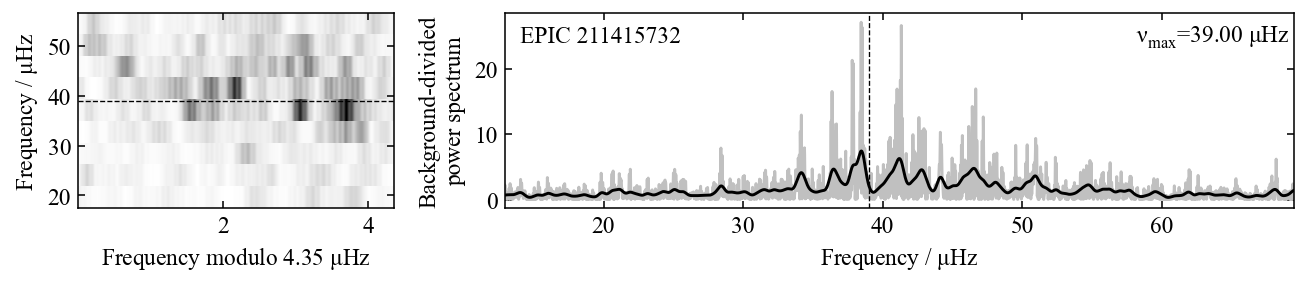}
    \includegraphics[width=.83\textwidth]{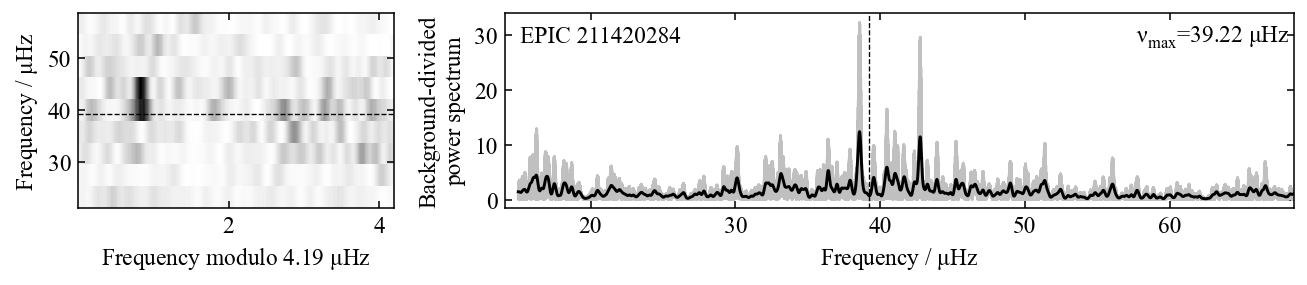}
    \includegraphics[width=.83\textwidth]{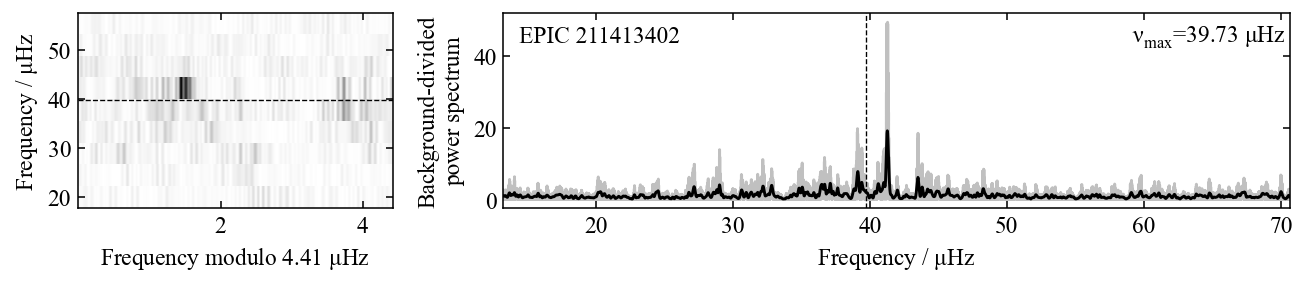}
    \includegraphics[width=.83\textwidth]{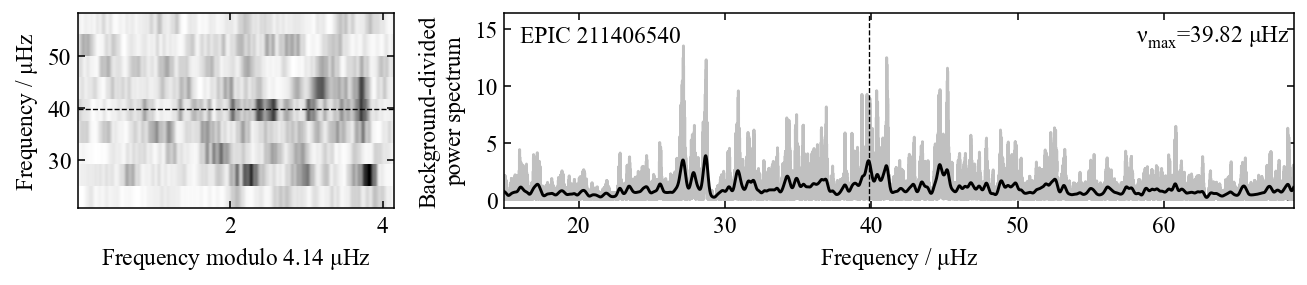}
    \includegraphics[width=.83\textwidth]{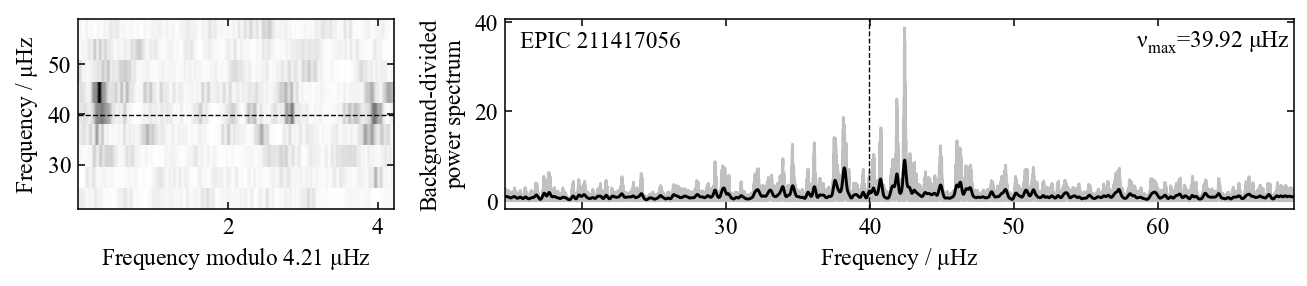}
    \caption{Echelle diagrams and power spectra of M67 red clump stars centred on their frequency of maximum oscillation power, indicated by a dashed black line on every panel. The radial ridge in the echelle diagrams is close to 4 \muhz\ (right side of the diagram) in all cases, except in the last star, EPIC 211417056, where the radial ridge is close to 0 (left side).}
    \label{fig:rc}
\end{center}
\end{figure*}

\bsp	
\label{lastpage}
\end{document}